\documentclass[10pt]{article}

\usepackage{a4wide,amsmath,amssymb,mathrsfs,verbatim,pgf,color}

\title{\textbf{On the construction of Hartle-Hawking-Israel states across a static
bifurcate Killing horizon}}
\author{Ko Sanders\thanks{E-mail: ko.sanders@itp.uni-leipzig.de}\\
Institut f\"ur Theoretische Physik\\ Universit\"at Leipzig\\
Br\"uderstra{\ss}e 16\\ D-04103 Leipzig}
\date{29 January 2015}

\newtheorem{definition}{Definition}[section]
\newtheorem{theorem}[definition]{Theorem}
\newtheorem{proposition}[definition]{Proposition}
\newtheorem{corollary}[definition]{Corollary}
\newtheorem{lemma}[definition]{Lemma}
\newenvironment{proof*}{\smallskip\par\noindent\emph{Proof: }
 \ignorespaces}{\hfill$\Box$\smallskip\par\ignorespaces}
\newtheorem{remark}[definition]{Remark}

\newcommand{\map}[3]{\ensuremath{#1\!:\!#2\!\rightarrow\!#3}}

\newcommand{\id}{\ensuremath{\mathrm{id}}}

\newcommand{\alg}[1]{\ensuremath{\mathcal{#1}}}

\begin{document}
\maketitle

\begin{abstract}
We consider a linear scalar quantum field propagating in a spacetime of
dimension $d\ge 2$ with a static bifurcate Killing horizon and a wedge
reflection. Under suitable conditions (e.g.\ positive mass) we prove the
existence of a pure Hadamard state which is quasi-free, invariant under the
Killing flow and restricts to a double $\beta_H$-KMS state on the union
of the exterior wedge regions, where $\beta_H$ is the inverse Hawking
temperature.

The existence of such a state was first conjectured by Hartle and Hawking
(1976) and by Israel (1976), in the more general case of a stationary black
hole spacetime. Jacobson (1994) has conjectured a similar state to exist
even for interacting fields in spacetimes with a static bifurcate Killing
horizon. The state can serve as a ground state on the entire spacetime
and the resulting situation generalises that of the Unruh effect in
Minkowski spacetime.

Our result complements a well-known uniqueness result of Kay and Wald
(1991) and Kay (1993), who considered a general bifurcate Killing horizon
and proved that a certain (large) subalgebra of the free field admits at
most one Hadamard state which is invariant under the Killing flow. This
state is pure and quasi-free and in the presence of a wedge reflection it
restricts to a $\beta_H$-KMS state on the smaller subalgebra associated to
one of the exterior wedge regions. Our result establishes the existence of
such a state on the full algebra, but only in the static case.

Our proof follows the arguments of Sewell (1982) and Jacobson (1994),
who exploited a Wick rotation in the Killing time coordinate to
construct a corresponding Euclidean theory. In particular we show that
for the linear scalar field we can recover a Lorentzian theory by Wick
rotating back. Because the Killing time coordinate is ill-defined on
the bifurcation surface we systematically replace it by a Gaussian
normal coordinate. A crucial part of our proof is to establish that
the Euclidean ground state satisfies the necessary analogues of
analyticity and reflection positivity with respect to this coordinate.
\end{abstract}

\section{Introduction}\label{Sec_Intro}

The equations that describe black hole physics have an uncanny similarity to the
laws of thermodynamics. This fact was gradually realised in the 1970s, starting
with the black hole area law and culminating in Hawking's discovery of black hole
radiation \cite{Bardeen+1973,Hawking1971,Hawing1974,Hawking1975,Hawking1975err,Wald1975}.
In the past few decades, much work has been devoted to investigating the fundamental
physics that underlies these striking similarities, which go under the name of
black hole thermodynamics \cite{Wald1}.

A closely related research effort has developed a rigorous mathematical framework
to describe quantum field theory (QFT) in general curved spacetimes in a generally
covariant way \cite{Brunetti+2003,Hollands+2001,Hollands+2008}. Largely motivated
by the desire to formulate the questions of black hole thermodynamics in a
precise and general setting, this area of research also has ramifications for our
wider understanding of QFT and its interaction with gravity. One of the main
breakthroughs in the development of generally covariant QFT was the introduction
of microlocal analysis as a mathematical tool to study and characterise the
singularities of $n$-point distributions of a quantum field \cite{Radzikowski1996}.
This has led to an easier and more illuminating characterisation of the important
class of Hadamard states.

In this paper, we will take advantage of these insights in QFT on curved
spacetimes and apply them to one of the questions of black hole thermodynamics.
Our goal is to prove the existence of a ground state for a linear scalar
quantum field that propagates in a spacetime with a static black hole (or a
generalisation thereof). The existence of the ground state in question was
first conjectured by Hartle and Hawking \cite{Hartle+1976} and by Israel
\cite{Israel1976}. They used a Wick rotation to argue that a ground state on
a stationary black hole spacetime can be defined by analytic continuation
from a Euclidean fundamental solution on a corresponding Riemannian
manifold. Whereas Hartle and Hawking were mostly interested in this state on
the physical exterior region of the black hole spacetime, Israel discussed
its extension to the wedge region on the other side of the black hole. A
mathematically rigorous construction of this so-called Hartle-Hawking-Israel
(HHI) state in the exterior regions of Kruskal spacetime was given by Kay
\cite{Kay1985}.

The HHI state was introduced to help understand the phenomenon of black hole
radiation \cite{Hartle+1976}. The black hole spacetime can be used to describe
the end state of the collapse of a massive object and one assumes, for the
sake of argument, that the quantum field will settle down in the HHI state,
which is the ground state. The fact that the restriction of the HHI state to
the physical exterior wedge is a thermal state at the Hawking temperature could
then be interpreted as the existence of Hawking radiation. In this way, the
HHI state establishes an interesting connection between black hole geometry
and thermality at the Hawking temperature. Moreover, the model is much simpler
than the more realistic description of \cite{Hawking1975,Wald1975,Fredenhagen+1990},
which describes the collapsing matter as a dynamical process. Unfortunately,
this argument is an oversimplification, as pointed out by Kay and Wald
\cite{Kay+1991}. It is sometimes difficult to imagine how a quantum field can
settle down in the HHI state by any physical process. In Kruskal spacetime, for
example, the HHI state shows a very high degree of correlation between the
thermal radiation coming in from past infinity and the state inside the white
hole region. Conversely, the absence of a HHI state would not invalidate the
analysis of \cite{Hawking1975,Wald1975,Fredenhagen+1990} that a black hole
radiates thermally. To understand black hole radiation as a dynamical process,
one needs a more suitable state, such as the Unruh state
\cite{Unruh1976,Dappiaggi+2011_3}. Moreover, there are recent and reliable
results indicating that Hawking radiation (as measured at future null
infinity) is a global consequence of a local physical phenomenon
(cf.\ \cite{Moretti+2010}). The global arguments involving the HHI state do not
seem appropriate (or even adequate) to address such local questions.

A further issue with the simplified model, which potentially undermines its
accuracy as a physical approximation, is the question whether the effect of
the quantum field on the metric can be neglected. In the light of the
semi-classical Einstein equation, one can justify this approximation by
showing that the expected (renormalised) energy density in the HHI state
remains bounded, so that large back reaction effects are avoided. In the
exterior regions this follows from the fact that the HHI state is Hadamard
(together with the generally covariant Hadamard regularisation scheme,
cf.\ \cite{Hollands+2001}). If the state is also Hadamard near the horizon,
or even just near the bifurcation surface, then this remains true throughout
the future and past regions, due to the propagation of singularities
(cf.\  \cite{Radzikowski1996,Duistermaat+1972}). However, the analysis near
the black hole horizon is more complicated.

The question whether the HHI state can be extended across the horizon of a
black hole spacetime was first addressed in a seminal paper by Kay and Wald
\cite{Kay+1991} (see also \cite{Kay1993} for an improved result). This paper
is remarkable, not only because of the uniqueness theorem that it proves,
but also because the assumptions of this theorem forced the authors to
introduce and refine several important notions. This includes the definition
of global Hadamard states and a criterion when a quasi-free Hadamard state
is pure. Furthermore, they gave a general description of the class of
spacetimes with a bifurcate Killing horizon (see also \cite{Boyer1969}), which
includes the non-extremal stationary black holes as well as Minkowski
spacetime with the Killing field of constantly accelerated observers (as it
appears in the Unruh effect \cite{Unruh1976}). The main result for a spacetime
with a bifurcate Killing horizon is that a certain subalgebra of the free
field algebra admits at most one state which is invariant under the Killing
field and Hadamard across the Killing horizon. Moreover, if the spacetime
admits a wedge reflection, then the restriction of this state to the physical
exterior wedge is a thermal (KMS) state at the Hawking temperature.

Unfortunately, the existence of such a state was not proved in \cite{Kay+1991}.
Besides, at a more technical level, the specification of the subalgebra
featuring in the uniqueness result is somewhat subtle, as it involves the
initial value problem of the Klein-Gordon equation on a null hypersurface (the
so-called Goursat, or characteristic Cauchy, problem). The null hypersurface
in question is a part of the Killing horizon, $\mathfrak{h}_A$, and Kay and
Wald consider solutions on the spacetime $M$ whose restriction to
$\mathfrak{h}_A$ is a given test-function $f\in C_0^{\infty}(\mathfrak{h}_A)$.
For the existence and uniqueness of such solutions, they refer to results and
techniques in \cite{Friedlander} and they recognise in a note added in proof
that such solutions may fail to be smooth across $\mathfrak{h}_A$.
Unfortunately, the only results proved in \cite{Friedlander} are of a local
nature and they apply only to null hypersurfaces which are the future null
cones of $\partial J^+(p)$ of some point $p$. It is to be expected that these
shortcomings can be overcome by a more detailed analysis of the Goursat
problem, e.g.\ along the lines of H\"ormander's remark \cite{Hoermander1990},
which seems to have gone unnoticed in much of the mathematical physics
literature. Such a more detailed analysis could also help to further
substantiate the claim of \cite{Kay+1991} that these solutions always
generate a large subalgebra of the Weyl algebra (see also footnote
\ref{ftn_KW} on page \pageref{ftn_KW}).

Making use of the notions and results of Kay and Wald, Jacobson
\cite{Jacobson1994} has argued that the original construction of HHI states
via a Wick rotation should work even across a bifurcate Killing horizon, at
least if this Killing horizon is static. Moreover, this construction should
also work for interacting QFT's. Earlier, Sewell had advanced similar
arguments to define the HHI state for interacting theories on the physical
exterior wedge only \cite{Sewell1980,Sewell1982}. In his sketch of a proof
Jacobson constructs a Euclidean theory on the associated Riemannian manifold
using path integral methods. He points out several properties of the
geometry that make it plausible that this theory can be Wick rotated back to
define a Lorentzian theory with a ground state. However, some doubt is cast
on this claim by the fact that the analytic continuation is defined in terms
of the Killing time coordinate, which is ill-defined at the bifurcation
surface. A detailed investigation near the bifurcation surface is therefore
necessary.

The purpose of this paper is to provide a mathematically complete and
rigorous construction of the HHI state for a linear scalar field, along the
lines set out by Jacobson. We will systematically replace the Killing time
coordinate by a Gaussian normal coordinate and we establish that the
Euclidean fundamental solution $G$ satisfies the necessary analogues of
analyticity and reflection positivity with respect to this coordinate. This
will lead to an HHI state, which we show to be pure, invariant under the Killing
flow and to restrict to a double $\beta_H$-KMS state in the exterior wedge
regions. At present, it is unclear whether our existence proof extends to
(perturbatively) interacting theories, e.g.\ using the arguments of
\cite{Gibbons+1978_2}. We will not investigate this question in detail,
nor will we consider fields with spin.

In general, analyticity of $G$ in the Gaussian normal coordinate may only
hold in an infinitesimal sense. By this we mean that the Cauchy Riemann
equations hold only when restricted to a hypersurface $\Sigma$, which can be
identified as a Cauchy surface for the Lorentzian spacetime. It follows that
the HHI state cannot be defined directly by analytic continuation in the
Gaussian normal coordinate, but we can use the Euclidean fundamental solution
$G$ to define initial data on the Cauchy surface, which in turn define the
HHI state. Similarly, the Hadamard property for the HHI state across the
Killing horizon does not follow from the fact that it is a boundary value of
an analytic function, but it must be established by investigating the
initial data on the Cauchy surface $\Sigma$. For this reason we have
included detailed results on the comparison between the geometry and the
Hadamard construction of both the Lorentzian spacetime and its Riemannian
counterpart near the surface $\Sigma$.

Our paper is organised as follows. In Section \ref{Sec_Geometry} we collect
all the geometric results that we need, including the analytic continuation.
In Section \ref{Sec_FField} we review the necessary theory of the linear
scalar field and its Wick rotation w.r.t.\ the Killing time parameter, which
leads to double $\beta$-KMS states on the exterior wedges. Section
\ref{Sec_Had} contains the details of the Hadamard construction in both the
Lorentzian and the Euclidean setting. One technical lemma has been deferred
to appendix \ref{App}. Section \ref{Sec_HHI} combines all these ingredients
to prove the existence of the HHI state across the Killing horizon and to
establish its main properties, namely its purity, invariance and the
$\beta_H$-KMS restriction.

\section{Geometric results}\label{Sec_Geometry}

A careful study of the behaviour of a quantum field near a bifurcate Killing
horizon requires a detailed understanding of the differential geometry of
the underlying spacetime. It is the purpose of this section to introduce the
class of spacetimes that we shall study and to present their relevant
features, referring the reader to the literature for proofs of known results.
Because our spacetimes of interest often have an exterior region which is
stationary or standard static, we refer in particular to the review
\cite{Sanders2012}, which describes thermal states for such spacetimes.

Our main technical tool for the purposes of this paper is contained in
Subsection \ref{SSec_infAC}, where we confront the problem that the Killing
time coordinate, used to define the analytic continuation in the static
case, breaks down at the bifurcation surface. We circumvent this problem by
introducing Gaussian normal coordinates near a suitable Cauchy surface and
by proving that all the relevant geometric quantities satisfy a certain
infinitesimal version of the Cauchy-Riemann equations w.r.t.\ these
coordinates. In addition, we consider Riemannian normal coordinates, which
are used to obtain the simplest coordinate expression for the Hadamard
series, and we express them in terms of the Gaussian normal coordinates.
These technical results will be crucial when showing that a double $\beta$-KMS
state at the Hawking temperature can be extended as a Hadamard state across
the Killing horizon.

Throughout this paper, we will use the following standard terminology:
\begin{definition}\label{Def_(GH)Spac}
By a \emph{spacetime} $M=(\mathcal{M},g_{ab})$ we will mean a smooth, oriented
manifold $\mathcal{M}$ of dimension $d\ge 2$ with a smooth Lorentzian metric
$g_{ab}$ of signature $(-+\ldots+)$.

A \emph{Cauchy surface} $\Sigma$ in $M$ is a subset $\Sigma\subset M$ that
is intersected exactly once by every inextendible timelike curve in $M$.
A spacetime is said to be \emph{globally hyperbolic} when it has a Cauchy
surface $\Sigma$.
\end{definition}
We adopt the convention that a spacetime is also connected, unless stated
otherwise. We are mainly interested in globally hyperbolic spacetimes and
we will only consider Cauchy surfaces that are smooth, spacelike
hypersurfaces \cite{Bernal+2003}. A globally hyperbolic spacetime is
automatically time orientable and we will always assume a choice of
time orientation has been fixed. It follows that any Cauchy surface
$\Sigma$ inherits a natural orientation. We let $h_{ij}$ denote the
Riemannian metric on $\Sigma$ induced by the Lorentzian metric $g_{ab}$ on
$M$.

\subsection{Spacetimes with a bifurcate Killing horizon}\label{SSec_StaticBH}

\begin{figure}[t!]
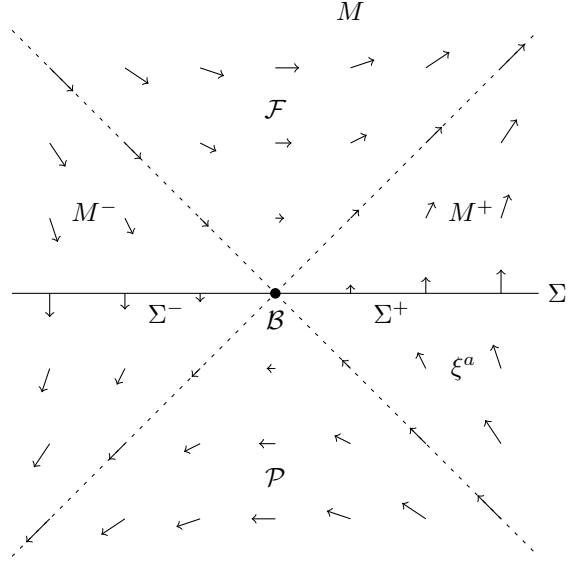

\begin{center}
\begin{pgfpicture}{0cm}{0cm}{7.5cm}{7.5cm}
\color{black}
\pgfline{\pgfxy(0,3.5)}{\pgfxy(7,3.5)}
\pgfsetdash{{0.05cm}{0.1cm}}{0cm}
\pgfline{\pgfxy(0,0)}{\pgfxy(7,7)}
\pgfline{\pgfxy(0,7)}{\pgfxy(7,0)}
\pgfsetdash{{1cm}{0cm}}{0cm}
\pgfcircle[fill]{\pgfxy(3.5,3.5)}{0.07cm}
\pgfsetendarrow{\pgfarrowto}
\pgfline{\pgfxy(6.5,0.5)}{\pgfxy(6.2,0.8)}
\pgfline{\pgfxy(5.5,0.5)}{\pgfxy(5.2,0.7)}
\pgfline{\pgfxy(4.5,0.5)}{\pgfxy(4.2,0.6)}
\pgfline{\pgfxy(3.5,0.5)}{\pgfxy(3.2,0.5)}
\pgfline{\pgfxy(2.5,0.5)}{\pgfxy(2.2,0.4)}
\pgfline{\pgfxy(1.5,0.5)}{\pgfxy(1.2,0.3)}
\pgfline{\pgfxy(0.5,0.5)}{\pgfxy(0.2,0.2)}

\pgfline{\pgfxy(6.5,1.5)}{\pgfxy(6.3,1.8)}
\pgfline{\pgfxy(5.5,1.5)}{\pgfxy(5.3,1.7)}
\pgfline{\pgfxy(4.5,1.5)}{\pgfxy(4.3,1.6)}
\pgfline{\pgfxy(3.5,1.5)}{\pgfxy(3.3,1.5)}
\pgfline{\pgfxy(2.5,1.5)}{\pgfxy(2.3,1.4)}
\pgfline{\pgfxy(1.5,1.5)}{\pgfxy(1.3,1.3)}
\pgfline{\pgfxy(0.5,1.5)}{\pgfxy(0.3,1.2)}

\pgfline{\pgfxy(6.5,2.5)}{\pgfxy(6.4,2.8)}
\pgfline{\pgfxy(5.5,2.5)}{\pgfxy(5.4,2.7)}
\pgfline{\pgfxy(4.5,2.5)}{\pgfxy(4.4,2.6)}
\pgfline{\pgfxy(3.5,2.5)}{\pgfxy(3.4,2.5)}
\pgfline{\pgfxy(2.5,2.5)}{\pgfxy(2.4,2.4)}
\pgfline{\pgfxy(1.5,2.5)}{\pgfxy(1.4,2.3)}
\pgfline{\pgfxy(0.5,2.5)}{\pgfxy(0.4,2.2)}

\pgfline{\pgfxy(6.5,3.5)}{\pgfxy(6.5,3.8)}
\pgfline{\pgfxy(5.5,3.5)}{\pgfxy(5.5,3.7)}
\pgfline{\pgfxy(4.5,3.5)}{\pgfxy(4.5,3.6)}
\pgfline{\pgfxy(2.5,3.5)}{\pgfxy(2.5,3.4)}
\pgfline{\pgfxy(1.5,3.5)}{\pgfxy(1.5,3.3)}
\pgfline{\pgfxy(0.5,3.5)}{\pgfxy(0.5,3.2)}

\pgfline{\pgfxy(6.5,4.5)}{\pgfxy(6.6,4.8)}
\pgfline{\pgfxy(5.5,4.5)}{\pgfxy(5.6,4.7)}
\pgfline{\pgfxy(4.5,4.5)}{\pgfxy(4.6,4.6)}
\pgfline{\pgfxy(3.5,4.5)}{\pgfxy(3.6,4.5)}
\pgfline{\pgfxy(2.5,4.5)}{\pgfxy(2.6,4.4)}
\pgfline{\pgfxy(1.5,4.5)}{\pgfxy(1.6,4.3)}
\pgfline{\pgfxy(0.5,4.5)}{\pgfxy(0.6,4.2)}

\pgfline{\pgfxy(6.5,5.5)}{\pgfxy(6.7,5.8)}
\pgfline{\pgfxy(5.5,5.5)}{\pgfxy(5.7,5.7)}
\pgfline{\pgfxy(4.5,5.5)}{\pgfxy(4.7,5.6)}
\pgfline{\pgfxy(3.5,5.5)}{\pgfxy(3.7,5.5)}
\pgfline{\pgfxy(2.5,5.5)}{\pgfxy(2.7,5.4)}
\pgfline{\pgfxy(1.5,5.5)}{\pgfxy(1.7,5.3)}
\pgfline{\pgfxy(0.5,5.5)}{\pgfxy(0.7,5.2)}

\pgfline{\pgfxy(6.5,6.5)}{\pgfxy(6.8,6.8)}
\pgfline{\pgfxy(5.5,6.5)}{\pgfxy(5.8,6.7)}
\pgfline{\pgfxy(4.5,6.5)}{\pgfxy(4.8,6.6)}
\pgfline{\pgfxy(3.5,6.5)}{\pgfxy(3.8,6.5)}
\pgfline{\pgfxy(2.5,6.5)}{\pgfxy(2.8,6.4)}
\pgfline{\pgfxy(1.5,6.5)}{\pgfxy(1.8,6.3)}
\pgfline{\pgfxy(0.5,6.5)}{\pgfxy(0.8,6.2)}

\pgfstroke
\pgfclearendarrow
\pgfputat{\pgfxy(3.5,3.15)}{\pgfbox[center,center]{$\mathcal{B}$}}
\pgfputat{\pgfxy(2.05,3.25)}{\pgfbox[center,center]{$\Sigma^-$}}
\pgfputat{\pgfxy(5.05,3.25)}{\pgfbox[center,center]{$\Sigma^+$}}
\pgfputat{\pgfxy(7.25,3.5)}{\pgfbox[center,center]{$\Sigma$}}
\pgfputat{\pgfxy(3.5,5.95)}{\pgfbox[center,center]{$\mathcal{F}$}}
\pgfputat{\pgfxy(3.5,1.05)}{\pgfbox[center,center]{$\mathcal{P}$}}
\pgfputat{\pgfxy(6.1,4.6)}{\pgfbox[center,center]{$M^+$}}
\pgfputat{\pgfxy(1.1,4.6)}{\pgfbox[center,center]{$M^-$}}
\pgfputat{\pgfxy(4.5,7.25)}{\pgfbox[center,center]{$M$}}
\pgfputat{\pgfxy(6,2.5)}{\pgfbox[center,center]{$\xi^a$}}
\end{pgfpicture}
\end{center}
\caption{The geometry of a bifurcate Killing horizon, as defined in
Definition \ref{Def_StaticBHSpac}, depicted in a spacetime
diagram.}\label{Fig_Bifurcation}
\end{figure}

We start with the definition of the class of spacetimes that we will
consider and that encompasses in particular the most common models of
black holes.
\begin{definition}\label{Def_StaticBHSpac}
A \emph{spacetime with a bifurcate Killing horizon} is a triple
$M=(\mathcal{M},g_{ab},\xi^a)$ such that
\begin{enumerate}
\item $(\mathcal{M},g_{ab})$ is a globally hyperbolic spacetime,
\item $\xi^a$ is a smooth, complete Killing vector field,
\item $\mathcal{B}:=\left\{x\in M|\ \xi^a(x)=0\right\}$ is a (not necessarily
connected), orientable, $(d-2)$-dimensional smooth submanifold of
$\mathcal{M}$, which is called the \emph{bifurcation surface},
\item there exists a Cauchy surface $\Sigma\subset M$ which contains
$\mathcal{B}$.\footnote{$\mathcal{B}$ is automatically a smooth submanifold
of $\Sigma$.}
\end{enumerate}
By a \emph{spacetime with a stationary}, resp.\ \emph{static}, \emph{bifurcate
Killing horizon} we will mean a spacetime $M$ with a bifurcate Killing horizon
for which $\Sigma$ can be chosen such that the Killing field $\xi^a$ is timelike
on $\Sigma\setminus\mathcal{B}$, resp.\ orthogonal to $\Sigma$.
\end{definition}
Our definition of bifurcate Killing horizons coincides with that of
\cite{Kay+1991}, except that we allow all dimensions $d\ge 2$ and disconnected
bifurcation surfaces $\mathcal{B}$. We refer to Figure \ref{Fig_Bifurcation}
for a depiction of a generic bifurcate Killing horizon and to \cite{Kay+1991}
for a more detailed description of this class of spacetimes.

Completeness of $\xi^a$ means that the corresponding flow
$\map{\Phi}{\mathbb{R}\times\mathcal{M}}{\mathcal{M}}$, defined by
$\Phi(0,x)=x$ and $\partial_t\Phi^a(t,x)|_{t=0}=\xi^a(x)$, yields a
well-defined diffeomorphism
$\map{\Phi_t}{\mathcal{M}}{\mathcal{M}}$ for all $t\in\mathbb{R}$,
defined by $\Phi_t(x):=\Phi(t,x)$. The fact that $\xi^a$ is a Killing
vector field means that $\Phi_t^*g_{ab}=g_{ab}$ for all $t\in\mathbb{R}$,
where $^*$ denotes the pull-back. Equivalently, it can be expressed in
terms of Killing's equation $\nabla_a\xi_b+\nabla_b\xi_a=0$.

From now on, we will assume that the bifurcate Killing horizon is at least
stationary.
Let us fix a Cauchy surface $\Sigma$ with the properties of Definition
\ref{Def_StaticBHSpac} and let $n^a$ denote the future pointing normal
vector field on $\Sigma$. We define the lapse function $v$ and the shift
vector field $w^a$ on $\Sigma$ by
\begin{equation}\label{Eq_Defv}
v:=-n_a\xi^a,\quad w^a:=\xi^a-vn^a,
\end{equation}
which means that $\xi^a=vn^a+w^a$ on $\Sigma$ and $w^a\in T\Sigma$. We
may decompose the Cauchy surface as
\[
\Sigma=\mathcal{B}\cup\Sigma^+\cup\Sigma^-,
\]
where $\Sigma^{\pm}$ are the sets where $\pm v>0$. We define the following
four globally hyperbolic regions of the spacetime $M$: the future
$\mathcal{F}:=I^+(\mathcal{B})$, the past $\mathcal{P}:=I^-(\mathcal{B})$
and the left ($-$) and right ($+$) wedge regions
$M^{\pm}:=D(\Sigma^{\pm})$.\footnote{\label{ftn_KW}These regions are
globally hyperbolic by Lemmas A.5.9 and A.5.12 of \cite{Baer+2007}. In
\cite{Kay+1991} Kay and Wald prefer to define left and right wedge regions
$\mathcal{L}$, $\mathcal{R}$ in terms of the chronological future and past
of portions of the Killing horizons of $\xi^a$. They then impose the
restriction that
$M=\overline{\mathcal{F}}\cup\overline{\mathcal{P}}\cup\mathcal{L}\cup
\mathcal{R}$, from which it follows that $\mathcal{L}=M^-$ and
$\mathcal{R}=M^+$. Note that in our case this restriction is not required,
so the wedge regions $M^{\pm}$ may be strictly larger than $\mathcal{R},\mathcal{L}$.}
Note in particular that $\Sigma^{\pm}$ is a Cauchy surface for $M^{\pm}$
and that we can partition $M$ as
\[
M=M^+\cup M^-\cup\overline{\mathcal{F}}\cup\overline{\mathcal{P}},
\]
where all sets are disjoint, except for
$\overline{\mathcal{F}}\cap\overline{\mathcal{P}}=\mathcal{B}$.
The region $\mathcal{F}$ may contain black holes. More precisely, each
connected component of $\mathcal{B}$ gives rise to a connected component
of $\mathcal{F}$, which, under suitable circumstances, may be a black
hole (cf.\ \cite{Wald} Sec.\ 12.1 for further discussion).

$\Sigma$ is a Cauchy surface on which the Killing field $\xi^a$ is
timelike or $0$, and the right wedge
$M^+=(\mathcal{M}^+,g_{ab}|_{\mathcal{M}^+},\xi^a|_{\mathcal{M}^+})$ is
a (possibly disconnected) stationary spacetime, as is the left wedge
$M^-=(\mathcal{M}^-,g_{ab}|_{\mathcal{M}^-},-\xi^a|_{\mathcal{M}^-})$ if
we change the sign of the Killing field to bring it in line with the
existing time orientation.\footnote{To see why this is the case, one may
pick an arbitrary, future pointing causal vector $v^a$ at an arbitrary
point $x\in M^+$. Let $\gamma$ denote the inextendible geodesic through
$(x,v^a)$ and let $\dot{\gamma}^a$ denote its derivative. Since $\xi^a$ is
a Killing field, the inner product $\xi_a\dot{\gamma}^a$ is constant along
$\gamma$ (cf.\ \cite{Wald} Proposition C.3.1). Note that $\gamma$
intersects $\Sigma$ at some point $y$ (\cite{Wald} Proposition 8.3.4.) and
that $\dot{\gamma}$ is future pointing and causal there, so
$\xi_av^a=\xi_a\dot{\gamma}^a(y)$ is negative. By varying $v^a$ and $x$
it follows that $\xi^a$ must be future pointing and timelike everywhere.}
The metric of $M^+$ can be written in terms of local coordinates on
$\Sigma^+$ and the Killing time coordinate $t$ as
\begin{equation}
g_{\mu\nu}=-v^2(dt^{\otimes 2})_{\mu\nu}+2w_{\mu}\otimes_sdt_{\nu}+h_{\mu\nu},\nonumber
\end{equation}
where $h_{\mu\nu}$ is the $t$-independent Riemannian metric on $\Sigma^+$
induced by $g_{\mu\nu}$ and the lapse $v$ and shift $w^a$ were defined
in Equation (\ref{Eq_Defv}). By the letter $\psi$ we will denote the
diffeomorphism
\begin{equation}\label{Eqn_psi}
\map{\psi}{\mathbb{R}\times(\Sigma\setminus\mathcal{B})}{M^+\cup M^-}:
(t,x)\mapsto \Phi(t,x),
\end{equation}
where we recall that $\Phi$ is the flow of the Killing field $\xi^a$.

If $\Sigma'$ is any other Cauchy surface containing $\mathcal{B}$, then
$\Sigma'\setminus\mathcal{B}$ must be a Cauchy surface for $M^+\cup M^-$
and hence $\xi^a$ is timelike on $\Sigma'\setminus\mathcal{B}$. In other
words, the stationary condition is independent of the choice of Cauchy
surface containing $\mathcal{B}$.

To preserve the logical order of our presentation we now give the
following geometric lemma, which will later reappear in Section
\ref{SSec_infAC}.
\begin{lemma}\label{Lem_CauchyGeodesics}
Let $\Sigma$ be a smooth spacelike hypersurface in a spacetime $M$ and
let $\xi^a$ be a timelike Killing vector field on $M$. Assume that $w^a$,
defined in Equation (\ref{Eq_Defv}), is a Killing field on $(\Sigma,h_{ij})$,
where $h_{ij}$ is the induced metric on $\Sigma$. For a smooth curve
$\gamma:[0,1]\rightarrow\Sigma$ the following statements are equivalent:
\begin{enumerate}
\item $\gamma$ is a geodesic for $(\Sigma,h_{ij})$,
\item $\gamma$ is a geodesic for $M$.
\end{enumerate}
\end{lemma}
The lemma applies in particular when $\xi^a$ is orthogonal to $\Sigma$.
\begin{proof*}
The statement is local, so we may introduce local coordinates $x^i$ on
$\Sigma$ and extend them to Gaussian normal coordinates on an open
neighbourhood of $M$. Using the special form
$g_{\mu\nu}dx^{\mu}dx^{\nu}=-(dx^0)^2+h_{ij}dx^idx^j$ of the metric in
these coordinates, the geodesic equation in $M$ for the curve $\gamma$
reduces to the geodesic equation in $(\Sigma,h_{ij})$ plus the equation
$0=(\partial_0h_{ij})\dot{\gamma}^i\dot{\gamma}^j$. We will show that
the latter is automatically satisfied, due to the assumption on $w^a$.
We may write $\xi_a=\xi^0n_a+w_a$ near $\Sigma$ with
$n^a:=(\partial_0)^a$ and consider the spatial components of Killing's
equation:
\[
0=2\nabla_{(i}\xi_{j)}=2\xi^0\nabla_{(i}n_{j)}+2n_{(i}\nabla_{j)}\xi^0
+2\nabla^{(h)}_{(i}w_{j)}.
\]
Here, the last two terms vanish on $\Sigma$, because $n^i|_{\Sigma}=0$
and $w^i$ is a Killing field on $\Sigma$. The first term can be written
using $\partial_0h_{ij}=\mathcal{L}_{n^a}g_{ij}=2\nabla_{(i}n_{j)}$.
Since $\xi^0\not=0$ on $\Sigma$ we find $\partial_0h_{ij}=0$, which
proves our claim.
\end{proof*}

\begin{remark}\label{Rem_UniqueCauchy}
If the bifurcation surface of $M$ is static, then the (possibly
disconnected) spacetimes $M^{\pm}$ are standard static globally
hyperbolic spacetimes (cf.\ \cite{Sanchez2005,Sanders2012}). If $\Sigma$
is a Cauchy surface satisfying the properties of the static case of
Definition \ref{Def_StaticBHSpac}, then the same is true for
$\Phi_T(\Sigma)$ for any $T\in\mathbb{R}$. Conversely, given any other
Cauchy surface $\Sigma'$ satisfying the properties of the definition,
we have $\Sigma'=\Phi_T(\Sigma)$ for some $T\in\mathbb{R}$. Indeed, for
any $x\in\Sigma\setminus\mathcal{B}$ the integral curve $t\mapsto\Phi_t(x)$
is smooth and it remains timelike (by Killing's equation). Since it is
inextendible (due to the completeness of $\xi^a$) there is a unique
$T(x)\in\mathbb{R}$ such that $\Phi_{T(x)}(x)\in\Sigma'$. Now note that
$\Sigma'$ and $\Phi_{T(x)}(\Sigma)$ both contain $x$ and that they are
both orthogonal to $\xi^a\not=0$. This shows that both surfaces coincide
near $x$ and hence that $T(x)$ is locally constant on
$\Sigma\setminus\mathcal{B}$.

To see that $T$ is even globally constant, we consider a geodesic segment
$\map{\gamma}{(-\epsilon,\epsilon)}{\Sigma}$ in $(\Sigma,h_{ij})$ which
intersects $\mathcal{B}$ only at $\gamma(0)=p$, where the intersection
is transversal. For $s\in(0,\epsilon)$ the points $\gamma(s)$ all lie
in the same connected component of $\Sigma\setminus\mathcal{B}$, so
there is a unique $T\in\mathbb{R}$ such that $\gamma':=\Phi_T\circ\gamma$
lies in $\Sigma'$ for $s\in(0,\epsilon)$. To see that $\gamma'$ lies
entirely in $\Sigma'$ we use the fact that $\gamma$ and $\gamma'$ are
also geodesics in $M$, by Lemma \ref{Lem_CauchyGeodesics}. Similarly, if
$h'_{ij}$ is the induced metric on $\Sigma'$ and $\tilde{\gamma}$ is the
unique inextendible geodesic in $(\Sigma',h'_{ij})$ which coincides up
to first-order with $\gamma'$ at $s=\frac12\epsilon$, then
$\tilde{\gamma}$ is also a geodesic in $M$. It follows that
$\tilde{\gamma}$ extends $\gamma'$, so $\gamma'$ lies entirely in
$\Sigma'$. Therefore, the locally constant function $T$ on
$\Sigma\setminus\mathcal{B}$ does not change value when we cross
$\mathcal{B}$. Since $\Sigma$ is connected, $T$ must be globally
constant and $\Phi_T(\Sigma)\subset\Sigma'$. The converse inclusion
follows by reversing the roles of $\Sigma$ and $\Sigma'$.
\end{remark}

A useful characteristic of the bifurcation surface is its surface
gravity, $\kappa>0$ (cf.\cite{Wald,Kay+1991}), which is a locally
constant function on $\mathcal{B}$ satisfying
\begin{equation}
\kappa^2\equiv \frac{-1}{2}(\nabla^b\xi^a)(\nabla_b\xi_a)|_{\mathcal{B}}.\nonumber
\end{equation}
(This equality follows from Equation (12.5.14) in \cite{Wald}.) It will
be convenient to know how the surface gravity can be computed from
geometric objects on the Cauchy surface $\Sigma$. The following lemma
answers this question.
\begin{lemma}\label{Lem_kappa}
$\kappa^2=h^{ij}(\partial_iv)(\partial_jv)|_{\mathcal{B}}-
\frac12 h^{ij}h^{kl}(\nabla^{(h)}_iw_k)(\nabla^{(h)}_jw_l)|_{\mathcal{B}}$.
\end{lemma}
\begin{proof*}
We use Gaussian normal coordinates $x^{\mu}=(x^0,x^i)$ near $\Sigma$ on a
neighbourhood of some arbitrary $p\in\mathcal{B}$. Combining the special
form of the metric in these coordinates with the fact that $\xi^a=vn^a+w^a$
vanishes on $\mathcal{B}$ and Killing's equation, we find
\begin{eqnarray}
\kappa^2&=&\frac{-1}{2}g^{\mu\nu}g^{\rho\sigma}(\nabla_{\mu}\xi_{\rho})
(\nabla_{\nu}\xi_{\sigma})|_{\mathcal{B}}
=\frac{-1}{2}g^{\mu\nu}g^{\rho\sigma}(\partial_{\mu}\xi_{\rho})
(\partial_{\nu}\xi_{\sigma})|_{\mathcal{B}}\nonumber\\
&=&\frac{-1}{2}h^{ij}h^{kl}(\partial_i\xi_k)(\partial_j\xi_l)|_{\mathcal{B}}
+h^{ij}(\partial_i\xi_0)(\partial_j\xi_0)|_{\mathcal{B}}\nonumber\\
&=&h^{ij}(\partial_iv)(\partial_jv)|_{\mathcal{B}}
-\frac12 h^{ij}h^{kl}(\nabla^{(h)}_iw_k)(\nabla^{(h)}_jw_l)|_{\mathcal{B}}.\nonumber
\end{eqnarray}
\end{proof*}

In the stationary case, the analysis of thermal (KMS) states of a quantum
field in the right wedge $M^+$ and the idea of purification of such states
naturally lead one to consider the case where $M^-$ is isomorphic to $M^+$,
except for a reversal of the time orientation \cite{Israel1976,Kay1985_2}.
We therefore introduce the following notions of wedge reflection\footnote{Our
definition of a wedge reflection is slightly less restrictive than that in
\cite{Kay+1991}, where $I$ is required to be a time orientation reversing
isometric diffeomorphism of the entire spacetime $M$, which leaves
$\mathcal{B}$ pointwise fixed and satisfies $I\circ I=\id$ and
$I^*\xi^a=\xi^a$ everywhere.}
\begin{definition}
A \emph{wedge reflection} $I$ for a spacetime $M$ with a stationary bifurcate
Killing horizon is a diffeomorphism $\map{I}{M^+\cup M^-\cup U}{M^+\cup M^-\cup U}$
for some open neighbourhood $U$ of $\mathcal{B}$, such that
\begin{enumerate}
\item $I$ is an isometry of $M^+\cup M^-$ onto itself, which reverses the
time orientation,
\item $I\circ I=\id$, the identity map,
\item $I$ leaves $\mathcal{B}$ pointwise fixed, and
\item $I^*\xi^a=\xi^a$ on $M^+\cup M^-$.
\end{enumerate}
A \emph{weak wedge reflection} is a pair $(\Sigma,\iota)$, where $\Sigma$ is
a Cauchy surface of $M$ as in Def.\ \ref{Def_StaticBHSpac} and
$\map{\iota}{\Sigma}{\Sigma}$ is a diffeomorphism such that
\begin{enumerate}
\item $\iota$ is an isometry for $(\Sigma,h_{ij})$,
\item $\iota\circ\iota=\id$,
\item $\iota$ leaves $\mathcal{B}$ pointwise fixed, and
\item $\iota^*v=-v$, $\iota^*w^i=w^i$.
\end{enumerate}
\end{definition}
We note that $\iota(\Sigma^{\pm})=\Sigma^{\mp}$. If the metric $g_{ab}$ is
analytic near $\mathcal{B}$, then the existence of $I$ on a neighbourhood
of $\mathcal{B}$ is guaranteed \cite{Kay+1991}. We now prove the following
additional results:
\begin{proposition}\label{Prop_WedgeRefl}
Let $M$ be a spacetime with a bifurcate Killing horizon.
\begin{enumerate}
\item All wedge reflections on $M$ agree on $M^+\cup M^-\cup\mathcal{B}$
and $I(M^{\pm})=M^{\mp}$.
\item A Cauchy surface $\Sigma$ admits at most one diffeomorphism $\iota$
such that $(\Sigma,\iota)$ is a weak wedge reflection.
\item If $(\Sigma,\iota)$ is a weak wedge reflection, then so is
$(\Phi_T(\Sigma),\Phi_T\circ\iota\circ\Phi_{-T})$.
\item Given a wedge reflection $I$, there is a weak wedge reflection
$(\Sigma,\iota)$ such that $\iota=I|_{\Sigma}$. In addition, if the
bifurcation surface is static, we may choose $\Sigma$ orthogonal to $\xi$.
\item In the stationary case, given a weak wedge reflection $(\Sigma,\iota)$
there exists a time orientation reversing isometric diffeomorphism
$\map{I}{M^+\cup M^-}{M^+\cup M^-}$ such that $I^*\xi^a=\xi^a$, $I\circ I=\id$
and $I|_{\Sigma\setminus\mathcal{B}}=\iota|_{\Sigma\setminus\mathcal{B}}$.
If $w^i$ is a Killing field for $(\Sigma,h_{ij})$ near $\mathcal{B}$, then
we may extend $I$ to a wedge reflection. This applies in particular in the
static case.
\end{enumerate}
\end{proposition}
\begin{proof*}
If $I$ is a wedge reflection on $M$ and $p\in\mathcal{B}$, then the
derivative $dI_p$ is an isomorphism of the tangent space $T_pM$, which
is isometric (by continuity). $dI_p$ acts trivially on tangent vectors
of $T_p\mathcal{B}$ and the orthogonal complement is spanned by two
future pointing null vectors $l^a,m^a$. Note that $\xi^a$ is timelike on a
neighbourhood of $p$ in $M^{+}\cup M^-$, where it is future pointing on
$M^+$ and past pointing on $M^-$ (cf.\ Figure \ref{Fig_Bifurcation}). Because
$I$ reverses the time orientation, it also reverses $M^+$ and $M^-$ on a
neighbourhood of $p$. Together with the fact that $I\circ I=\id$ this implies
that $dI_p(l^a)=- l^a$ and $dI_p(m^a)=- m^a$.

If $I$ and $I'$ are two wedge reflections of $M$, then
$\psi:=I'\circ I$ is a diffeomorphism of $M^+\cup M^-\cup U$ into $M$,
for some neighbourhood $U$ of $\mathcal{B}$. Note that $d\psi$ acts as
the identity on $TM|_{\mathcal{B}}$ and that it is isometric on
$\overline{M^+\cup M^-}$. We may use the exponential map to show that
$\psi=\id$ on $(M^+\cup M^-)\cap V$ for some open neighbourhood $V$ of
$\mathcal{B}$. Because $M^+\cup M^-\cup\mathcal{B}$ is connected, we may
continue the result $\psi=\id$ to this entire set, so that $I=I'$ on
$M^+\cup M^-\cup\mathcal{B}$.
The fact that $I(M^{\pm})=M^{\mp}$ will follow from statement 4 and the
facts that $M^{\pm}=D(\Sigma^{\pm})$ and $\iota(\Sigma^{\pm})=\Sigma^{\mp}$.

Now let $\Sigma$ be any Cauchy surface as in Def.\ \ref{Def_StaticBHSpac}.
If $(\Sigma,\iota)$ is a weak wedge reflection and $p\in\mathcal{B}$, then
$\iota(p)=p$ and $d\iota_p$ acts on $T_p\Sigma$ as the orthogonal reflection
(w.r.t.\ $h_{ij}$) in the linear subspace $T_p\mathcal{B}$. The uniqueness
of $\iota$ is then shown by the same argument as in the previous paragraph.
The fact that $(\Phi_T(\Sigma),\Phi_T\circ\iota\circ\Phi_{-T})$ is also a
weak wedge reflection is straightforward.

Now suppose that $I$ is a wedge reflection and $\Sigma'$ is any Cauchy surface
as in Def.\ \ref{Def_StaticBHSpac}. By the results of \cite{Bernal+2006} there
exists a smooth function $T'$ on $M$ whose gradient $\nabla_a T'$ is everywhere
timelike and past pointing and such that $\Sigma'=(T')^{-1}(0)$. Now set
$T:=T'-I^*T'$, which is again a smooth function with a past pointing, timelike
gradient. $\Sigma:=T^{-1}(0)$ is a smooth, spacelike hypersurface. On every
inextendible timelike curve $\gamma$ in $M$ we can find points $p_{\pm}$ such
that $\pm T'(p_{\pm})>0$ and $\mp T'(I(p_{\pm}))>0$. Hence $\pm T(p_{\pm})>0$,
so that $\gamma$ must intersect $\Sigma$ and $\Sigma$ is a Cauchy surface.
Furthermore, $I(\Sigma)=\Sigma$, so that $\iota:=I|_{\Sigma}$ is well-defined.
It is immediately verified that $(\Sigma,\iota)$ is a weak wedge reflection.

In the static case, the Cauchy surface $\Sigma$ constructed above may fail
to be orthogonal to $\xi^a$. However, if $\Sigma$ is any Cauchy surface
orthogonal to $\xi^a$, as in Def.\ \ref{Def_StaticBHSpac}, then
$(\Phi_T\circ I)(\Sigma)=\Sigma$ for some $T\in\mathbb{R}$, by Remark
\ref{Rem_UniqueCauchy}. On $T_pM$ with $p\in\mathcal{B}$ we may consider the
linear isomorphism $d(\Phi_T\circ I)_p=(d\Phi_T)_p\circ dI_p$, which acts
trivially on vectors in $T_p\mathcal{B}$. On the normal vector $r^a$ to
$\mathcal{B}$ in $\Sigma$ we have $dI_p(r^a)=-r^a$ by the first paragraph of
this proof and by considering the action of $\Phi_T$ we see that
$d(\Phi_T\circ I)_p(r^a)$ can only lie in $T_p\Sigma$ if and only if $T=0$.
It follows that we must have $I(\Sigma)=\Sigma$, so taking
$\iota:=I|_{\Sigma}$ we find the weak wedge reflection $(\Sigma,\iota)$ with
$\Sigma$ orthogonal to $\xi$.

Finally, let $(\Sigma,\iota)$ be a weak wedge reflection. If the
bifurcation surface is stationary, we can define $I$ on $M^+\cup M^-$ by
\[
I\circ\psi(t,x):=\psi(t,\iota(x))
\]
with $\psi$ as in Equation (\ref{Eqn_psi}). It is clear that $I\circ I=\id$,
so $I$ is a diffeomorphism, and
$I|_{\Sigma\setminus\mathcal{B}}=\iota|_{\Sigma\setminus\mathcal{B}}$ by
construction. Furthermore, $I^*\xi^a=\xi^a$ and $I(M^{\pm})=M^{\mp}$, so $I$
must reverse the time orientation.

To see that $I$ is isometric we fix a $p\in\Sigma\setminus\mathcal{B}$ and
we note that $I^*(n^a)=I^*(\frac{1}{v}(\xi^a-w^a))=\frac{1}{-v}(\xi^a-w^a)=-n^a$,
because $I^*(w^a)=\iota^*(w^a)=w^a$. Decomposing any tangent vector
$\nu^a\in T_pM$ as $\nu^a=\alpha n^a+\mu^a$ with $\mu^a\in T_p\Sigma$ we
find $I^*(\nu^a)=-\iota^*(\alpha)n^a+\iota^*(\mu^a)$. As $\iota$ is an
isometry of $T_p\Sigma$, it follows that the same is true for $I$ on
$T_pM$. Because $I$ commutes with the isometries $\Phi_t$, it must then
be isometric on $M^+\cup M^-$.

Let $N\mathcal{B}\subset TM|_{\mathcal{B}}$ be the normal bundle to
$\mathcal{B}$ in $M$. There is a neighbourhood $V$ of the zero section of
this bundle on which the exponential map is a diffeomorphism
$V\simeq\exp(V)=:\tilde{V}$. Without loss of generality we may assume that
$V$ has a convex intersection with each fibre of $N\mathcal{B}$ and that
$V=-V$, where $-1$ is the fibre-wise multiplication by $-1$ on $N\mathcal{B}$.
Then $I':=\exp\circ -1\circ\exp^{-1}$ is a diffeomorphism of $\tilde{V}$ onto
itself. Any wedge reflection must coincide with $I'$ on a neighbourhood of
$\mathcal{B}$ in $M^+\cup M^-$. Conversely, if $I'$ coincides with $I$ on
such a neighbourhood, then we can extend $I$ to a wedge reflection.
Now, given any $x\in\tilde{V}\cap\Sigma$, let $\gamma(s):=\exp^{(h)}_p(sv)$
be the geodesic in $(\tilde{V}\cap\Sigma,h_{ij})$ from $I(x):=\gamma(-1)$ to
$x=\gamma(1)$. If $w^i$ is a Killing field for $(\Sigma,h_{ij})$ on
$\tilde{V}\cap\Sigma$ we know from Lemma \ref{Lem_CauchyGeodesics} that
$\gamma$ is also a geodesic in $M$, which entails that
$I'(\gamma(s))=\gamma(-s)$. Therefore $I$ and $I'$ coincide on
$\tilde{V}\cap\Sigma$. Because both commute with the flow of
$\xi^a$ they even coincide on a neighbourhood of $\mathcal{B}$ in $M^+\cup M^-$,
so we may extend $I$ by $I'$ to $M^+\cup M^-\cup\tilde{V}$, making it into a
wedge reflection.
\end{proof*}
Note in particular that in the static case, a wedge reflection is
equivalent to a weak wedge reflection $(\Sigma,\iota)$ with $\Sigma$
orthogonal to $\xi^a$. Furthermore, for any two weak wedge reflections
$(\Sigma,\iota)$ and $(\Sigma',\iota')$ with both $\Sigma$ and $\Sigma'$
orthogonal to $\xi^a$ we must have $\Sigma'=\Phi_T(\Sigma)$ and
$\iota'=\Phi_T\circ\iota\circ\Phi_{-T}$ for some $T\in\mathbb{R}$.
Hence, both weak wedge reflections give rise to the same map $I$ on
$M^+\cup M^-$. (Whether an equivalence of weak and strong wedge
reflections holds in the general stationary case is unclear.)

\subsection{Complexification beyond the horizon and the Hawking
temperature}\label{SSec_Complexify}

If $M$ is a spacetime with a static bifurcate Killing horizon, then
$M^+$ is a (possibly disconnected) standard static spacetime and we
may define complexifications and Riemannian manifolds with a compactified
imaginary time variable (cf.\ \cite{Sanders2012}). For $R>0$ we
define the cylinder
\[
\mathcal{C}_R:=\mathbb{C}/\sim,\quad z\sim z'\Leftrightarrow
z-z'\in 2\pi iR\, \mathbb{Z}.
\]
Under this equivalence relation, the imaginary axis of $\mathbb{C}$
becomes compactified to the circle $\mathbb{S}_R$ of radius $R$.
The complexification $(M^+)^c_R$ is then defined as a real manifold,
endowed with a symmetric, complex-valued tensor field:
\[
(M^+)^c_R=\mathcal{C}_R\times\Sigma^+,
\]
\[
(g^c_R)_{\mu\nu}=-v^2(dz^{\otimes 2})_{\mu\nu}+h_{\mu\nu},
\]
where $v$ and $h_{\mu\nu}$ are independent of the coordinate $z=t+i\tau$
on $\mathcal{C}_R$. Using the diffeomorphism $\psi$ of Equation
(\ref{Eqn_psi}), restricted to $\mathbb{R}\times\Sigma^+$, we can
embed $M^+$ into $(M^+)^c_R$ as the $\tau=0$ surface. $(g^c_R)_{\mu\nu}$
is the analytic continuation of $g_{\mu\nu}$ in $z$. We may also consider
the associated Riemannian manifold, endowed with the pull-back metric:
\[
M^+_R:=\left\{(z,x)\in (M^+)^c_R|\ t=0\right\},
\]
\[
(g_R)_{\mu\nu}=v^2(d\tau^{\otimes 2})_{\mu\nu}+h_{\mu\nu}.
\]
Note that $M^+_R\simeq\mathbb{S}_R\times\Sigma^+$ as a manifold.
We can identify the surface $\Sigma^+\simeq M^+\cap M^+_R$ in $(M^+)^c_R$
also with the $\tau=0$ surface in $M^+_R$. Furthermore, $M^+_R$ has
a Killing field $(\xi_R)^a\partial_a=\partial_{\tau}$, which can be
viewed as the analytic continuation of $\xi^a\partial_a=\partial_t$.

\begin{figure}[t!]
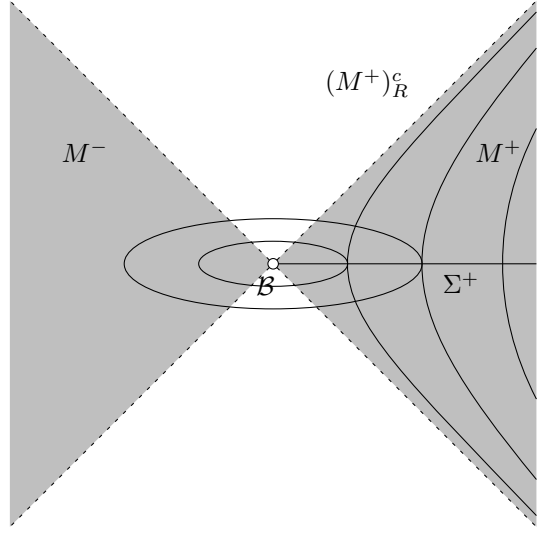

\begin{center}
\begin{pgfpicture}{0cm}{0cm}{7cm}{7cm}
\color{lightgray}
\pgfmoveto{\pgfxy(3.55,3.45)}
\pgflineto{\pgfxy(7,0)}
\pgflineto{\pgfxy(7,7)}
\pgflineto{\pgfxy(3.55,3.55)}
\pgfcurveto{\pgfxy(3.58,3.55)}{\pgfxy(3.58,3.45)}{\pgfxy(3.55,3.45)}
\pgffill
\pgfmoveto{\pgfxy(3.45,3.45)}
\pgflineto{\pgfxy(0,0)}
\pgflineto{\pgfxy(0,7)}
\pgflineto{\pgfxy(3.45,3.55)}
\pgfcurveto{\pgfxy(3.42,3.55)}{\pgfxy(3.42,3.45)}{\pgfxy(3.45,3.45)}
\pgffill
\color{black}
\pgfline{\pgfxy(3.57,3.5)}{\pgfxy(7,3.5)}
\pgfputat{\pgfxy(6,3.25)}{\pgfbox[center,center]{$\Sigma^+$}}
\pgfputat{\pgfxy(6.5,5)}{\pgfbox[center,center]{$M^+$}}
\pgfputat{\pgfxy(4.75,5.9)}{\pgfbox[center,center]{$(M^+)^c_R$}}\pgfputat{\pgfxy(1,5)}{\pgfbox[center,center]{$M^-$}}
\pgfsetdash{{0.05cm}{0.1cm}}{0cm}
\pgfline{\pgfxy(3.45,3.45)}{\pgfxy(0,0)}
\pgfline{\pgfxy(3.55,3.55)}{\pgfxy(7,7)}
\pgfline{\pgfxy(3.45,3.55)}{\pgfxy(0,7)}
\pgfline{\pgfxy(3.55,3.45)}{\pgfxy(7,0)}
\pgfsetdash{{1cm}{0cm}}{0cm}
\pgfmoveto{\pgfxy(7,0.15)}
\pgfcurveto{\pgfxy(3.65,3.65)}{\pgfxy(3.65,3.35)}{\pgfxy(7,6.85)}
\pgfstroke
\pgfmoveto{\pgfxy(7,0.63)}
\pgfcurveto{\pgfxy(4.97,3.1)}{\pgfxy(4.97,3.9)}{\pgfxy(7,6.37)}
\pgfstroke
\pgfmoveto{\pgfxy(7,1.7)}
\pgfcurveto{\pgfxy(6.4,2.87)}{\pgfxy(6.4,4.13)}{\pgfxy(7,5.3)}
\pgfstroke
\pgfcircle[stroke]{\pgfxy(3.5,3.5)}{0.07cm}
\pgfputat{\pgfxy(3.4,3.2)}{\pgfbox[center,center]{$\mathcal{B}$}}
\pgfellipse[stroke]{\pgfxy(3.5,3.5)}{\pgfxy(0.99,0)}{\pgfxy(0,0.3)}
\pgfellipse[stroke]{\pgfxy(3.5,3.5)}{\pgfxy(1.98,0)}{\pgfxy(0,0.6)}
\end{pgfpicture}
\end{center}
\caption{The embedding of $M^-$ into $(M^+)^c_R$. Depicted are $M^+$,
some integral curves of the Killing field $\xi^a$, and their
complexifications, which are compactified to circles. $(M^+)^c_R$ is
obtained by rotating $M^+$ around a vertical axis through $\mathcal{B}$
and $M^-\simeq M^+$ is embedded on the opposite side of the circle.
Note, however, that the isomorphism $M^-\simeq M^+$ reverses the time
orientation when compared to $M^-$ in
Figure \ref{Fig_Bifurcation}.}\label{Fig_Complexify}
\end{figure}

If $M$ has a wedge reflection $I$, and hence a weak wedge reflection
$(\Sigma,\iota)$, we can extend the embedding of $M^+$ into $(M^+)^c_R$
to an embedding
\begin{equation}\label{Def_chi}
\map{\chi}{M^+\cup M^-}{(M^+)^c_R}:
\left\{
\begin{array}{ll}
\chi\circ\psi(t,x)=(t,x)&\psi(t,x)\in M^+\\
\chi\circ\psi(t,x)=(t+i\pi R,\iota(x))&\psi(t,x)\in M^-
\end{array}\right.,
\end{equation}
where we used the diffeomorphism $\psi$ of Equation (\ref{Eqn_psi}).
In other words, if $x\in M^-$, then $\psi(x)$ is obtained by composing
the wedge reflection $I$, the embedding of $M^+$ into $(M^+)^c_R$ and
a rotation over the angle $\pi$. (See Figure \ref{Fig_Complexify}.)

$\chi$ restricts to an embedding $\mu:=\chi|_{\Sigma\setminus\mathcal{B}}$
of $\Sigma\setminus\mathcal{B}$ into the Riemannian manifold $M^+_R$,
so that $\Sigma^-$ is embedded as the $\tau=\pi R$ hypersurface.
We now wish to consider whether this embedding can be extended to
an embedding $\overline{\mu}$ of all of $\Sigma$ into some
extension $\overline{M^+_R}$ of the manifold $M^+_R$, and whether
the Riemannian metric $(g_R)_{ab}$ can be extended to
$\overline{M^+_R}$ as well.

A suitable extension $\overline{M^+_R}$ of $M^+_R$ can readily be
obtained by a standard gluing technique. To see how this works, we let
$\map{\pi_{N\mathcal{B}}}{N\mathcal{B}}{\mathcal{B}}$ denote the
normal bundle of $\mathcal{B}$ in $\Sigma$ with zero section
$\mathcal{Z}$. Note that $N\mathcal{B}\subset T\Sigma|_{\mathcal{B}}$
and since both $\Sigma$ and $\mathcal{B}$ are orientable,
$N\mathcal{B}\simeq\mathcal{B}\times\mathbb{R}$ is a trivial bundle.
We may introduce the normal vector field $n^a$ to $\mathcal{B}$ in
$\Sigma$, which points towards $\Sigma^+$. This determines
an orthonormal frame and an orientation on $N\mathcal{B}$. There is a
neighbourhood $U\subset N\mathcal{B}$ of $\mathcal{Z}$ on which the
exponential map $\map{\exp}{U}{\Sigma}$ defines a diffeomorphism.
Next we consider the bundle $\mathcal{B}\times\mathbb{R}^2$ with the
canonical Euclidean inner product in each fibre and a fixed
orthonormal frame. We introduce the subbundles
\[
X:=\left\{(x,v)\in\mathcal{B}\times\mathbb{R}^2|\ (x,|v|n^a)\in U\right\},
\quad \dot{X}:=\left\{(x,v)\in X|\ v\not=0\right\}
\]
and we may embed $\dot{X}$ into $M^+_R$ by
\[
\map{\eta}{\dot{X}}{M^+_R}:(x,re^{i\phi})\mapsto (R\phi,\exp_x(r)),
\]
where $\phi$ is defined with respect to the fixed orthonormal frame.
The extended spacetime can then be defined by gluing $X$ against
$M^+_R$ along $\dot{X}$, i.e.
\[
\overline{M^+_R}:=(M^+_R\cup X)/\sim,
\]
where $\sim$ indicates that we identify the domain and range of
$\eta$.

On $\exp(U\setminus\mathcal{Z})$ the embedding $\mu$ is given by
$\eta^{-1}\circ\mu(\exp_x(s))=(x,(s,0))$. This may be checked
separately for the cases $s>0$ and $s<0$, using the properties
of $\iota$, which imply $\exp_x(-s)=\iota(\exp_x(s))$ for
$x\in\mathcal{B}$. The extension
$\map{\overline{\mu}}{\Sigma}{\overline{M^+_R}}$ can then be
defined by taking $\overline{\mu}(\exp_x(s))=(x,(s,0))$ also
when $s=0$.

Now that we have defined the extended manifold $\overline{M^+_R}$ we
wish to investigate whether the Riemmannian metric $(g_R)_{ab}$ on
$M^+_R$ can be extended too. This is where a particular value of the
radius $R$ is singled out, which corresponds to the Hawking temperature
(cf.\ \cite{Jacobson1994,Fulling+1987} and references therein).
\begin{lemma}\label{Lem_HawkingR}
The components of the metric $(g_R)_{ab}$ can be extended to
$\overline{M^+_R}$ as bounded functions. A continuous extension
$(\bar{g}_R)_{ab}$ exists if and only if $R\equiv\kappa^{-1}$, in which
case the extension is even smooth.
\end{lemma}
\begin{proof*}
To prove this lemma, we work in suitably chosen local coordinates. First,
we introduce local coordinates $x^i$, $i=2,\ldots, d-1$ on $\mathcal{B}$
and we let $r$ denote the Gaussian normal coordinate near $\mathcal{B}$ on
$\Sigma$, with $r>0$ on $\Sigma^+$. As before, we let $t$ denote the Killing
time on $M^+$, so for some $\rho>0$ the local coordinates
$(t,r,x^i)\in\mathbb{R}\times(0,\rho)\times\mathcal{B}$ describe an open
region in $M^+$ whose boundary in $M$ contains $\mathcal{B}$. After
complexification and restriction to the Riemannian manifold, we have local
coordinates $(\tau,r,x^i)\in\mathbb{S}_R\times(0,\rho)\times\mathcal{B}$.
Expressed in these local coordinates the Riemannian metric $h_{ij}$ on
$\Sigma$ takes the form
\[
h_{ij}=(dr^{\otimes 2})_{ij}+k_{ij}(r,x^l),
\]
where $k_{ij}(r,x^l)$ denotes the Riemannian metric induced on $\mathcal{B}$.
Correspondingly, the Lorentzian metric on $M^+$ and the Riemannian metric
on $M^+_R$ take the form
\begin{eqnarray}\label{Eqn_MetricPolar}
g_{\mu\nu}&=&-v^2(dt^{\otimes 2})_{\mu\nu}+(dr^{\otimes 2})_{\mu\nu}+k_{\mu\nu}(r,x^i),\nonumber\\
(g_R)_{\mu\nu}&=&v^2(d\tau^{\otimes 2})_{\mu\nu}+(dr^{\otimes 2})_{\mu\nu}+k_{\mu\nu}(r,x^i).
\end{eqnarray}
Changing coordinates $(\tau,r)\rightarrow(X,Y)$ with
\[
X:=r\cos\left(\frac{\tau}{R}\right),\quad Y:=r\sin\left(\frac{\tau}{R}\right)
\]
the metric on $M^+_R$ takes the form
\[
(g_R)_{\mu\nu}=(1-\alpha Y^2)(dX^{\otimes 2})_{\mu\nu}+(1-\alpha X^2)(dY^{\otimes 2})_{\mu\nu}
+\alpha XY(dX\otimes dY+dY\otimes dX)_{\mu\nu}+k_{\mu\nu}(X,Y,x^i),
\]
where the function $\alpha$ is defined by $\alpha:=r^{-2}-r^{-4}R^2v^2$.

By construction of the Gaussian normal coordinate $r$ we have $\iota^*r=-r$.
As $\iota^*k_{ij}=k_{ij}$ it follows that $k_{ij}$ is an even tensor in $r$
and its Taylor expansion around $r=0$ only contains even powers. This
suffices to show that $k_{ij}$ depends smoothly on $X$ and $Y$, since
$r^2=X^2+Y^2$. Hence, $k_{ij}$ extends smoothly to all of $\overline{M^+_R}$.

The functions $X^2\alpha$, $Y^2\alpha$ and $XY\alpha$ remain bounded near the
set $\mathcal{B}$, where $r=0$, but if we take the limit $r\rightarrow 0^+$
we find that e.g.\ $X^2\alpha$ approaches a value that may in general depend
on $\tau$ as well as $x^i$. To eliminate this dependence and to get a
continuous extension, it is necessary and sufficient to impose
\[
\lim_{r\rightarrow 0^+}r^2\alpha(r,x^i)=0.
\]
In order to analyse this limit, we first prove that
$v(r,x^i)=\kappa r+r^3\beta(r,x^i)$ for some smooth $\beta$ near $\mathcal{B}$.
To see this, we use a Taylor expansion around $r=0$. As $\iota^*v=-v$ we cannot
have any even terms in $r$, so the constant and second-order terms vanish.
Since $\nabla^{(h)}_av$ and $\nabla^{(h)}_ar$ are both normal to $\mathcal{B}$
the first-order term is fixed by Lemma \ref{Lem_kappa}. The term with $\beta$
is just the remainder.\footnote{$\beta$ is smooth by e.g.\ \cite{Lang} Ch.13,
\S 6 and Theorem 8.1.} Now the vanishing of the limit above simply means
\[
R^{-1}\equiv\lim_{r\rightarrow 0^+}\frac{v(r,x^i)}{r}=\kappa.
\]
To see that the extension $(\bar{g}_R)_{ab}$ is even smooth when this holds,
we note that $\alpha$ takes the form $\alpha=-2R\beta-r^2R^2\beta^2$ near
$\mathcal{B}$, which is smooth.
\end{proof*}

In order to satisfy the condition of Lemma \ref{Lem_HawkingR} it is necessary
for the surface gravity $\kappa$ to be constant, because $R$ is constant too.
If $\mathcal{B}$ is connected this is no additional assumption, but in general
it may fail (cf.\ \cite{Kay+1991} for further discussion and examples).
Anticipating the relation between the radius $R$ and the temperature, we define
the Hawking radius to be
\[
R_H:=\kappa^{-1}
\]
whenever $\kappa$ is constant.

The Killing field
$\xi_R^{\mu}\partial_{\mu}=\partial_{\tau}=\frac{1}{R}(X\partial_Y-Y\partial_X)$
always admits a smooth extension to $\overline{M^+_R}$, which we will denote by
$\overline{\xi_R}^{\mu}$. Furthermore, we wish to record the following lemma,
whose proof is closely related to that of Lemma \ref{Lem_HawkingR}:
\begin{lemma}\label{Lem_SmoothV'}
If $V\in C^{\infty}(\Sigma)$ satisfies $\iota^*V=V$ and $R>0$, then there exists
a unique smooth extension $W$ of $V$ to $\overline{M^+_R}$ such that
$\overline{\xi_R}^{\mu}\partial_{\mu}W=0$.
\end{lemma}
\begin{proof*}
On $M^+_R\simeq\mathbb{S}^1\times\Sigma^+$ there is exactly one smooth function
$W$ such that $\xi_R^{\mu}\partial_{\mu}W=0$ and $W|_{\Sigma^+}=V|_{\Sigma^+}$.
It is given by $W(\tau,x^i)=V(x^i)$. Note that $W(\pi R,x^i)=V(x^i)=V(\iota(x^i))$,
so $W|_{\Sigma^-}=V|_{\Sigma^-}$ too. We now define $W|_{\mathcal{B}}:=V|_{\mathcal{B}}$
and it remains to prove that $W$ is smooth. For this purpose we use again local
coordinates $(\tau,r,x^i)$ and $(X,Y,x^i)$ near $\mathcal{B}$, as in the proof of
Lemma \ref{Lem_HawkingR}. We have $W(X,Y,x^i)=V(\sqrt{X^2+Y^2},x^i)$, so $W$ is
continuous at $\mathcal{B}$. Moreover, the Taylor series of $V$ at $r=0$ is even
in $r$, because $\iota^*V=V$. $V$ therefore only depends on $r^2$ and $W$ depends
only on $X$ and $Y$ through $X^2+Y^2$, so the extension is smooth.
\end{proof*}

\subsection{Analytic continuation beyond the horizon}\label{SSec_infAC}

The Killing time coordinate on $M^+$ is used to define the complexification
$(M^+)^c_R$ and the Riemannian manifold $M^+_R$, but it becomes a bad choice
of coordinate near the boundary of $M^+\subset M$. This is particularly
inconvenient when we wish to study the behaviour near the bifurcation surface
$\mathcal{B}$. For that reason, we now consider Gaussian normal coordinates
instead and study their properties regarding the complexification procedure
above. Furthermore, we will consider Riemannian normal coordinates, which
are the most convenient choice of coordinates when describing the Hadamard
parametrix construction in Section \ref{Sec_Had} below. In order to
investigate this construction in the light of our complexification procedure,
we will also establish some results on the relation between Riemannian and
Gaussian normal coordinates.

We consider a spacetime $M$ with a static bifurcate Killing horizon, with a
wedge reflection and with a surface gravity $\kappa>0$ which is globally
constant. Let $x^i$ denote local coordinates on a neighbourhood $U$ in a
Cauchy surface $\Sigma$ with the properties of Definition
\ref{Def_StaticBHSpac}. We let $x^{\mu}=(x^0,x^i)$ denote corresponding Gaussian
normal coordinates on a portion $V$ of $M$ containing $U$. Furthermore, we
will write $M':=\overline{M^+_{R_H}}$ and we let $(x')^{\mu}=((x')^0,(x')^i)$
be Gaussian normal coordinates on a region $V'\subset M'$, containing
$U':=\overline{\mu}(U)$, such that $x^i=\overline{\mu}^*(x')^i$. We choose the
Gaussian normal coordinates in such a way that $\partial_{x^0}$ and
$\partial_{(x')^0}$ point in the same direction as $\pm\partial_t$ and
$\pm\partial_{\tau}$ on $\Sigma^{\pm}$ and $\overline{\mu}(\Sigma^{\pm})$,
respectively. This determines them uniquely.

\begin{remark}
The results of this subsection focus specifically on the case of the
Hawking radius, $M'$, but analogous results hold for $M^+_R$ with any $R>0$,
when $U$ is a coordinate neighbourhood of $\Sigma\setminus\mathcal{B}$.
\end{remark}

\begin{proposition}\label{Prop_complexGnormal2}
Expressing the metrics $g_{ab}$ and $(\bar{g}_R)_{ab}$ in these
Gaussian normal coordinates as
\[
g_{\mu\nu}dx^{\mu}dx^{\nu}=-(dx^0)^{\otimes 2}+h_{ij}dx^idx^j,\quad
(\bar{g}_R)_{\mu\nu}=(d(x')^0)^{\otimes 2}+h'_{ij}d(x')^id(x')^j,
\]
we have for $1\le i,j\le d-1$ and $n\ge 0$:
\begin{eqnarray}\label{Eqn_Infholo3}
\partial_0^{2n}h_{ij}|_U&=&i^{2n}\overline{\mu}^*\left(
(\partial'_0)^{2n}h'_{ij}|_{U'}\right),\\
\partial_0^{2n+1}h_{ij}|_U&=&i^{2n+1}
\overline{\mu}^*\left((\partial'_0)^{2n+1}h'_{ij}|_{U'}\right)=0.\nonumber
\end{eqnarray}
\end{proposition}
\begin{proof*}
On $U\cap\Sigma^+$ and $U'\cap\overline{\mu}(\Sigma^+)=\mu(U\cap\Sigma^+)$
this follows directly from Proposition 3.3 in \cite{Sanders2012}. The
same is then seen to be true on $U\cap \Sigma^-$, after applying the
isomorphism $I$ to $M^-\cup M^+$ and the isomorphism
$(\tau,x)\mapsto (i\pi R+\tau,x)$ to $M'$. The result extends by continuity
to $U\cap\mathcal{B}$ and $\overline{\mu}(U\cap\mathcal{B})$.
\end{proof*}
In \cite{Sanders2012} we argued that Equation (\ref{Eqn_Infholo3}) on
$\Sigma^+$ can be interpreted as an infinitesimal analytic continuation
in the Gaussian normal coordinates. Proposition \ref{Prop_complexGnormal2}
shows that this infinitesimal analytic continuation still works fine
across the bifurcation surface $\mathcal{B}$, where the Killing time
coordinate is no longer a good coordinate.

The information of Proposition \ref{Prop_complexGnormal2} allows us to
prove analogous statements for various objects which can be constructed
from the metric:
\begin{corollary}\label{Cor_CCurvature}
Expressing the Killing fields, metric, inverse metric, Christoffel symbol and
Riemann curvature of $g_{ab}$ and $(\bar{g}_R)_{ab}$ in Gaussian normal
coordinates we have for all $n\ge 0$:
\begin{eqnarray}\label{Eqn_Infholo4}
\partial_0^ng^{\mu\nu}|_U&=&i^{n+c}\overline{\mu}^*\left(
(\partial'_0)^n\bar{g}_R^{\mu\nu}|_{U'}\right),\nonumber\\
\partial_0^n(\Gamma^{\mu}_{\nu\rho})|_U&=&i^{n+c}\overline{\mu}^*\left(
(\partial'_0)^n(\bar{\Gamma}_R)^{\mu}_{\nu\rho}|_{U'}\right),\\
\partial_0^n(R_{\mu\nu\rho}^{\phantom{\mu\nu\rho}\alpha})|_U&=&
i^{n+c}\overline{\mu}^*\left((\partial'_0)^n
(\bar{R}_R)_{\mu\nu\rho}^{\phantom{\mu\nu\rho}\alpha}|_{U'}\right),\nonumber\\
\partial_0^n\xi^{\mu}|_U&=&i^{n+c+1}\overline{\mu}^*\left(
(\partial'_0)^n(\overline{\xi_R})^{\mu}|_{U'}\right)\nonumber
\end{eqnarray}
where $c$ is the number of lower indices equal to zero, minus the number
of upper indices equal to zero.
\end{corollary}
Whereas the left-hand side of all these equations is always real, the right-hand
side is real or purely imaginary, depending on whether $n+c$ is even or odd. In
this way we see that the expressions on both sides vanish when $n+c$ is odd.
\begin{proof*}
The first statement is obvious when one or both of the indices are $0$, because
the inverse metric components are then constantly $0$, $1$ or $-1$. For the
remaining indices this can be proven by induction by taking normal derivatives of
the equality $\delta^i_{\phantom{i}j}=h^{il}h_{jl}$ and its Euclidean counterpart
and using the results of Proposition \ref{Prop_complexGnormal2}.

The Christoffel symbol $\Gamma^{\mu}_{\nu\rho}$ vanishes when at least two of the
indices $\mu,\nu,\rho$ are zero, since $\partial_{\mu}g_{00}=0$. The analogous
statement in the Euclidean case is also true. For the remaining choices of
indices we can express the Christoffel symbol in terms of $h_{ij}$ and its inverse,
so the result follows from Proposition \ref{Prop_complexGnormal2} and the first
line of Equation (\ref{Eqn_Infholo4}) in a straightforward manner. The claim for
the Riemann curvature follows from its expression in terms of the Christoffel
symbols.

Finally we note that the Killing fields are uniquely determined by their initial
values on $\Sigma$ and Killing's Equation. In particular, $\partial_0\xi^0\equiv0$
and hence $\partial_0^n\xi^0|_U=0$ when $n\ge 1$ and similarly for
$\overline{\xi_R}^0$. Since $\xi^0|_U=v=\overline{\mu}^*(\overline{\xi_R}^0|_{U'})$
this proves the claim for $\mu=0$. For a detailed proof concerning the spatial
components we refer to the proof of Proposition 3.3 in \cite{Sanders2012}.
\end{proof*}

The following corollary is a related result on the geometry of the Cauchy
surface $\Sigma$ (see also Lemma \ref{Lem_CauchyGeodesics}):
\begin{corollary}\label{Cor_CauchyGeodesics}
For a smooth curve $\map{\gamma}{[0,1]}{\Sigma}$ the following
statements are equivalent:
\begin{enumerate}
\item $\gamma$ is a geodesic in $(\Sigma,h_{ij})$,
\item $\gamma$ is a geodesic in $M$,
\item $\overline{\mu}\circ\gamma$ is a geodesic in $M'$.
\end{enumerate}
\end{corollary}
The proof is the same as for Corollary 3.13 in \cite{Sanders2012}.

To extend the comparison of the geometry near $\Sigma$ in $M$ and
$\overline{\mu}(\Sigma)$ in $M'$ further we will now consider Riemannian
normal coordinates. These can be defined locally on any pseudo-Riemannian
manifold and for the purposes of defining them we will consider this general
setting.

Let $O$ be a convex normal neighbourhood of a pseudo-Riemannian manifold
$N=(\mathcal{N},g_{ab})$. We may introduce local coordinates on
$O^{\times 2}$ as follows. Define the embedding $O^{\times 2}\rightarrow TO$
by $(x,y)\mapsto(\exp_y^{-1}(x),y)$, where $\exp_y$ is the exponential map,
which defines a diffeomorphism from a neighbourhood of $0\in T_yN$ onto $O$.
Next, we introduce an arbitrary frame $(e_{\mu})^a$ of $TO$ to identify
$TO\simeq \mathbb{R}^d\times O$, with standard Cartesian coordinates
$\tilde{v}^{\mu}$ on $\mathbb{R}^d$ and arbitrary coordinates
$\tilde{y}^{\mu}$ on $O$. The composition of these two maps is an embedding
$\map{\rho}{O^{\times 2}}{\mathbb{R}^d\times O}$. The desired
coordinates on $O^{\times 2}$ are then given by
\begin{equation}\label{Eqn_SpecialCoords}
(v^{\mu},y^{\nu})=\rho^*(\tilde{v}^{\mu},\tilde{y}^{\nu}).
\end{equation}
For any fixed $y\in O$, the coordinates $v^{\mu}$ are Riemannian normal
coordinates on $O$, centred on $y$ and satisfying
$g_{\mu\nu}(0)=(e_{\mu})^a(y)(e_{\nu})_a(y)$. With a slight abuse of language
we will also refer to the coordinates $(v,y)$ as Riemannian normal coordinates
on $O^{\times 2}$.

We now return to the geometry of spacetimes with a static bifurcate Killing
horizon. For any point $y\in\Sigma$ we can choose convex normal neighbourhoods
$V\subset M$ and $V'\subset M'$ such that $y\in V$ and $\overline{\mu}(y)\in V'$.
The sets $V$ and $V'$ do not contain any pair of points which are conjugate
along the unique geodesic that connects them (cf.\ \cite{ONeill} Proposition
10.10 and the comments below it). We may also choose a convex normal
neighbourhood $U\subset\Sigma$ containing $y$ and such that
$U\subset V\cap\Sigma$ and $U':=\overline{\mu}(U)\subset V'$.

We let $x^{\mu}$ and $y^{\mu}$ be Gaussian normal coordinates on a
neighbourhood of $U$ and we let $(v^{\mu},y^{\mu})$ be Riemannian normal
coordinates on $U^{\times 2}$, defined using the frame $\partial_{y^{\mu}}$
associated to the coordinates $y^{\mu}$. Similarly, let $(x')^{\mu}$, $(y')^{\mu}$
be Gaussian normal coordinates near $U'$ such that $x^i=\overline{\mu}^*(x')^i$
and $y^i=\overline{\mu}^*(y')^i$ on $U$ and let $((v')^{\mu},(y')^{\mu})$ be
Riemannian normal coordinates defined using the frame $\partial_{(y')^{\mu}}$.
\begin{proposition}\label{Prop_CRiemann}
On $U^{\times 2}$ we have, in the coordinates introduced above:
\begin{eqnarray}
\partial_{x^0}^k\partial_{y^0}^lv^j&=&i^{k+l}(\overline{\mu}^{\times 2})^*
\left(\partial_{(x')^0}^k\partial_{(y')^0}^l(v')^j\right),\nonumber\\
\partial_{x^0}^k\partial_{y^0}^lv^0&=&i^{k+l-1}(\overline{\mu}^{\times 2})^*
\left(\partial_{(x')^0}^k\partial_{(y')^0}^l(v')^0\right).\nonumber
\end{eqnarray}
\end{proposition}
\begin{proof*}
For $x,y\in V$, $v^{\mu}(x,y)\in T_yV$ is the unique vector such that
$[0,1]\ni t\mapsto \exp_y(tv^{\mu}(x,y))$ is the unique geodesic in $V$ from
$y$ to $x$, where the index $\mu$ refers to the frame $\partial_{y^{\mu}}$.
For $x,y\in U$ we note that $v^0\equiv 0$, by Corollary \ref{Cor_CauchyGeodesics}.
Similarly, $(v')^0\equiv 0$ on $(U')^{\times 2}$. Furthermore, the relations
$x^i=\overline{\mu}^*(x')^i$ and $y^i=\overline{\mu}^*(y')^i$ on $U$ and the
fact that $\overline{\mu}$ is an isometry entail that
\[
v^j=(\overline{\mu}^{\times 2})^*(v')^j
\]
on $U^{\times 2}$, which proves the desired equality in the absence of normal
derivatives.

Let us now fix $x,y\in U$ and write $x=(0,x^i)$ and $y=(0,y^i)$. For sufficiently
small $s$ the curves $\gamma_0(s)$ and $\gamma_1(s)$ in $V$, defined in Gaussian
normal coordinates by $\gamma_0^{\mu}(s):=(s,y^i)$ and $\gamma_1^{\mu}(s):=(s,x^i)$,
are geodesics with tangent vector $n^{\mu}$ at $y$, resp.\ $x$. For some
sufficiently small $\epsilon>0$ we may then define the map
$\map{\gamma^{\mu}}{(-\epsilon,\epsilon)^{\times 2}\times[0,1]}{V}$ such that
$t\mapsto\gamma^{\mu}(r,s,t)$ is the unique geodesic in $V$ between $\gamma_0(r)$ and
$\gamma_1(s)$. Note that $\gamma$ is uniquely determined by the choice of $x,y$
and that $v^{\mu}(\gamma_1(s),\gamma_0(r))=\partial_t\gamma^{\mu}(r,s,t)|_{t=0}$.

We will now derive an equation for
$X^{\mu}_{k,l}(t):=\partial_r^k\partial_s^l\gamma^{\mu}(0,0,t)$ for all $k,l\ge 0$,
in analogy with the Jacobi equation (also known as the geodesic deviation equation).
We start with the geodesic equation for fixed $r,s$:
\[
\partial_t^2\gamma^{\mu}+\Gamma^{\mu}_{\nu\rho}(\gamma)\partial_t\gamma^{\nu}
\partial_t\gamma^{\rho}=0.
\]
Taking partial derivatives with respect to $r$ and $s$ and evaluating on $r=s=0$
then yields:
\begin{eqnarray}\label{Eqn_JacobiLor}
0=\partial_t^2X^{\mu}_{k,l}&+&\sum_{k'=0}^k\sum_{k''=0}^{k'}\sum_{l'=0}^l\sum_{l''=0}^{l'}
\left(\begin{array}{c}k\\ k'\end{array}\right)
\left(\begin{array}{c}k'\\ k''\end{array}\right)
\left(\begin{array}{c}l\\ l'\end{array}\right)
\left(\begin{array}{c}l'\\ l''\end{array}\right)\nonumber\\
&&\left(\partial_r^{k''}\partial_s^{l''}\Gamma^{\mu}_{\nu\rho}(\gamma)\right)
\partial_tX^{\nu}_{k'-k'',l'-l''}\partial_tX^{\rho}_{k-k',l-l'}.
\end{eqnarray}

Similarly, we consider the map
$\map{(\gamma')^{\mu}}{(-\epsilon,\epsilon)^{\times 2}\times[0,1]}{V'}$ such that
$t\mapsto(\gamma')^{\mu}(r,s,t)$ is the geodesic between $(\gamma')^{\mu}(r,0,0)$ and
$(\gamma')^{\mu}(0,s,1)$, where $r\mapsto(\gamma')^{\mu}(r,s,0)$ and
$s\mapsto(\gamma')^{\mu}(r,s,1)$
are geodesics through $x$ and $y$ with tangent vector $n^{\mu}$. Defining
$(X')^{\mu}_{k,l}(t):=\partial_r^k\partial_s^l(\gamma')^{\mu}(0,0,t)$ in
Gaussian normal coordinates one derives the equation
\begin{eqnarray}\label{Eqn_JacobiEucl}
0=\partial_t^2(X')^{\mu}_{k,l}&+&\sum_{k'=0}^k\sum_{k''=0}^{k'}\sum_{l'=0}^l\sum_{l''=0}^{l'}
\left(\begin{array}{c} k\\ k'\end{array}\right)
\left(\begin{array}{c} k'\\ k''\end{array}\right)
\left(\begin{array}{c} l\\ l'\end{array}\right)
\left(\begin{array}{c} l'\\ l''\end{array}\right)\nonumber\\
&&\left(\partial_r^{k''}\partial_s^{l''}(\bar{\Gamma}_R)^{\mu}_{\nu\rho}(\gamma')\right)
\partial_t(X')^{\nu}_{k'-k'',l'-l''}\partial_t(X')^{\rho}_{k-k',l-l'}
\end{eqnarray}
in analogy to Equation (\ref{Eqn_JacobiLor}).

Define $Y^{\mu}_{k,l}:=X^{\mu}_{k,l}-i^{k+l+c}(X')^{\mu}_{k,l}$ as a function
of $t$, where $c=-1$ if $\mu=0$ and $c=0$ else. We will prove by induction
over $N=k+l$ that $Y^{\mu}_{k,l}\equiv 0$. If $k=l=0$ we have
\[
Y^j_{0,0}=\gamma^j-(\gamma')^j\equiv0,\quad
Y^0_{0,0}=\gamma^0+i(\gamma')^0=0+i0=0.
\]
Now assume that the claim holds for all $(k',l')$ with $k'+l'\le N$ for some
$N\ge 0$ and consider $k,l$ with $k+l=N+1$. We may use Equations
(\ref{Eqn_JacobiLor},\ref{Eqn_JacobiEucl}) to write
\begin{eqnarray}
0=\partial_t^2Y^{\mu}_{k,l}&+&\sum_{k'=0}^k\sum_{k''=0}^{k'}\sum_{l'=0}^l\sum_{l''=0}^{l'}
\left(\begin{array}{c} k\\ k'\end{array}\right)
\left(\begin{array}{c} k'\\ k''\end{array}\right)
\left(\begin{array}{c} l\\ l'\end{array}\right)
\left(\begin{array}{c} l'\\ l''\end{array}\right)\nonumber\\
&&\left(\partial_r^{k''}\partial_s^{l''}\Gamma^{\mu}_{\nu\rho}(\gamma)\right)
\partial_tX^{\nu}_{k'-k'',l'-l''}\partial_tX^{\rho}_{k-k',l-l'}\nonumber\\
&&-i^{k+l+c}
\left(\partial_r^{k''}\partial_s^{l''}(\bar{\Gamma}_R)^{\mu}_{\nu\rho}(\gamma')\right)
\partial_t(X')^{\nu}_{k'-k'',l'-l''}\partial_t(X')^{\rho}_{k-k',l-l'}.\nonumber
\end{eqnarray}
If we use the chain rule to expand the derivatives acting on the Christoffel symbols,
then any normal derivative acting on $\Gamma^{\mu}_{\nu\rho}$ is accompanied by a factor
$X^0$. By the induction hypothesis and Corollary \ref{Cor_CCurvature} we therefore see
that all terms in the sum vanish, except those involving $X^{\mu}_{k,l}$ and
$(X')^{\mu}_{k,l}$ with $k+l=N+1$. This leads to
\begin{eqnarray}
0&=&\partial_t^2Y^{\mu}_{k,l}+
2\Gamma^{\mu}_{\nu\rho}(\gamma)\partial_tX^{\nu}_{0,0}\partial_tX^{\rho}_{k,l}
-2i^{k+l+c}(\bar{\Gamma}_R)^{\mu}_{\nu\rho}(\gamma)\partial_t(X')^{\nu}_{0,0}
\partial_t(X')^{\rho}_{k,l}\nonumber\\
&&+(\partial_{\alpha}\Gamma^{\mu}_{\nu\rho})(\gamma)\partial_tX^{\nu}_{0,0}
\partial_tX^{\rho}_{0,0}X^{\alpha}_{k,l}
-i^{k+l+c}(\partial'_{\alpha}(\bar{\Gamma}_R)^{\mu}_{\nu\rho})(\gamma)
\partial_t(X')^{\nu}_{0,0}\partial_t(X')^{\rho}_{0,0}(X')^{\alpha}_{k,l}\nonumber\\
&=&\partial_t^2Y^{\mu}_{k,l}+
2\Gamma^{\mu}_{\nu\rho}(\gamma)\partial_tX^{\nu}_{0,0}\partial_tY^{\rho}_{k,l}
+(\partial_{\alpha}\Gamma^{\mu}_{\nu\rho})(\gamma)\partial_tX^{\nu}_{0,0}
\partial_tX^{\rho}_{0,0}Y^{\alpha}_{k,l},\nonumber
\end{eqnarray}
which is the Jacobi equation for the vector field $Y^{\mu}_{k,l}$ on
$\gamma(0,0,.)$.\footnote{This equation is more commonly written using the covariant
derivative
$D_tY^{\mu}:=\partial_tY^{\mu}+\Gamma^{\mu}_{\nu\rho}(\gamma)\partial_tX^{\nu}_{0,0}Y^{\rho}$,
in terms of which the Jacobi equation reads
\[
D_t^2Y^{\mu}=-R_{\nu\alpha\beta}^{\phantom{\nu\alpha\beta}\mu}
\partial_tX^{\alpha}_{0,0}\partial_tX^{\beta}_{0,0}Y^{\nu},
\]
cf.\ \cite{Wald} Eq.\ (3.3.18).}
The values of $Y^{\mu}_{k,l}$ at the endpoints $x$ and $y$ of the geodesic
$\gamma(0,0,.)$ are easily determined by the fact that
$\gamma(r,s,0)=\gamma_0(r)=(r,\gamma_0^i(0))$ and
$\gamma(r,s,1)=\gamma_1(s)=(s,\gamma_1^i(0))$ and similarly for the Euclidean
case. Taking derivatives with respect to $r$ and $s$ one easily finds that
$Y^{\mu}_{k,l}(x)=Y^{\mu}_{k,l}(y)=0$ for all $k+l\ge 0$ and all $\mu$. Recall
that the points $x$ and $y$ are not conjugate along the unique geodesic
$\gamma(0,0,.)$ in $V$ that connects them, so the Jacobi vector field
$Y^{\mu}_{k,l}$ which vanishes at the boundaries must vanish identically.
Hence, $Y^{\mu}_{k,l}=0$ for all $k,l$. This result on $Y^{\mu}_{k,l}=0$
implies the proposition.
\end{proof*}

In our discussion of the Hadamard series, it will be convenient to consider
Riemannian normal coordinates based on an orthonormal frame $(e_{\mu})^a$,
rather than a coordinate frame. We will now discuss the modifications
that this entails for the above results. We may first choose an orthonormal
frame $(e_i)^a$ of $TU$, with a corresponding frame
$(e'_i)^a:=\overline{\mu}_*(e_i)^a$ of $TU'$. These frames can be extended
to orthonormal frames of $TM|_U$ and $TM'|_{U'}$, respectively, by including
the normal vector field $e_0^a:=n^a$, resp.\ $(e'_0)^a:=n^a$. Furthermore,
the frames can be extended to a neighbourhood of $U$, resp.\ $U'$, by
parallel transporting them along the geodesics whose tangent vectors are
$e_0$ on $U$, resp.\ $e'_0$ on $U'$.

Using these orthonormal frames we have
\begin{lemma}\label{Lem_CFrame}
Expressing all components and derivatives in terms of the Gaussian normal
coordinates $x^{\mu}$, resp.\ $(x')^{\mu}$, the orthonormal frames
$(e_{\alpha})^{\mu}$ and $(e'_{\alpha})^{\mu}$ satisfy
\begin{eqnarray}
(e_0)^{\mu}&=&\delta^{\mu}_0\ =\ (e'_0)^{\mu}\nonumber\\
(e_i)^0&=&0\ =\ (e_i')^0\nonumber\\
\partial_{x^0}^k(e_{\alpha})^{\mu}&=&i^k\overline{\mu}^*
(\partial_{(x')^0}^k(e'_{\alpha})^{\mu})\nonumber
\end{eqnarray}
on $U$ for all $k\ge 0$.
\end{lemma}
\begin{proof*}
By definition we have $e_0=\partial_{x^0}$, which means that $(e_0)^i\equiv 0$
and $(e_0)^0\equiv 1$. Similarly, $(e'_0)^i\equiv 0$ and $(e'_0)^0\equiv 1$,
from which the statement for $\alpha=0$ follows. For $\alpha>0$ the vanishing
of $(e_{\alpha})^0$ and $(e'_{\alpha})^0$ follows from the orthonormality of
the frames. Furthermore, the last equality holds for $k=0$, by definition of
$e'_i$ in terms of $e_i$ and by the fact that $x^i=\overline{\mu}^*(x')^i$.
The extension away from $U$, resp.\ $U'$, is then defined by the parallel
transport, which is expressed by the equations
\[
\partial_{x^0}(e_i)^{\mu}=-\frac12 (e_i)^{\nu}g^{\mu\rho}\partial_{x^0}g_{\nu\rho},
\]
resp.\
\[
\partial_{(x')^0}(e'_i)^{\mu}=-\frac12 (e'_i)^{\nu}(\bar{g}_R)^{\mu\rho}\partial_{x^0}(\bar{g}_R)_{\nu\rho},
\]
where we used the fact that the relevant components of the Christoffel symbols
simplify in Gaussian normal coordinates. For the components $(e_i)^0$ and
$(e'_i)^0$ the right-hand side vanishes identically, so these components vanish
identically. For the other components we may prove the desired equality by
induction over $k\ge 0$, by applying $(k-1)$ normal derivatives on both sides and
noting that the factors of $i$ are due to Proposition \ref{Prop_complexGnormal2},
Corollary \ref{Cor_CCurvature} and the induction hypothesis.
\end{proof*}

When using the frames $e_{\alpha}$ and $e'_{\alpha}$ to define Riemannian normal
coordinates $(w^{\mu},y^{\mu})$ and $((w')^{\mu},(y')^{\mu})$, the corresponding
statement of Proposition \ref{Prop_CRiemann} remains valid. To see this, we introduce
the dual frames $(f^{\alpha})_{\mu}:=g_{\mu\nu}\eta^{\alpha\beta}(e_{\beta})^{\nu}$
of $(e_{\alpha})^{\mu}$ and similarly for $(e'_{\alpha})^{\mu}$. Note that
$(f^{\alpha})_{\mu}(e_{\beta})^{\mu}=\delta^{\alpha}_{\phantom{\alpha}\beta}$ and
$(f^{\alpha})_{\mu}(e_{\alpha})^{\nu}=\delta^{\nu}_{\phantom{\nu}\mu}$. (The first
follows directly from the fact that the $(e_{\alpha})^{\mu}$ are orthonormal. The
second follows from the fact that the $(e_{\alpha})^{\mu}$ are a frame, because it
holds when contracted with any $(e_{\beta})^{\mu}$.) We may now write
$w^{\mu}(x,y)=(f^{\mu})_{\nu}(y)v^{\nu}(x,y)$. Using the definition of the dual
frame and Lemma \ref{Lem_CFrame} it follows that the desired equalities for
$w^{\mu}$ and $(w')^{\mu}$ are equivalent to those of Proposition
\ref{Prop_CRiemann}. This proves
\begin{proposition}\label{Prop_CRiemann'}
On $U^{\times 2}$ we have, in the coordinates introduced above:
\begin{eqnarray}
\partial_{x^0}^k\partial_{y^0}^lw^j&=&i^{k+l}(\overline{\mu}^{\times 2})^*
\left(\partial_{(x')^0}^k\partial_{(y')^0}^l(w')^j\right),\nonumber\\
\partial_{x^0}^k\partial_{y^0}^lw^0&=&i^{k+l-1}(\overline{\mu}^{\times 2})^*
\left(\partial_{(x')^0}^k\partial_{(y')^0}^l(w')^0\right).\nonumber
\end{eqnarray}
\end{proposition}

To close this section, we consider the squared geodesic distance of a
pseudo-Riemannian manifold, which is also known as Synge's world function
in the Lorentzian case. It is defined as
\begin{equation}
\sigma(x,y):=\frac12 \|\exp_y^{-1}(x)\|^2_{g(y)},\nonumber
\end{equation}
and in general it may take both positive and negative values. In the
Riemannian normal coordinates $v^{\mu}$ (defined using the frame
$\partial_{y^{\mu}}$) it takes the form
\[
\sigma(v,y)=\frac12 (\exp_y^*g)_{\mu\nu}(0)v^{\mu}v^{\nu}.
\]
As the map $t\mapsto\exp_y(tv^{\mu})$ is a geodesic, by definition of
the exponential map, one may use the geodesic equation and a partial
integration to show that
\[
\sigma(v,y)=\frac12\int_0^1(\exp_y^*g)_{\mu\nu}(tv)v^{\mu}v^{\nu}dt.
\]
In other words, $\sigma(x,y)$ is the length squared of the unique
geodesic in $V$ which connects $x$ to $y$ in unit parameter time.
We therefore have $\sigma(x,y)=\sigma(y,x)$ for all $x,y\in V$ and
one can also show that
\begin{eqnarray}\label{Eqn_SyngeProp}
(\exp_y^*g)_{\mu\nu}(0)v^{\nu}&=&\partial_{\mu}\sigma(v,y)
=(\exp_y^*g)_{\mu\nu}(v)v^{\nu},\nonumber\\
(\exp_y^*g)^{\mu\nu}(v)\partial_{\nu}\sigma(v,y)&=&v^{\mu}
=(\exp_y^*g)^{\mu\nu}(0)\partial_{\nu}\sigma(v,y),\nonumber\\
2\sigma(v,y)&=&(\exp_y^*g)^{\mu\nu}(v)\partial_{\mu}\sigma(v,y)
\partial_{\nu}\sigma(v,y),\\
\sigma(0,y)&=&\partial_{\mu}\sigma(0,y)=0,\nonumber\\
\partial_{\mu}\partial_{\nu}\sigma(0,y)&=&(\exp_y^*g)_{\mu\nu}(0),\nonumber
\end{eqnarray}
where all derivatives refer to the coordinates $v^{\mu}$.

A comparison of $\sigma$ in the Euclidean and Lorentzian case yields:
\begin{corollary}\label{Cor_CSynge}
Let $\sigma$ be Synge's world function on $V$ and let $\bar{\sigma}_R$ be
the squared geodesic distance on $V'$. For all $k,l\ge 0$ we have
\begin{equation}
\partial_{x^0}^k\partial_{y^0}^l\sigma=i^{k+l}(\overline{\mu}^{\times 2})^*
(\partial_{(x')^0}^k\partial_{(y')^0}^l\bar{\sigma}_R)\nonumber
\end{equation}
on $U^{\times 2}$.
\end{corollary}
\begin{proof*}
This is a direct consequence of Propositions \ref{Prop_CRiemann} and
\ref{Prop_complexGnormal2} and the fact that
\[
\sigma=-\frac12(v^0)^2+\frac12h_{ij}v^iv^j,\quad
\bar{\sigma}_R=-\frac{i^2}{2}((v')^0)^2+\frac12h'_{ij}(v')^i(v')^j,
\]
where $h_{ij}$ and $h'_{ij}$ are evaluated at $y$ and $y'$, respectively.
\end{proof*}

\section{The linear scalar quantum field}\label{Sec_FField}

In this section, we introduce the linear scalar field, its quantisation
in a spacetime with a bifurcate Killing horizon and the class of
quasi-free Hadamard states. We apply the initial value formulation of the
field equation to two-point distributions, which yields a convenient
setting to discuss the local aspects of the Wick rotation in the static
case. We also briefly review how a Wick rotation can be used to obtain double
$\beta$-KMS states in the disconnected spacetime $M^+\cup M^-$ (and hence
$\beta$-KMS states in $M^+$). For the purpose of this Wick rotation, we use
global methods as in \cite{Sanders2012}, which complement the local
description that is used throughout most of this paper.

As a matter of convention, we will identify distribution densities on
$M,\ M^+_R,\ \Sigma$ etc.\ with distributions, using the respective
volume forms $d\mathrm{vol}_g$, $d\mathrm{vol}_{g_R}$ and $d\mathrm{vol}_h$.
To unburden our notation we will often leave the volume form implicit, which
should not lead to any confusion. However, we point out that the volume form
is important when restricting to submanifolds, because in that case a change
in volume form is involved.

\subsection{Initial value formulation of the linear scalar field}\label{SSec_InitValFF}

We recall that it is well understood how to quantise a linear scalar field
on any globally hyperbolic spacetime $M$
(cf.\ e.g.\ \cite{Baer+2009,Dimock1980,Brunetti+2003,Baer+2007}). At the
classical level the theory is described by the (modified) Klein-Gordon
operator
\[
K:=-\Box+V,
\]
where the real-valued function $V\in C^{\infty}(M,\mathbb{R})$ serves as a
potential. In any globally hyperbolic spacetime, the operator $K$ has unique
advanced ($-$) and retarded ($+$) fundamental solutions $E^{\pm}$ and we
define $E:=E^--E^+$. We describe the quantum theory by the Weyl $C^*$-algebra
$\alg{A}$, generated by the operators $W(f)$ with
$f\in C_0^{\infty}(M,\mathbb{R})$ satisfying the Weyl relations
\begin{enumerate}
\item $W(f)^*=W(-f)$,
\item $W(Kf)=I$,
\item $W(f)W(f')=e^{\frac{-i}{2}E(f,f')}W(f+f')$.
\end{enumerate}
Note that $W(f)$ and $W(f')$ are linearly dependent if and only if
they are equal, which is the case if and only if
$f'\in f+KC_0^{\infty}(M,\mathbb{R})$.\footnote{Proof: if $W(f)=\lambda W(f')$
for some $\lambda\in\mathbb{C}$, then we may use the fact that $W(-f)=W(f)^{-1}$
and compute for all $\chi\in C_0^{\infty}(M,\mathbb{R})$:
\[
I=\lambda W(\chi)W(f')W(-f)W(-\chi)=\lambda e^{-iE(\chi,f'-f)+\frac{i}{2}E(f',f)}W(f'-f).
\]
Comparing a general $\chi$ with $\chi=0$ we can eliminate the Weyl operators to
find $1=e^{-iE(\chi,f'-f)}$ for all $\chi$, which means that $E(f'-f)=0$. By a
standard result \cite{Dimock1980} it follows that
$f'-f\in K C_0^{\infty}(M,\mathbb{R})$, which in turn implies
$W(f)=W(f-f')W(f')=W(f')$.}

An algebraic state $\omega$ on the Weyl algebra $\alg{A}$ gives rise to a
representation $\pi_{\omega}$ of the algebra on a Hilbert space
$\mathcal{H}_{\omega}$ by the GNS-construction. We will mostly consider
states for which the maps
\[
\omega_n(f_1,\ldots,f_n):=(-i)^n\partial_{s_1}\cdots\partial_{s_n}
\omega(W(s_1f_1)\cdots W(s_nf_n))|_{s_1=\ldots=s_n=0}
\]
are distributions on $M^{\times n}$ for all $n\ge 1$: the $n$-point
distributions. In fact, our primary interest is in quasi-free states, for
which all $n$-point distributions can be expressed in terms of the two-point
distribution via Wick's Theorem. We mention without proof the following
well-known result:
\begin{proposition}
The two-point distribution $\omega_2\in\mathcal{D}(M^{\times 2})$ of any
state $\omega$ has the following properties:
\begin{enumerate}
\item $\omega_2(x,y)$ solves the Klein-Gordon equation in both variables,
\item $2\omega_{2-}(x,y):=\omega_2(x,y)-\omega_2(y,x)=iE(x,y)$,
\item $\omega_2(\overline{f},f)\ge 0$ for all $f\in C_0^{\infty}(M)$.
\end{enumerate}
Furthermore, any distribution $\omega_2$ with these properties is the
two-point distribution of a unique quasi-free state.
\end{proposition}
For quasi-free states it only remains to analyse the distributions
$\omega_2$ with these three properties. Equivalently, we can study
one-particle structures:
\begin{definition}
A \emph{one-particle structure} for $K$ on $M$ is a pair
$(p,\mathcal{H})$, which consists of a Hilbert space $\mathcal{H}$ and
an $\mathcal{H}$-valued distribution $p$ on $M$ such that
\begin{enumerate}
\item $p$ has a dense range,
\item $p(Kf)=0$ for all $f\in C_0^{\infty}(M)$,
\item $\langle p(\bar{f}),p(f')\rangle-\langle p(\bar{f'}),p(f)\rangle=iE(f,f')$.
\end{enumerate}
\end{definition}
The bijective relationship between one-particle structures and two-point
distributions is given by
\begin{equation}\label{Eqn_StateOPS}
\omega_2(f,f')=\langle p(\bar{f}),p(f')\rangle.
\end{equation}
Note that any two-point distribution $\omega_2$ determines a one-particle Hilbert
space $\mathcal{K}_{\omega_2}$, which is defined as the Hilbert space completion
of $C_0^{\infty}(M)$ after dividing out the linear space of vectors of zero norm
in the semi-definite inner product $\langle f,h\rangle:=\omega_2(\overline{f},h)$.
The map $\map{K_{\omega_2}}{C_0^{\infty}(M)}{\mathcal{K}_{\omega_2}}$ defined by
$K_{\omega_2}(f):=[f]$ is a Hilbert space-valued distribution
(cf.\ \cite{Strohmaier+2002}), which may be interpreted as
$K_{\omega_2}(f)=\pi_{\omega}(\Phi(f))\Omega_{\omega}$, where
$\Omega_{\omega}\in\mathcal{H}_{\omega}$ is the GNS-vector in the
GNS-representation $\pi_{\omega}$ of the quasi-free state $\omega$ determined
by $\omega_2$.

Let us now recall the initial value formulation of the Klein-Gordon equation
in a globally hyperbolic spacetime $M$ on a Cauchy surface $\Sigma\subset M$
with future pointing normal vector field $n^a$.
If $\omega_2$ is the two-point distribution of a state, then it is completely
determined by its initial data\footnote{To analyse the singularities and
restrictions of distributions we freely make use of basic notions and results
from microlocal analysis, referring the reader to \cite{Hoermander} for
details.}
on $\Sigma^{\times 2}$, namely
\begin{eqnarray}
\omega_{2,00}:=\omega_2|_{\Sigma^{\times 2}},&&
\omega_{2,01}:=(1\otimes n^a\nabla_a)\omega_2|_{\Sigma^{\times 2}}\nonumber\\
\omega_{2,10}:=(n^a\nabla_a\otimes 1)\omega_2|_{\Sigma^{\times 2}},&&
\omega_{2,11}:=(n^a\nabla_a\otimes n^b\nabla_b)\omega_2|_{\Sigma^{\times 2}}.\nonumber
\end{eqnarray}
These distributional restrictions are well-defined by a microlocal argument
and for their definition we treat $\omega_2$ as a distribution, not a
distribution density. To see how these initial data determine $\omega_2$ we
let $f,f'\in C_0^{\infty}(M)$ and we introduce the initial data
$f_0:=Ef|_{\Sigma}$, $f_1:=n^a\nabla_aEf|_{\Sigma}$ and similarly for $f'$.
By a standard computation (analogous to Lemma A.1 of \cite{Dimock1980}) one
may show that
\begin{eqnarray}\label{Eqn_InitValState}
\omega_2(f,f')&=&\sum_{m,n=0}^1(-1)^{m+n}\omega_{2,mn}(f_{1-m},f'_{1-n}),
\end{eqnarray}
where we used the fact that $\omega_2$ is a distributional bi-solution to
the Klein-Gordon equation. (Recall that the volume forms of $M$, respectively
$\Sigma$, are implicit on the left, respectively right-hand side of this
equation.)

There is a preferred class of states, called Hadamard states, which are
characterised by the fact that their two-point distribution has a singularity
structure at short distances that is of the same form as that of the Minkowski
vacuum state. To put it more precisely, $\omega_2$ is of Hadamard form if and
only if
\cite{Radzikowski1996}
\begin{eqnarray}\label{Eqn_DefHad}
WF(\omega_2)&=&\left\{(x,k;y,l)\in T^*M^{\times 2}|\ l\not=0 \mathrm{\ is\
future\ pointing\ and\ lightlike\ and\ } (y,l)\right.\\
&&\left.\mathrm{\ generates\ a\ geodesic\ } \gamma
\mathrm{\ which\ goes\ through\ } x \mathrm{\ with\ tangent\
vector\ } -k\right\}.\nonumber
\end{eqnarray}

By the Propagation of Singularities Theorem and the fact that $\omega_2$
solves the Klein-Gordon equation in both variables it suffices to check the
condition in Equation (\ref{Eqn_DefHad}) on a Cauchy surface $\Sigma$:
\[
WF(\omega_2)|_{\Sigma}\subset\left\{(x,-k;x,k)|\ (x,k)\in V^+M|_{\Sigma}\right\},
\]
where $V^+M$ denotes the fibre bundle of future pointing covectors.
Unfortunately, it is somewhat complicated to see whether a state $\omega_2$ is
Hadamard by inspecting its initial data on a Cauchy surface $\Sigma$. The
initial data of $\omega_2$ should be smooth away from the diagonal in
$\Sigma^{\times 2}$, so it suffices to characterise the singularities near
the diagonal.\footnote{Conversely, if $\omega_2$ has the correct singularity
structure near the diagonal on $\Sigma$, then it follows essentially from
\cite{Radzikowski1996_2} and the propagation of singularities that $\omega_2$
is Hadamard and hence smooth away from the diagonal in $\Sigma^{\times 2}$.}
However, for the singularities near the diagonal we are not aware of any
general argument that avoids the use of the Hadamard parametrix construction,
which involves the Hadamard series after which Hadamard states are
named.\footnote{The recent work \cite{Gerard+2012} presents a more elegant
procedure, but it makes additional assumptions on the Cauchy surface that we
wish to avoid.} We will explain this construction in detail for both the
Lorentzian and Euclidean setting in Section \ref{Sec_Had} below.

\subsection{Double $\beta$-KMS states on $M^+\cup M^-$ in the stationary case}\label{SSec_DoubleKMS}

We consider the Klein-Gordon equation on a spacetime $M$ with a stationary
bifurcate Killing horizon. Because the right wedge $M^+$ is a (possibly
disconnected) stationary, globally hyperbolic spacetime we can apply the
analysis of \cite{Sanders2012} to obtain ground and $\beta$-KMS states under
suitable circumstances. We will briefly review these results and show how
they can be extended to the disconnected spacetime $M^+\cup M^-$.

In order to apply the results of \cite{Sanders2012}, we assume that the potential
$V$ is stationary and positive on the right wedge:
\[
\xi^a\nabla_aV|_{M^+}=\partial_tV|_{M^+}\equiv 0,\quad V|_{M^+}>0.
\]
On $M^+$ the Klein-Gordon operator can be written in terms of the Killing time
coordinate $t$ and the induced metric $h_{ij}$ on the Cauchy surface $\Sigma^+$:
\begin{eqnarray}\label{Eqn_DefC}
v^{\frac32}Kv^{\frac12}&=&\partial_t^2+D\partial_t+C,\nonumber\\
D&:=&-(\nabla^{(h)}_iw^i+w^i\nabla^{(h)}_i),\\
C&:=&-v^{\frac12}\nabla^{(h)}_i(vh^{ij}-v^{-1}w^iw^j)\nabla^{(h)}_j
v^{\frac12}+Vv^2.\nonumber
\end{eqnarray}
The operator $K$ is a symmetric operator on $L^2(M^+)$ defined on the dense
domain $C_0^{\infty}(M^+)$.

We now formulate the fundamental result on ground and $\beta$-KMS states on
$M^+$ (\cite{Sanders2012} Theorems 5.1 and 6.2, which may be generalised to
spacetimes which are not necessarily connected). For an overview of further
properties of the ground and $\beta$-KMS states, we refer to
\cite{Sanders2012} and references therein.
\begin{theorem}\label{Thm_KMS}
Consider a linear scalar field on $M^+$ with a stationary potential $V$ such
that $V>0$.
\begin{enumerate}
\item There exists a unique extremal ground state $\omega^0$ with a
well-defined, vanishing one-point distribution.
\item For every $\beta>0$ there exists a unique extremal $\beta$-KMS state
$\omega^{(\beta)}$ with a well-defined, vanishing one-point distribution.
\end{enumerate}
All these states are quasi-free and Hadamard.
\end{theorem}

\begin{remark}\label{Rem_GroundKMS}
Other ground and $\beta$-KMS states can be obtained as follows. Firstly,
one may replace the quantum field $\Phi(x)$ by $\Phi(x)+\phi(x)I$ (a gauge
transformation of the second kind), where $\phi(x)$ is a real-valued, Killing
time independent (weak) solution of the Klein-Gordon equation, if such
solutions exist. More precisely, we replace $W(f)$ by $e^{i\phi(f)}W(f)$, where
$\phi$ is interpreted as a distribution density. This defines an automorphism
of the Weyl algebra and the pull-back of the states in Theorem \ref{Thm_KMS}
under this isomorphism remain extremal ground, resp.\ $\beta$-KMS states.
Furthermore, one may take mixtures of such ground or $\beta$-KMS states to
obtain non-extremal ones. It can be shown that all ground and $\beta$-KMS
states are of this form \cite{Sanders2012} and that their two-point
distributions $\omega_2$ majorise those of Theorem \ref{Thm_KMS}
(i.e.\ $\omega_2(\overline{f},f)\ge\omega^{(\beta)}_2(\overline{f},f)$ and
similarly for ground states). If any solutions $\phi(x)$ exist at all, the
corresponding ground and $\beta$-KMS states are often discarded, because the
one-point distribution $\phi(x)$ grows exponentially near spatial infinity.
Restricting attention e.g.\ to tempered $n$-point distributions in Minkowksi
spacetime one disqualifies all states other than the ones in Theorem
\ref{Thm_KMS}.
\end{remark}

We will now describe how the one-particle structure
$(p_{(\beta)},\mathcal{H}_{(\beta)})$, which gives rise to the two-point
distribution $\omega^{(\beta)}_2$ of the $\beta$-KMS state on $M^+$, can
be obtained from the classical Hilbert space of finite energy solutions
(cf.\ \cite{Kay1985_1}).

We let $\mathcal{H}_e$ be the Hilbert space of finite energy solutions $\phi$
of the Klein-Gordon equation on $M^+$, where the norm is given by the square root
of the energy. $\mathcal{H}_e$ contains a dense subset of spacelike compact,
smooth solutions, whose energy may be obtained by integrating the energy density
over any Cauchy surface (cf.\ \cite{Sanders2012}). Complex conjugation on these
spacelike compact solutions preserves the energy, so it can be extended to a
complex conjugation $C$ on $\mathcal{H}_e$ (i.e.\ a complex anti-linear
involution). There is an $\mathcal{H}_e$-valued distribution
\[
p_{cl}:C_0^{\infty}(M^+)\rightarrow\mathcal{H}_e:f\mapsto Ef,
\]
which satisfies $Cp_{cl}(f)=p_{cl}(\bar{f})$ and solves the Klein-Gordon
equation in the sense that $p_{cl}(Kf)=0$ for all $f\in C_0^{\infty}(M^+)$.
The Killing time evolution is implemented on $\mathcal{H}_e$ by a strongly
continuous unitary group $e^{itH_e}$ , where the Hamiltonian $H_e$ is an
invertible self-adjoint operator. We note that the range of $p_{cl}$ is a
core for all powers of $H_e$ and for $|H_e|^{-1}$ (cf.\ \cite{Sanders2012}
Thm.\ 4.2) and we let $P_{\pm}$ denote the spectral projections onto the
positive and negative spectrum of $H_e$.

The one-particle structure $(p_{(\beta)},\mathcal{H}_{(\beta)})$ can now be
expressed as (cf.\ \cite{Sanders2012} Thm.\ 4.3):
\begin{eqnarray}
p_{(\beta)}(f)&:=&\sqrt{2}P_-|H_e|^{-\frac12}(I-e^{-\beta|H_e|})^{-\frac12}p_{cl}(f)
\nonumber\\
&&\oplus\sqrt{2}P_+|H_e|^{-\frac12}e^{-\frac{\beta}{2}|H_e|}
(I-e^{-\beta|H_e|})^{-\frac12}p_{cl}(f),\nonumber
\end{eqnarray}
which is a distribution on $M^+$ with values in the Hilbert space
$P_-\mathcal{H}_e\oplus P_+\mathcal{H}_e\simeq\mathcal{H}_e$. Note that
$p_{(\beta)}$ has a dense range, so
$H_{(\beta)}=\mathcal{H}_e$.\footnote{Proof: Given any $\psi\in\mathcal{H}_e$
we define $\psi_{\pm}:=P_{\pm}\psi$. For a dense set of such $\psi$ the vector
$\tilde{\psi}:=|H_e|^{\frac12}(I-e^{-\beta|H_e|})^{\frac12}
(\psi_-\oplus e^{\frac{\beta}{2}|H_e|}\psi_+)$ is well-defined. Because the
range of $p_{cl}$ is a core for $|H_e|^{-\frac12}(I-e^{-\beta|H_e|})^{-\frac12}$,
we can find a sequence $f_n\in C_0^{\infty}(M^+)$ such that
$p_{cl}(f_n)$ converges to $\tilde{\psi}$ and
$|H_e|^{-\frac12}(I-e^{-\beta|H_e|})^{-\frac12}p_{cl}(f_n)$ converges to
$\psi_-\oplus e^{\frac{\beta}{2}|H_e|}\psi_+$. Because
$e^{-\frac{\beta}{2}|H_e|}$ is bounded it follows that
$p_{(\beta)}(f_n)$ converges to $\sqrt{2}\psi$ and therefore that
$p_{(\beta)}$ has a dense range.} The Killing time evolution
is now implemented by $H=|H_e|\oplus -|H_e|=-H_e$. A similar,
but simpler, description holds for the ground state.

We now assume that $M$ admits a wedge reflection and we wish to extend
the states above from $M^+$ to the union $M^+\cup M^-$. More precisely, in
this section we will only assume that there is an isometric, involutive
diffeomorphism
$I$ of $M^+\cup M^-$ which reverses the time orientation and which satisfies
$I^*\xi^a=\xi^a$. This assumption is even weaker than the existence of a
weak wedge reflection, but it suffices for the purposes of this section,
because we are not yet investigating extensions across the Killing horizon.
Note that a Cauchy surface $\Sigma^+$ of $M^+$ maps to a Cauchy surface
$\Sigma^-:=I(\Sigma^+)$ of $M^-$.

The quotient space $C_0^{\infty}(M^+\cup M^-,\mathbb{R})/KC_0^{\infty}(M^+\cup M^-,\mathbb{R})$
is a symplectic space with the symplectic form $E$. If we also assume
\[
I^*V=V,
\]
then it naturally carries the structure of a double linear dynamical
system, in the sense of \cite{Kay1985_1}. This means that it is a direct
sum of two symplectic spaces,
\[
(C_0^{\infty}(M^+,\mathbb{R})/KC_0^{\infty}(M^+),\mathbb{R})\oplus
(C_0^{\infty}(M^-,\mathbb{R})/KC_0^{\infty}(M^-,\mathbb{R})),
\]
each of which is preserved under the Killing time evolution, and there is
a linear involution, namely $I^*$, which maps the symplectic subspace of
$M^+$ onto that of $M^-$ and vice versa, which commutes with the Killing
time evolution and which is anti-symplectic in the sense that
$E(I^*f,I^*f')=-E(f,f')$. To see how this last property of $I^*$ arises
we only need to fix a Cauchy surface $\Sigma^+$ of $M^+$ and to express
the symplectic form $E$ in terms of initial data on
$\Sigma:=\Sigma^+\cup I(\Sigma^+)$:
\begin{equation}\label{Eqn_SymplData}
E(f,f')=\int_{\Sigma}Ef\cdot n^a\nabla_aEf'-Ef'\cdot n^a\nabla_aEf.
\end{equation}
To compute $E(I^*f,I^*f')$, the data of $EI^*f$ and $EI^*f'$ are
expressed as the pull-backs of the data of $Ef$ and $Ef'$ by
$I|_{\Sigma}$, where the normal derivatives get an additional sign,
because $I$ reverses the time orientation.

The Weyl algebra of the scalar quantum field on $M^+\cup M^-$ is
the spatial tensor product of the algebras on $M^+$ and $M^-$ and
the wedge reflection $I$ gives rise to a complex anti-linear involution
$\tau_I:zW(f)\mapsto \overline{z}W(I^*f)$, which preserves
products and the $^*$-operation and which commutes with the Killing
time evolution. We will call a state $\omega$ on $M^+\cup M^-$ a
double $\beta$-KMS state when its restriction to $M^+$ is a
$\beta$-KMS state $\omega^{(\beta)}$ and when
\begin{eqnarray}\label{Eqn_DefdKMS}
\omega(W(f^++f^-))=F_{W(I^*f^-),W(f^+)}\left(\frac{i\beta}{2}\right)
\end{eqnarray}
for all $f^{\pm}\in C_0^{\infty}(M^{\pm})$, where $F_{W(I^*f^-)W(f^+)}$
is the bounded continuous function on
$\overline{\mathcal{S}}_{\beta}:=\{z\in\mathbb{C}|\ \mathrm{Im}(z)\in[0,\beta]\}$
which is holomorphic on its interior and which satisfies
$F_{W(I^*f^-)W(f^+)}(t)=\omega^{(\beta)}(W(I^*f^-)W(\Phi_{-t}(f^+)))$
and $F_{W(I^*f^-)W(f^+)}(t+i\beta)=\omega^{(\beta)}(W(\Phi_{-t}(f^+))W(I^*f^-))$.
This function exists by the definition of $\beta$-KMS states. We note
that any double $\beta$-KMS state is invariant under the wedge
reflection in the sense that
\begin{equation}\label{Eqn_WedgeInv}
\omega(\tau_I(A))=\overline{\omega(A)},
\end{equation}
because $F_{W(I^*f^-),W(f^+)}\left(\frac{i\beta}{2}\right)=
F_{W(f^+),W(I^*f^-)}\left(\frac{i\beta}{2}\right)$.

For any $\beta>0$ the one-particle structure $(p_{(\beta)},\mathcal{H}_e)$
on $M^+$ also determines a double $\beta$-KMS one-particle structure in
the sense of \cite{Kay1985_1} on the double linear dynamical system of
$M^+\cup M^-$. This is a one-particle structure $(p,\mathcal{H})$ on
$M^+\cup M^-$ such that $p$ has a dense range on each of $M^+$ and $M^-$;
the Killing time evolution is implemented by a strongly continuous unitary
group $e^{itH}$, where $H$ has no zero eigenvalue; there is a complex
conjugation $C$ on $\mathcal{H}$ such that $Cp(f)=-p(I^*\bar{f})$ for all
$f\in C_0^{\infty}(M^+\cup M^-)$; and $p(C_0^{\infty}(M^{\pm}))$ is in the
domain of $e^{\mp\frac{\beta}{2}H}$ with
\[
e^{\mp\frac{\beta}{2}H}p(f)=-p(I^*f),\quad
f\in C_0^{\infty}(M^{\pm}).
\]
To obtain this double $\beta$-KMS structure we take
$(p^d_{(\beta)},\mathcal{H}_e)$ with
\[
p^d_{(\beta)}(f^++f^-):=p_{(\beta)}(f^+)-e^{-\frac{\beta}{2}H}
p_{(\beta)}(I^*f^-),
\]
where $f^{\pm}\in C_0^{\infty}(M^{\pm})$. It can be verified that this
is well-defined and it has all the desired properties, where the
Killing time evolution is again implemented by $H=-H_e$. The complex
conjugation on $\mathcal{H}_e$ is the given conjugation $C$, which
satisfies $CH=-HC$, so it exchanges the negative and positive frequency
subspaces of $H_e$. We denote by $\omega^{(\beta),d}_2$ the two-point
distribution on $M^+\cup M^-$ determined by $p^d_{(\beta)}$.

Kay has shown that this double $\beta$-KMS one-particle structure is
unique \cite{Kay1985_1} and he considered corresponding quasi-free
double $\beta$-KMS states on double wedge algebras in
\cite{Kay1985_2,Kay1985}. In our case, we may obtain the following
result:
\begin{theorem}\label{Thm_DoubleKMS}
Let $M$ be a globally hyperbolic spacetime with a stationary
bifurcate Killing horizon and assume that there is an isometric,
involutive diffeomorphism
$I$ of $M^+\cup M^-$ onto itself which reverses the time
orientation and satisfies $I^*\xi^a=\xi^a$. We consider the
Klein-Gordon equation with a stationary potential $V$ such that $V>0$
and $I^*V=V$. For each $\beta>0$ there exists a unique double
$\beta$-KMS state $\omega^{(\beta),d}$ on $M^+\cup M^-$ whose
restriction to $M^+$ is $\omega^{(\beta)}$. This state is pure,
quasi-free, Hadamard, it has the Reeh-Schlieder property and its
two-point distribution is given by $\omega^{(\beta),d}_2$.
\end{theorem}
\begin{proof*}
By Equation (\ref{Eqn_DefdKMS}) there is at most one double
$\beta$-KMS state on $M^+\cup M^-$ which restricts to a given
$\beta$-KMS state on $M^+$. It is clear that the quasi-free
state $\omega^{(\beta),d}$ with two-point distribution
$\omega^{(\beta),d}_2$ restricts to $\omega^{(\beta)}$ on
$M^+$ and we will show this is a double $\beta$-KMS state and
prove its properties.

Using the complex conjugation $C$ and the properties of
$p_{(\beta)}^d$ we find
\begin{eqnarray}\label{Eqn_Iinv2pt}
\omega^{(\beta),d}_2(I^*f,I^*f')&=&\langle p_{(\beta)}^d(I^*\overline{f}),
p_{(\beta)}^d(I^*f')\rangle
=\langle -Cp_{(\beta)}^d(f),-Cp_{(\beta)}^d(\overline{f'})\rangle\\
&=&\langle p_{(\beta)}^d(\overline{f'}),p_{(\beta)}^d(f)\rangle
=\omega^{(\beta),d}_2(f',f)\nonumber
\end{eqnarray}
Because $\omega^{(\beta),d}$ restricts to $\omega^{(\beta)}$ on
$M^+$ it is Hadamard there. The symmetry property above proves
that $\omega^{(\beta),d}$ is Hadamard on $M^-$ as well, because
$I$ reverses the time orientation and it interchanges the
arguments of $\omega^{(\beta),d}_2$. Furthermore, we may use
$Cp_{(\beta)}^d(f^{\pm})=-p_{(\beta)}^d(I^*f^{\pm})=e^{\mp\frac{\beta}{2}H}
p_{(\beta)}^d(f^{\pm})$ for any
$f^{\pm}\in C_0^{\infty}(M^{\pm},\mathbb{R})$ to see that
\begin{eqnarray}
\omega^{(\beta),d}_2(f^+,f^-)&=&\langle p_{(\beta)}^d(f^+),p_{(\beta)}^d(f^-)\rangle
=\langle Cp_{(\beta)}^d(f^-),Cp_{(\beta)}^d(f^+)\rangle\nonumber\\
&=&\langle e^{\frac{\beta}{2}H}p_{(\beta)}^d(f^-),
e^{-\frac{\beta}{2}H}p_{(\beta)}^d(f^+)\rangle
=\omega^{(\beta),d}_2(f^-,f^+)\nonumber\\
&=&-\langle e^{-\frac{\beta}{2}H}p_{(\beta)}(I^*f^-),p_{(\beta)}(f^+)\rangle.
\nonumber
\end{eqnarray}
It follows that
\[
\omega^{(\beta),d}(W(f^-)W(f^+))=e^{-\frac12\omega^{(\beta),d}_2(f^++f^-,f^++f^-)}
=\omega(W(I^*f^-))\omega(W(f^+))e^{-\omega^{(\beta),d}_2(f^-,f^+)}.
\]
On the other hand, we can use the $\beta$-KMS condition to find
$F_{W(I^*f^-),W(f^+)}$, which may be written as
\[
F_{W(I^*f^-),W(f^+)}(z)=\omega(W(I^*f^-))\omega(W(f^+))
e^{-F_{I^*f^-,f^+}(z)},
\]
where the function $F_{I^*f^-,f^+}(z)$ on $\overline{\mathcal{S}}_{\beta}$
is holomorphic on the interior and is given by
\[
F_{I^*f^-,f^+}(z)=\langle p_{(\beta)}(I^*f^-),e^{izH}p_{(\beta)}(f^+)\rangle.
\]
Evaluating this function at $z=\frac{i\beta}{2}$ we see that $\omega^{(\beta),d}$
is a double $\beta$-KMS state.

We now prove that $p^d_{(\beta)}$ has a dense range on
$C_0^{\infty}(M^+\cup M^-,\mathbb{R})$. We use the complex
conjugation $C$ to write $\mathcal{H}_e$ as a direct sum of
the real Hilbert spaces of real vectors, with $C\psi=\psi$,
and imaginary ones, with $C\psi=-\psi$. Taking the time
derivative of $Ce^{itH_e}C=e^{itH_e}$ at $t=0$ we find
$CH_eC=-H_e$ and therefore $C(P_+-P_-)C=-(P_+-P_-)$. This
means that the linear involution $Q:=P_+-P_-$ maps real vectors
to imaginary ones and vice versa. The operators
$X^{\pm}:=|H_e|^{-\frac12}\tanh(\frac{\beta}{4}|H_e|)^{\pm\frac12}$
are strictly positive for each choice of the sign and the range
of $p_{cl}$ on $C_0^{\infty}(M^+)$ is a core for both of these
operators (cf.\ Thm.\ 4.2 in \cite{Sanders2012}). The range
of $X^{\pm}p_{cl}$ on $C_0^{\infty}(M^+)$ is therefore dense.
Furthermore, the complex conjugation $C$ commutes with $|H_e|$
and $Cp_{cl}(f)=p_{cl}(\bar{f})$, which means that $X^{\pm}p_{cl}$
has a dense range in the real subspace of $\mathcal{H}_e$ if $f$
ranges over the real-valued test-functions
$C_0^{\infty}(M^+,\mathbb{R})$. A straightforward computation
shows that
\begin{eqnarray}
p^d_{(\beta)}(f-I^*f)&=&\sqrt{2}X^-p_{cl}(f),\nonumber\\
p^d_{(\beta)}(f+I^*f)&=&-\sqrt{2}QX^+p_{cl}(f)\nonumber
\end{eqnarray}
with $f\in C_0^{\infty}(M^+,\mathbb{R})$. By varying $f$, the
arguments on the left remain in
$C_0^{\infty}(M^+\cup M^-,\mathbb{R})$, and the ranges on the
right are dense in the space of real and imaginary vectors,
respectively. Therefore $p^d_{(\beta)}$ has a dense range on
$C_0^{\infty}(M^+\cup M^-,\mathbb{R})$. The fact that
$p^d_{(\beta)}$ has a dense range already on
$C_0^{\infty}(M^+\cup M^-,\mathbb{R})$ entails that
$\omega^{(\beta),d}$ is pure \cite{Kay+1991} and that it is
the unique state with this two-point distribution \cite{Kay1993}.

Finally, note that the
GNS-representation space of $\omega^{(\beta),d}$ is the same
as for $\omega^{(\beta)}$. Since the latter already has the
Reeh-Schlieder property (cf.\ \cite{Strohmaier2000}), the
same is true for the former, at least on $M^+$. That this also
holds on $M^-$ follows from the symmetry with respect to the
wedge reflection $I$ at the one-particle level.
\end{proof*}

\begin{remark}\label{Rem_DKMS}
In analogy to Remark \ref{Rem_GroundKMS} one may obtain additional,
pure, double $\beta$-KMS states by applying an automorphism of the
Weyl algebra determined by $W(f)\mapsto e^{i\phi(f)}W(f)$, where
$\phi(x)$ is now a real-valued (weak) solution to the Klein-Gordon
equation on $M^+\cup M^-$ which is independent of the Killing time
and satisfies $I^*\phi=\phi$. Subsequently one may obtain mixed
double $\beta$-KMS states by taking mixtures of these pure ones. It
is straightforward to verify that the double $\beta$-KMS condition
is invariant under these automorphisms and under taking mixtures.
Note that any double $\beta$-KMS state is uniquely determined by
its $\beta$-KMS restriction to $M^+$ and Equation
(\ref{Eqn_DefdKMS}). Conversely, any $\beta$-KMS state $\omega$ on
$M^+$ has a double $\beta$-KMS extension. To see this we note that
$\omega$ can be obtained from $\omega^{(\beta)}$ by applying suitable
automorphisms and mixing. These operations can be extended to
$\omega^{(\beta),d}$ on $M^+\cup M^-$, by requiring each $\phi(x)$
to be symmetric under the wedge reflection $I$. This yields a
double $\beta$-KMS state $\omega^d$ with the prescribed restriction
$\omega$.
\end{remark}

\subsection{Double $\beta$-KMS states in the static case and Wick rotation}\label{SSec_GlobalWick}

Let us now consider the Klein-Gordon equation on a spacetime $M$ with a static
bifurcate Killing horizon. Because the right wedge $M^+$ is a (possibly
disconnected) standard static spacetime we can obtain the two-point distributions
of its ground and $\beta$-KMS states from a Wick rotation. We will briefly review
this procedure and show how it can be extended to the disconnected spacetime
$M^+\cup M^-$.

In the static case, we have $D=0$ in Equation (\ref{Eqn_DefC}) and we quote the
following properties of $C$ from \cite{Sanders2012}, Proposition 4.3:
\begin{proposition}\label{Prop_C}
Consider the partial differential operator $C$ of Equation (\ref{Eqn_DefC}),
defined on the dense domain
$C_0^{\infty}(\Sigma^+)$ of $L^2(\Sigma^+)$ (in the metric volume form
$d\mathrm{vol}_h$). $C$ preserves its domain and all integer powers of $C$ are
essentially self-adjoint on this domain. The self-adjoint operator
$\overline{C}$ is strictly positive (i.e.\ positive and injective) and it
satisfies $C\ge Vv^2$. Finally, $C_0^{\infty}(\Sigma^+)$ is in the domain of
$\overline{C}^{\pm\frac12}$ for both signs.
\end{proposition}
From now on, we shall use $C$ to denote the unique self-adjoint extension.

In analogy to the Lorentzian theory on $M^+$ one considers a Euclidean
theory on $M^+_R$ for any given $R>0$. This theory is defined by the
Euclidean version
\[
K_R:=-\Box_{g_R}+V
\]
of the Klein-Gordon operator, which satisfies
\[
v^{\frac32}K_Rv^{\frac12}=-\partial_{\tau}^2+C.
\]
Here, $\tau$ is the imaginary Killing time, compactified to a circle of
radius $R$, and the function $v$ and the operator $C$ depend only on
the spatial coordinates on $\Sigma^+$. The operator $K_R$ is symmetric and
positive on the dense domain $C_0^{\infty}(M^+_R)$ of $L^2(M^+_R)$. We let
$\hat{K}_R$ be the self-adjoint Friedrichs extension of $K_R$, which
satisfies $\hat{K}_R\ge V$ on the domain of $\hat{K}_R^{\frac12}$, so that
$\hat{K}_R$ is strictly positive and the domain of $\hat{K}_R^{-\frac12}$
contains $C_0^{\infty}(M^+_R)$ (cf.\ \cite{Sanders2012} Lemma A.6). Hence,
the operator
\[
G_R:=\hat{K}_R^{-1}
\]
defines a distribution density on $(M^+_R)^{\times 2}$
(loc.cit.\ Theorem A.1), which is the Euclidean Green's function.

\begin{remark}
In \cite{Sanders2012} we used a different, but equivalent, definition
of the Euclidean Green's function. There we noted that $vK_Rv$ is
essentially self-adjoint on the domain of test-functions, that the
closure $\overline{vK_Rv}\ge Vv^2$ is strictly positive and that
the domain of $(\overline{vK_Rv})^{-\frac12}$ contains the space of
all test-functions $C_0^{\infty}(M^+_R)$, which is a core. We then set
$G_R:=v(\overline{vK_Rv})^{-1}v$, which again defines a distribution
density on $(M^+_R)^{\times 2}$. To see that both definitions are
equivalent we argue as follows. Define the operator
$X:=(\overline{vK_Rv})^{\frac12}v^{-1}$ on the domain of test-functions.
$X^*$ extends $v^{-1}(\overline{vK_Rv})^{\frac12}$, which is defined
on the range of $(\overline{vK_Rv})^{-\frac12}$, acting on the domain of
test-functions. This means that $X^*$ is densely defined, so $X$ is
closable. Now $|\overline{X}|^2=X^*\overline{X}$ extends $K_R$. Note
that the domain of $|X|$ equals the form domain of $\hat{K}_R$, which
implies that $\hat{K}_R=X^*\overline{X}=v^{-1}(\overline{vK_Rv})v^{-1}$.
Taking the inverses, we see that both definitions of the Euclidean Green's
function are equivalent.
\end{remark}

The dependence of $G_R$ on the imaginary time $\tau$ can be determined
explicitly, leading to a continuous function from $\mathbb{S}_R^{\times 2}$
into the the distribution densities on $(\Sigma^+)^{\times 2}$ which is
given by
\begin{equation}\label{Eqn_euclideanG}
G_R(\tau,\tau';\overline{f},f')=\langle C^{-\frac12}vf,
\frac{\cosh((\tau-\tau'+\pi R)\sqrt{C})}{2\sinh(\pi R\sqrt{C})}vf'\rangle
\end{equation}
when $\tau-\tau'\in[-2\pi R,0]$. We now note the following result, which
is familiar from Wick rotations in Minkowski spacetime
\cite{Osterwalder+1973,Osterwalder+1975}.
\begin{proposition}[Reflection positivity]\label{Prop_ReflPos}
Consider the open region $V'\subset M^+_R$ defined by
$V':=\left\{(\tau,x)\in M^+_R|\ \tau\in(0,\pi R)\right\}$
and let $\map{R_{\tau}}{M^+_R}{M^+_R}$ be the imaginary time reflection
$R_{\tau}(\tau,x):=(-\tau,x)$. For every $\phi\in C_0^{\infty}(V')$ we then
have
\[
G_R(\overline{R_{\tau}^*\phi},\phi)\ge 0.
\]
\end{proposition}
\begin{proof*}
Without the imaginary time reflection $R_{\tau}$ this formula would be clear
from the positivity of $G_R=\hat{K}_R^{-1}$. To see that the positivity remains
valid in the presence of the reflection we note that it suffices to consider
test-functions of the form $\phi(\tau,x)=\chi(\tau)f(x)$, by Schwartz Kernels
Theorem. In that case, we may use Proposition \ref{Prop_C} to introduce the
vector $\psi\in L^2(\Sigma^+)$ defined by
\[
\psi:=(I-e^{-2\pi R\sqrt{C}})^{-\frac12}C^{-\frac14}vf
\]
and we note that
\begin{eqnarray}
G_R(\overline{R_{\tau}^*\phi},\phi)&=&\frac12\int_{\mathbb{S}_R^{\times 2}}
(\langle e^{-\tau\sqrt{C}}\psi,e^{-\tau'\sqrt{C}}\psi\rangle
+\langle e^{(\tau-\pi R)\sqrt{C}}\psi,e^{(\tau'-\pi R)\sqrt{C}}\psi\rangle)
\overline{\chi}(\tau)\chi(\tau')d\tau\ d\tau',\nonumber
\end{eqnarray}
where we used the support properties of $\chi$ and Equation
(\ref{Eqn_euclideanG}). Performing the integrations we end up with a
sum of squared norms of vectors in $L^2(\Sigma^+)$, which is clearly
non-negative.
\end{proof*}

$G_R$ can be analytically continued to $z=t+i\tau$
(\cite{Sanders2012} Theorem 6.4). In this way we find a holomorphic
function $G^c_R(z,z')$ from $(z,z')\in\mathcal{C}_R^{\times 2}$ with
$\mathrm{Im}(z-z')\in(-2\pi R,0)$ into the distribution densities on
$(\Sigma^+)^{\times 2}$:
\begin{equation}\label{Eqn_complexG}
G^c_R(z,z';\overline{f},f')=\langle C^{-\frac12}vf,
\frac{\cos((z-z'+i\pi R)\sqrt{C})}{2\sinh(\pi R\sqrt{C})}vf'\rangle
\end{equation}
for $f,f'\in C_0^{\infty}(\Sigma^+)$. This function has continuous
boundary values at $\mathrm{Im}(z-z')\in[-2\pi R,0]$. Restricting $G^c_R$ to real
times $t,t'$ with $\mathrm{Im}(z-z')\rightarrow 0^-$ yields the two-point
distribution $\omega^{(\beta)}_2$ of the quasi-free $\beta$-KMS state
$\omega^{\beta}$ on $M^+$ with $\beta=2\pi R$. A similar result holds
for the ground state, in the degenerate case $R=\infty$. That
$\omega^{\beta}$ is Hadamard follows from the fact that it is a
boundary value of a holomorphic function and positivity can be shown
using the initial data formulation and reflection positivity.

After this review of the Wick rotation for $\beta$-KMS states on $M^+$ it
is now easy to describe a corresponding result for the disconnected
spacetime $M^+\cup M^-$, if the spacetime $M$ has a (weak) wedge
reflection.\footnote{As in Section \ref{SSec_DoubleKMS} it suffices to
assume that there is an isometric, involutive diffeomorphism $I$ of $M^+\cup M^-$ which
reverses the time orientation and which satisfies $I^*\xi^a=\xi^a$, because
we are not yet investigating the behaviour near the Killing horizon.}
Indeed, due to the wedge reflection there is an embedding
$\map{\chi}{M^+\cup M^-}{(M^+)^c_R}$ (cf.\ Equation (\ref{Def_chi})), such that
the complement of its range in $(M^+)^c_R$ is the union of the two regions where
$\mathrm{Im}(z)\in (-\pi R,0)$, respectively $\mathrm{Im}(z)\in (-2\pi R,-\pi R)$.
Analogously, the image of $\chi^{\times 2}$ in $(M^c_R)^{\times 2}$
is the boundary of the region
\[
\left\{(z,x;z',x')|\ \mathrm{Im}(z)\in(-\pi R,0),\ \mathrm{Im}(z')\in(0,\pi R)\right\}
\]
where $\mathrm{Im}(z-z')\in(-2\pi R,0)$. Taking the continuous extension of
$G^c_R$ to this boundary (cf.\ Equation (\ref{Eqn_complexG})) defines
a distribution density $\omega^{(\beta),d}_2$ on $M^+\cup M^-$, which
extends the $\beta$-KMS two-point distribution on $M^+$.

To see what this boundary value looks like we proceed as follows. For
any test-function $f\in C_0^{\infty}(\Sigma\setminus\mathcal{B})$ we
may use the wedge reflection $\iota$ to write $f=f^++\iota^*f^-$ with
unique $f^+,f^-\in C_0^{\infty}(\Sigma^+)$. It is then easy to see that
$\omega^{(\beta),d}_2$ takes the form
\begin{eqnarray}
\omega^{(\beta),d}_2(t,f;t',f')&=&
\langle C^{-\frac12}|v|\overline{f^+},
\frac{\cos((t-t'+i\pi R)\sqrt{C})}{2\sinh(\pi R\sqrt{C})}|v|(f')^+\rangle
\nonumber\\
&+&\langle C^{-\frac12}|v|\overline{f^-},
\frac{\cos((t-t')\sqrt{C})}{2\sinh(\pi R\sqrt{C})}|v|(f')^+\rangle
\nonumber\\
&+&\langle C^{-\frac12}|v|\overline{f^+},
\frac{\cos((t-t')\sqrt{C})}{2\sinh(\pi R\sqrt{C})}|v|(f')^-\rangle
\nonumber\\
&+&\langle C^{-\frac12}|v|\overline{f^-},
\frac{\cos((t-t'-i\pi R)\sqrt{C})}{2\sinh(\pi R\sqrt{C})}|v|(f')^-\rangle
\nonumber
\end{eqnarray}
for any test-functions $f,f'\in C_0^{\infty}(\Sigma\setminus\mathcal{B})$.
$\omega^{(\beta),d}_2$ is a bi-solution to the Klein-Gordon equation on
$M^+\cup M^-$, given by the Klein-Gordon operator
\begin{eqnarray}
K&=&\partial_t^2+A\nonumber\\
A&=&-|v|^{\frac12}\nabla^{(h)}_i|v|h^{ij}\nabla^{(h)}_j|v|^{\frac12}+Vv^2\nonumber
\end{eqnarray}
with the potential function $V$ extended from $M^+$ to $M^+\cup M^-$ such
that
\[
\iota^*V=V.
\]
Because $V$ is stationary this implies that $I^*V=V$, $A|_{M^+}=C$,
$A|_{M^-}=\iota^*C$ and $I^*K=K$. Note that $\omega^{(\beta),d}_2$ is
again Hadamard, because on $M^-$ the reversed Killing time orientation is
compensated for by taking the boundary value of a holomorphic function
from the opposite imaginary direction when compared to $M^+$.

To close this section, we wish to show that $\omega^{(\beta),d}_2$ is
indeed the double $\beta$-KMS state on $M^+\cup M^-$ as defined in
Section \ref{SSec_DoubleKMS}. For this purpose, we first compute the
initial data of $\omega^{(\beta),d}_2$ on $\Sigma$. Note that
$L^2(\Sigma\setminus\mathcal{B})\simeq L^2(\Sigma^+)\oplus L^2(\Sigma^-)$.
The weak wedge reflection $\iota$ gives rise to a unitary involution $T_{\iota}$
of $L^2(\Sigma\setminus\mathcal{B})$ defined by
$T_{\iota}(\psi^+\oplus\psi^-):=\iota^*\psi^-\oplus \iota^*\psi^+$, which
shows in particular that $L^2(\Sigma^-)\simeq L^2(\Sigma^+)$. From
Proposition \ref{Prop_C} and the definition of $C$ we immediately conclude
the following:
\begin{corollary}\label{Prop_A}
Consider the partial differential operator
\[
A:=-|v|^{\frac12}\nabla^{(h)}_i|v|h^{ij}\nabla^{(h)}_j
|v|^{\frac12}+Vv^2
\]
on $\Sigma\setminus\mathcal{B}$, where $V$ satisfies $\iota^*V=V$ and $A$ is
defined on the dense domain $C_0^{\infty}(\Sigma\setminus\mathcal{B})$ of
$L^2(\Sigma\setminus\mathcal{B})$ (in the metric volume form $d\mathrm{vol}_h$).
$A$ preserves this dense domain and all integer powers of $A$ are essentially
self-adjoint on it. The self-adjoint operator $\overline{A}$ is strictly
positive and it satisfies $A\ge Vv^2$. Finally,
$C_0^{\infty}(\Sigma\setminus\mathcal{B})$ is in the domain of
$\overline{A}^{\pm\frac12}$ for both signs.
\end{corollary}
From now on we will use $A$ to denote the unique self-adjoint extension.

The initial data of $\omega^{(\beta),d}_2$ can be conveniently expressed in terms
of $A$ and $T_{\iota}$ as:
\begin{eqnarray}
\omega^{(\beta),d}_{2,00}(f,f')&=&\frac12\langle A^{-\frac12}|v|^{\frac12}\overline{f},
\coth(\pi R\sqrt{A})|v|^{\frac12}f'\rangle
+\frac12\langle A^{-\frac12}|v|^{\frac12}T_{\iota}\overline{f},
\sinh(\pi R\sqrt{A})^{-1}|v|^{\frac12}f'\rangle\nonumber\\
\omega^{(\beta),d}_{2,10}(f,f')&=&\frac{-i}{2}\langle\overline{f},f'\rangle\nonumber\\
\omega^{(\beta),d}_{2,01}(f,f')&=&\frac{i}{2}\langle\overline{f},f'\rangle\nonumber\\
\omega^{(\beta),d}_{2,11}(f,f')&=&\frac12\langle A^{\frac12}|v|^{-\frac12}\overline{f},
\coth(\pi R\sqrt{A})|v|^{-\frac12}f'\rangle
-\frac12\langle A^{\frac12}|v|^{-\frac12}T_{\iota}\overline{f},
\sinh(\pi R\sqrt{A})^{-1}|v|^{-\frac12}f'\rangle,\nonumber
\end{eqnarray}
where we used the fact that $\partial_t=v\partial_{x^0}=\pm|v|\partial_{x^0}$ on
$\Sigma^{\pm}$, where $x^0$ is the Gaussian normal coordinate, and the restriction
of a distribution density from $M^+\cup M^-$ to $\Sigma\setminus\mathcal{B}$
involves a change of measure, which yields a factor $|v|^{-\frac12}$ for every
test-function on $\Sigma\setminus\mathcal{B}$. The distributions
$\omega^{(\beta),d}_{2,10}$ and $\omega^{(\beta),d}_{2,01}$ show that the
anti-symmetric part of $\omega^{(\beta),d}_2$ is indeed the canonical commutator. We
may use the reflection positivity of Proposition \ref{Prop_ReflPos} to show that
$\omega^{(\beta),d}_2$ is of positive type, so it defines a quasi-free state.

To see that the two-point distribution $\omega^{(\beta),d}_2$ determined by the
initial data above corresponds to the distribution of Section \ref{SSec_DoubleKMS}
we proceed as follows. In the static case, the one-particle structure
$(p_{(\beta)},\mathcal{H}_{(\beta)})$ on $M^+$ can be given explicitly in terms of
initial data (cf.\ \cite{Sanders2012} Proposition 4.3\footnote{Note that the
statement of the proposition has a sign error, which can be corrected by changing
the sign of each $f_1$. The error enters in the proof of loc.cit.\ via erroneous
expressions for $V^*P_{\pm}V$. Here we use the corrected expression.}),
namely $\mathcal{H}_{(\beta)}=L^2(\Sigma^+)^{\oplus 2}$ and
\begin{eqnarray}
p_{(\beta)}(f)&=&\frac{1}{\sqrt2}\left(I-e^{-\beta\sqrt{C}}\right)^{-\frac12}
\left(C^{\frac14}v^{-\frac12}f_0+iC^{-\frac14}v^{\frac12}f_1\right)\nonumber\\
&&\oplus\frac{1}{\sqrt2}e^{-\frac{\beta}{2}\sqrt{C}}
\left(I-e^{-\beta\sqrt{C}}\right)^{-\frac12}
\left(C^{\frac14}v^{-\frac12}f_0-iC^{-\frac14}v^{\frac12}f_1\right),\nonumber
\end{eqnarray}
where $f_0:=Ef|_{\Sigma}$ and $f_1:=n^a\nabla_aEf|_{\Sigma}$. The Killing time
evolution is implemented by $H=\sqrt{C}\oplus -\sqrt{C}$. This expression can be
rewritten in a nicer way by using $\iota^*$ on the second summand of
$\mathcal{H}_{(\beta)}$ to identify it with $L^2(\Sigma\setminus\mathcal{B})$
and by exploiting the fact that $A=C\oplus 0+T_{\iota}(C\oplus 0)T_{\iota}$. After
some straightforward computations one finds
\begin{eqnarray}
p^d_{(\beta)}(f)&=&\frac{1}{\sqrt{2}}(I-e^{-\beta\sqrt{A}})^{-\frac12}
\left((I+T_{\iota}e^{-\frac{\beta}{2}\sqrt{A}})A^{\frac14}|v|^{-\frac12}(f_0\oplus 0)
\right.\nonumber\\
&&\left.+i(I-T_{\iota}e^{-\frac{\beta}{2}\sqrt{A}})A^{-\frac14}|v|^{\frac12}(f_1\oplus 0)\right).\nonumber
\end{eqnarray}
We may now obtain the double $\beta$-KMS one-particle structure from Section
\ref{SSec_DoubleKMS}. Keeping in mind that the wedge reflection $I$ reverses the time
orientation, so that $(I^*f^-)_1=-\iota^*(f^-_1)$, we find that
\begin{eqnarray}
p^d_{(\beta)}(f)&=&\frac{1}{\sqrt{2}}(I-e^{-\beta\sqrt{A}})^{-\frac12}
\left((I-T_{\iota}e^{-\frac{\beta}{2}\sqrt{A}})A^{\frac14}|v|^{-\frac12}f_0\right.\\
&&\left.+i(I+T_{\iota}e^{-\frac{\beta}{2}\sqrt{A}})A^{-\frac14}|v|^{\frac12}f_1\right)\nonumber
\end{eqnarray}
(up to unitary equivalence). It is a straightforward exercise to verify the initial
data of $\omega^{(\beta),d}_2$ from this expression, using Equations
(\ref{Eqn_StateOPS}) and (\ref{Eqn_InitValState}).

\section{Hadamard's parametrix construction}\label{Sec_Had}

The definition of the HHI state, and the verification that it is a
Hadamard state, will involve a detailed comparison of the Euclidean
Green's function $G_R$ and its Wick rotation to the Lorentzian spacetime
$M$. In this section, we will focus on the local singularity structures
in this comparison. The local singularities of a fundamental solution of
a second-order operator can nicely be characterised using Hadamard's
parametrix construction. Here we will describe this construction in some
detail for both the Euclidean and the Lorentzian setting. Our
presentation is essentially an expanded version of Section 17.4 of
\cite{Hoermander3} (see also \cite{Baer+2007,Baer+2009} for a more
detailed description in the case of the advanced and retarded
fundamental solutions in a Lorentzian setting).

There is no harm in considering the more general situation of a
pseudo-Riemannian manifold $N=(\mathcal{N},g_{ab})$ on which we consider
a partial differential operator $P$ given in local coordinates as
\[
P=-\partial_{\mu}g^{\mu\nu}\partial_{\nu}+b^\mu\partial_{\mu}+c,
\]
where $g^{\mu\nu}$ is the inverse of the pseudo-Riemannian metric
$g_{\mu\nu}$ and $b^{\mu}$, $c$ are a smooth vector field and function,
respectively, on $\mathcal{N}$. Consider any $y\in\mathcal{N}$ and
choose coordinates $x^{\mu}$ near $y$ such that
$g_{\mu\nu}(y)=\eta_{\mu\nu}$, where $\eta_{\mu\nu}$ is a real-valued
diagonal matrix with eigenvalues contained in $\left\{+1,-1\right\}$.
We will denote the inverse of $\eta_{\mu\nu}$ by $\eta^{\mu\nu}$. The
basic idea of Hadamard's construction is to approximate the operator
$P$ near $y$ by $P_0:=-\partial_{\mu}\eta^{\mu\nu}\partial_{\nu}$ and
to make sense of the formal geometric series
$P^{-1}=(P_0+(P-P_0))^{-1}=\sum_{k=0}^{\infty}(P_0-P)^kP_0^{-k-1}$.

\subsection{The Hadamard coefficients}\label{SSec_HadCoeff}

To see how this approximation works, we first consider the
operator $P_0=-\partial_{\mu}\eta^{\mu\nu}\partial_{\nu}$ on
$\mathbb{R}^d$. The formal geometric series for $P^{-1}$ motivates
us to consider fundamental solutions $F_k$ of
$\frac{1}{k!}P_0^{k+1}$, $k\ge 0$, which can be studied using Fourier
analysis. If we define the principal symbol
$p(\xi):=\eta^{\mu\nu}\xi_{\mu}\xi_{\nu}$ of $P_0$ with
characteristic set $C:=p^{-1}(0)$, then the Fourier transforms
$\widehat{F}_k$ of $F_k$ are given by $k!p(\xi)^{-k-1}$ on
$\mathbb{R}^d\setminus C$. Furthermore, they are homogeneous of
degree $-2(k+1)$ and they satisfy
\begin{eqnarray}
p(\xi)\widehat{F}_k(\xi)&=&k\widehat{F}_{k-1}(\xi)\nonumber\\
-2\eta^{\mu\nu}\xi_{\nu}\widehat{F}_k(\xi)&=&
\partial_{\xi_{\mu}}\widehat{F}_{k-1}(\xi)\nonumber
\end{eqnarray}
outside $C$ for $k\ge 1$, where the bottom equality essentially
expresses the invariance of $\widehat{F}_k$ under the symmetry
group of $(\mathbb{R}^d,\eta_{\mu\nu})$. Similarly, in the case
$k=0$:
\[
p(\xi)\widehat{F}_0(\xi)=1.
\]

We now assume that we can find tempered distributions
$\widehat{F}_k(\xi)$, $k\ge 0$, on all of $\mathbb{R}^d$
which extend the distributions $p(\xi)^{-k-1}$, which have
inverse Fourier transforms\footnote{Because $\widehat{F}_k(\xi)$
falls off like $\|\xi\|^{-2k-2}$ in the Euclidean setting, it
seems plausible that we can require the distributions $F_k$ to
be even more regular, e.g.\ $F_k\in C^{k+1-d}(\mathbb{R}^d)$.
As this extra regularity only occurs in the Euclidean setting
it will not be essential to our arguments and we will not pursue it.}
$F_k\in C^{k+1-d}(\mathbb{R}^d)$ if $2k\ge d-1$ and which still
satisfy the two relations above for $k\ge 0$, modulo an
additional term with a smooth inverse Fourier transform. To
find such extensions is a non-trivial issue, which depends on
the signature of the $\eta_{\mu\nu}$. In Subsection \ref{SSec_Fk}
below we will comment on the existence
and uniqueness aspects for the Euclidean and Lorentzian case, but
for now we will simply assume that distributions with these
properties are given. This means that the tempered distributions
$F_k$ satisfy
\begin{eqnarray}\label{Eqn_Fk1}
P_0F_k&\sim&\left\{
\begin{array}{ll}
kF_{k-1}&k\ge 1\\
\delta_0&k=0
\end{array}
\right.,\\
\partial_{\mu}F_k&\sim&\frac{-1}{2}(\partial_{\mu}s)
F_{k-1},\ k\ge 1,\nonumber
\end{eqnarray}
where $s(x):=\frac12\eta_{\mu\nu}x^{\mu}x^{\nu}$ and $\sim$
means equality modulo a smooth function.

In order to fully exploit the properties of the $F_k$ on the
pseudo-Riemannian manifold $N$, we need to choose Riemannian normal
coordinates on $U^{\times 2}$, where $U\subset N$ is a convex normal
neighbourhood. More precisely, we will use arbitrary
coordinates $\tilde{y}^{\mu}$ on $U$ and an orthonormal frame
$(e_{\mu})^a$ of $TU$ in order to describe the Riemannian
normal coordinates in terms of an embedding
$\map{\rho}{U^{\times 2}}{\mathbb{R}^{\times d}\times U}$
as in Section \ref{SSec_infAC} (cf.\ Equation
(\ref{Eqn_SpecialCoords})). We then define the pull-backs
$\tilde{F}_k:=\rho^*(F_k\otimes 1)$ as distributions on
$U^{\otimes 2}$, i.e.\ $\tilde{F}_k(v,y)=F_k(v)$. In the
following discussion, we continue to work in the coordinates
$(v,y)$ and all derivatives will be taken with respect to
$v$. From Equation (\ref{Eqn_SyngeProp}) and the last line of
(\ref{Eqn_Fk1}) we then find
\begin{equation}
(\exp_y^*g)^{\mu\nu}(v)\partial_{\nu}\tilde{F}_k(v,y)\ \sim\ g^{\mu\nu}(y)
\partial_{\nu}\tilde{F}_k(v,y)\nonumber
\end{equation}
for $k\ge 1$. Equation (\ref{Eqn_Fk1}) then leads to
\begin{eqnarray}\label{Eqn_Fk3}
-\partial_{\mu}g^{\mu\nu}(v)\partial_{\nu}\tilde{F}_k(v,y)&\sim&
P_0\tilde{F}_k(v,y)\ \sim\ k\tilde{F}_{k-1}(v,y),\\
g^{\mu\nu}(v)\partial_{\nu}\tilde{F}_k(v,y)&\sim&
\frac{-1}{2}v^{\mu}\tilde{F}_{k-1}(v,y)\nonumber
\end{eqnarray}
for $k\ge 1$. Note that $-\partial_{\mu}g^{\mu\nu}(x)\partial_{\nu}$
is the principal part of $P$. It is the special virtue of the
Riemmanian normal coordinates $v^{\mu}$, centred on $y$,
which allowed us to replace $g^{\mu\nu}(0)$ by $g^{\mu\nu}(v)$,
leading to agreement between the highest-order part of $P$ and
$P_0$ at any $y$.  In addition to these properties for
$\tilde{F}_k$, $k\ge 1$, we will assume that $\tilde{F}_0$
satisfies
\begin{eqnarray}\label{Eqn_Fk2}
-\partial_{\mu}g^{\mu\nu}(v)\partial_{\nu}\tilde{F}_0(v,y)
&\sim&\delta_0(v)\\
g^{\mu\nu}(v)\partial_{\nu}\tilde{F}_0(v,y)&=&\frac{-1}{2}v^{\mu}\tilde{F}_{-1}(v,y)
\nonumber
\end{eqnarray}
for some distribution $\tilde{F}_{-1}$.

Returning to the formal geometric series for $P^{-1}$, the
idea is now to approximate a right parametrix for $P$ on $U$ by
\begin{equation}\label{Eqn_HadSeries}
\mathcal{R}^{(N)}:=\sum_{k=0}^Nu_k\tilde{F}_k
\end{equation}
for some smooth coefficients $u_k\in C^{\infty}(U^{\times 2})$.
A straightforward computation using Equation (\ref{Eqn_Fk3})
shows that for $k\ge 1$
\begin{eqnarray}
P(u_k\tilde{F}_k)&\sim&(Pu_k)\tilde{F}_k+ku_k\tilde{F}_{k-1}
+u_kb^{\mu}\partial_{\mu}\tilde{F}_k-2g^{\mu\nu}
(\partial_{\nu}u_k)\partial_{\mu}\tilde{F}_k\nonumber\\
&=&(Pu_k)\tilde{F}_k+ku_k\tilde{F}_{k-1}
-\frac12(g_{\mu\nu}v^{\nu}b^{\mu}u_k-2v^{\nu}
\partial_{\nu}u_k)\tilde{F}_{k-1},\nonumber
\end{eqnarray}
where $b^{\mu}$ and $g^{\mu\nu}$ are evaluated at $v$. For $k=0$
we have
\begin{eqnarray}
P(u_0\tilde{F}_0)&\sim&(Pu_0)\tilde{F}_0+u_0\rho^*(\delta_0\otimes 1)
+u_0b^{\mu}\partial_{\mu}\tilde{F}_0
-2g^{\mu\nu}(\partial_{\nu}u_0)\partial_{\mu}\tilde{F}_0\nonumber\\
&=&(Pu_0)\tilde{F}_0+u_0\rho^*(\delta_0\otimes 1)-\frac12
(g_{\mu\nu}v^{\nu}b^{\mu}u_0-2v^{\mu}\partial_{\mu}u_0)\tilde{F}_{-1}.
\nonumber
\end{eqnarray}
Adding these equations together we find
\begin{equation}\label{Eqn_PHadSeries}
P\mathcal{R}^{(N)}\sim u_0\rho^*(\delta_0\otimes 1)+(Pu_N)\tilde{F}_N
+\sum_{k=0}^N(Pu_{k-1}+ku_k
-\frac12(g_{\mu\nu}v^{\nu}b^{\mu}u_k-2v^{\nu}\partial_{\nu}u_k))
\tilde{F}_{k-1}
\end{equation}
where we set $u_{-1}\equiv 0$. The factor $Pu_N$ is smooth by
assumption and $\tilde{F}_N\in C^{N+1-d}(U^{\times 2})$ if $2N\ge d-1$,
so the second term becomes more regular as $N$ increases. Moreover, the
terms in the sum can be made to vanish by choosing the coefficients $u_k$
appropriately (cf.\ \cite{Hoermander3} Lemma 17.4.1):
\begin{lemma}\label{Lem_HadCoeff}
There are unique functions $u_k\in C^{\infty}(U^{\times 2})$,
$k\ge 0$, such that $u_0(0,y)=1$ and
\[
2ku_k-g_{\mu\nu}v^{\nu}b^{\mu}u_k+2v^{\nu}\partial_{\nu}u_k=-2Pu_{k-1}
\]
in the coordinates $(v,y)$. In coordinates $x,y\in U$ these functions
are given recursively by
\begin{eqnarray}
u_0(x,y)&:=&\exp\left(\frac12\int_0^1 g_{\mu\nu}(x(t))b^{\mu}(x(t))\dot{x}^{\nu}(t)
dt\right),\nonumber\\
u_k(x,y)&:=&-u_0(x,y)\int_0^1t^{k-1}\frac{(Pu_{k-1})(x(t),y)}{u_0(x(t),y)}dt,\nonumber
\end{eqnarray}
where $t\mapsto x(t)$ is the unique geodesic in $U$ with $x(0)=y$ and $x(1)=x$.
\end{lemma}

In the special case that $P=-\Box+c$ we have $b^{\mu}=-\frac12g^{\mu\nu}\partial_{\nu}\log|g|$,
where $g:=\det g_{\alpha\beta}$, and one may show that
\begin{eqnarray}\label{Eqn_VVM}
u_0(x,y)^2&=&(-1)^d\frac{\det g(x)}{|\det g(x)|}\frac{1}{\sqrt{g(x)g(y)}}
\det\left(\partial_{x^{\mu}}\partial_{y^{\nu}}\sigma(x,y)\right)\\
&=&\Delta_{VVM}(x,y),\nonumber
\end{eqnarray}
the Van Vleck-Morette determinant, because both sides satisfy the same
differential equation in the coordinate $x$, which may be integrated
along geodesics:
\[
\nabla^{\mu}\sigma\cdot\nabla_{\mu}\log(\Delta_{VVM})=d-\Box\sigma
\]
(cf.\ \cite{Poisson+2011,Camporesi1990,Moretti1999}). In the case of
$u_0^2$ this equation may be verified by using Riemannian normal
coordinates around $y$ and using Lemma \ref{Lem_HadCoeff}. We
therefore find:
\begin{definition}\label{Def_HadCoeff}
Let $U$ be a convex normal neighbourhood in a pseudo-Riemannian
manifold $N=(\mathcal{N},g_{ab})$. The \emph{Hadamard coefficients}
$u_k$, $k\ge 0$, on $U$ for the operator $P=-\Box+c$ are the
smooth functions defined by
\begin{eqnarray}
u_0(x,y)&:=&\sqrt{\Delta_{VVM}(x,y)},\nonumber\\
u_{k+1}(x,y)&:=&-u_0(x,y)\int_0^1t^k
\frac{(Pu_k)(x(t),y)}{u_0(x(t),y)}dt,\nonumber
\end{eqnarray}
where $t\mapsto x(t)$ is the unique geodesic segment in $U$
from $x(0)=y$ to $x(1)=x$.
\end{definition}
Our Hadamard coefficients $u_k$ equal Moretti's heat kernel
coefficients $a_k$ \cite{Moretti1999,Moretti2000} (at least
when $c$ is real-valued) and they differ from the Hadamard
coefficients $V^k$ in \cite{Baer+2007} by $V^k=k!u_k$.

The approximate parametrices $\tilde{\mathcal{R}}^{(N)}$ can
be used to construct an exact local right parametrix using
Borel's Lemma and from that one can construct a local right
fundamental solution for the operator $P$. Different choices
of extensions $\widehat{F}_k$ may give rise to different
approximate fundamental solutions, but the Hadamard
coefficients only depend on the operator $P$ and the geometry
of the pseudo-Riemannian manifold. We refer the reader to
\cite{Baer+2007} for an elaboration of this procedure in the
case of advanced and retarded fundamental solutions for wave
equations.

Under suitable circumstances a right parametrix is also a
left parametrix. This follows from
\begin{theorem}[Moretti's Theorem]\label{Thm_Moretti}
If $M=(\mathcal{M},g_{ab})$ is a Riemannian or Lorentzian
manifold and $P=-\Box+c$ with real-valued $c$, then the
Hadamard coefficients are symmetric: $u_k(x,y)=u_k(y,x)$
for all $k\ge 0$.
\end{theorem}
See \cite{Moretti1999,Moretti2000} for a proof.\footnote{Presumably
this result also holds in other pseudo-Riemannian manifolds,
because the procedure of \cite{Moretti2000} can be used
inductively to change the signature of the metric by changing
the sign of one eigenvalue at a time \cite{MorettiPC}.}

\subsection{The distributions $F_k$}\label{SSec_Fk}

To complete our discussion of the Hadamard parametrix construction
we return to the issue of finding suitable distributions $F_k$ such
that $F_k\in C^{k+1-d}(\mathbb{R}^d)$ if $2k\ge d-1$ and such that
Equations (\ref{Eqn_Fk1}) and (\ref{Eqn_Fk2}) hold.  We do this first
for the Euclidean and then for the Lorentzian case.

\subsubsection{The Euclidean case}\label{SSSec_FkEucl}

In the Euclidean case, the characteristic set $C$ reduces to the
origin, so we need to extend the homogeneous distributions
$k!p(\xi)^{-k-1}$ and $-\log|p(\xi)|$ (for $k=-1$) from
$\mathbb{R}^d\setminus\left\{0\right\}$ to $\mathbb{R}^d$. For each
$k$ such an extension always exists (cf.\ \cite{Hoermander} Section
3.2 and 7.1). The difference of two extensions is supported at the
origin, so its inverse Fourier transform is a polynomial, which does
not contribute to the singularities appearing in Equation
(\ref{Eqn_HadSeries}). In particular, Equation (\ref{Eqn_Fk1}) is
automatically satisfied.

A convenient explicit expression for such extensions is given in
the following result:
\begin{proposition}\label{Prop_FkEucl}
For a fixed $d\ge 2$ and any $k\in\mathbb{N}_0$ we define the
distributions
\begin{equation}\label{Eqn_HadFk1}
F^e_k(x):=s(x)^{k+1-\frac{d}{2}}(c_k+d_k\log(s(x))),\quad k\ge 0,
\end{equation}
viewed as locally integrable functions, where the constants
$c_k,d_k$, $k\ge 0$, are given by
\begin{eqnarray}\label{Eqn_ck}
c_k&=&\left\{\begin{array}{ll}
0&k+1-\frac{d}{2}\in\mathbb{N}_0\\
(4\pi)^{-\frac{d}{2}}2^{\frac{d}{2}-1-k}\Gamma(\frac{d}{2}-1-k)
&\rm{otherwise}
\end{array}\right.\\
d_k&=&\left\{\begin{array}{ll}
-(4\pi)^{-\frac{d}{2}}(-2)^{\frac{d}{2}-1-k}\Gamma(k+2-\frac{d}{2})^{-1}
&k+1-\frac{d}{2}\in\mathbb{N}_0\\
0&\rm{otherwise}
\end{array}\right..\nonumber
\end{eqnarray}
Then $F^e_k\in C^{k+1-d}(\mathbb{R}^d)$ if $2k\ge d-1$ and the
$F^e_k$ satisfy Equation (\ref{Eqn_Fk1}).
\end{proposition}
\begin{proof*}
The constants are chosen such that $(2k+2-d)c_k=-c_{k-1}$ and
$(2k+2-d)d_k=-d_{k-1}$ for $k\ge 1$ and $k\not=\frac{d}{2}-1$.
For $k=\frac{d}{2}-1\ge 1$ we have $2d_k=-c_{k-1}$ while $c_k=d_{k-1}=0$.
Outside $x=0$ we may then compute
\begin{equation}\label{Eqn_DerivativeFk}
\partial_{\mu}F^e_k=\left\{\begin{array}{ll}
\frac{-1}{2}(\partial_{\mu}s)F^e_{k-1}
+d_k(\partial_{\mu}s)s^{k-\frac{d}{2}}& k\ge 1,\ k\not=\frac{d}{2}-1\\
\frac{-1}{2}(\partial_{\mu}s)F^e_{k-1}& k=\frac{d}{2}-1\ge 1
\end{array}\right..
\end{equation}
These equations extend to $x=0$ in the distributional sense,
because all functions and distributions involved are locally integrable.
Applying this result twice, or using similar computations, we find
\[
P_0F^e_k=\left\{\begin{array}{ll}
kF^e_{k-1}+(d-2-4k)d_ks^{k-\frac{d}{2}}&k\ge 1,\ k\not=\frac{d}{2}-1\\
kF^e_{k-1}&k=\frac{d}{2}-1\ge 1
\end{array}\right..
\]
For $k=0$ we have $P_0F^e_0=\delta_0$, by \cite{Hoermander} Theorem 3.3.2
and the fact that the volume of the unit sphere in $\mathbb{R}^d$ is given
by $\mathrm{vol}(\mathbb{S}^{d-1})=2\pi^{\frac{d}{2}}\Gamma\left(\frac{d}{2}\right)^{-1}$.
(This fixes $c_0$ and $d_0$. The other constants are determined by
the recursion relations above.)

Equation (\ref{Eqn_Fk1}) follows once it is realised that the extra terms
with $s^{k-\frac{d}{2}}$ vanish unless $k-\frac{d}{2}\in\mathbb{N}_0$,
in which case they are polynomials and hence smooth. $F^e_k$ is continuous
for $k\ge\frac{d-1}{2}$, both for even and odd dimensions $d$. The
claimed regularity for $2k\ge d-1$ then follows by induction from
Equation (\ref{Eqn_DerivativeFk}).
\end{proof*}

The corresponding distribution $\tilde{F}^e_0$ does not satisfy the bottom
line of Equation (\ref{Eqn_Fk2}) as stated, but
\[
\partial_{\mu}F^e_0=\frac{c_{-1}}{2}(\partial_{\mu}s)s^{-\frac{d}{2}}
\]
as a product is a locally integrable function (cf.\ \cite{Hoermander}
Theorem 3.3.2). From this it follows that
\begin{eqnarray}\label{Eqn_ProptildeFek}
g^{\mu\nu}(v)\partial_{\nu}\tilde{F}^e_0(v,y)&=&
g^{\mu\nu}(0)\partial_{\nu}\tilde{F}^e_0(v,y)\ =\ \frac{c_{-1}}{2}v^{\mu}
s(v)^{-\frac{d}{2}}\nonumber\\
-\partial_{\mu}g^{\mu\nu}(v)\partial_{\nu}\tilde{F}^e_0(v,y)&=&
P_0\tilde{F}^e_0(v,y)\ =\ \delta_0(v),\nonumber
\end{eqnarray}
which suffices to show that the $k=0$ term in the summation of
Equation (\ref{Eqn_PHadSeries}) vanishes when $u_0$ solves the
transport equation of Lemma \ref{Lem_HadCoeff}.

\subsubsection{The Lorentzian case}\label{SSSec_FkLor}

In the Lorentzian case different choices of extensions can
lead to different fundamental solutions. The easiest choice
is to take $F_k=E^{\pm}_k$ with
\[
\widehat{E}^{\pm}_k(\xi):=\lim_{V^+\ni\epsilon\rightarrow 0^+}
k!(\eta^{\mu\nu}(\xi_{\mu}\mp i\epsilon_{\mu})
(\xi_{\nu}\mp i\epsilon_{\nu}))^{-k-1},
\]
where we use a fixed choice of the sign for all $k$. These distributions
are well-defined by \cite{Hoermander} Theorem 3.1.15 and they are tempered,
because they are homogeneous (loc.cit.\ Theorem 7.1.18). Also note
that they are invariant under the proper, orthochronous Lorentz group.
It is not hard to show that the inverse Fourier transforms
$E^{\pm}_k(x)$ are supported in the future (+) or past (-) light cone
(cf.\ \cite{Fulling1991} Ch.\ 4) and that they satisfy Equation
(\ref{Eqn_Fk1}) with equalities. In fact, one may prove by induction
over $k$ that $E^{\pm}_k(x)$ is uniquely determined by the top line of
(\ref{Eqn_Fk1}) with equality and by its support property. Indeed, the
difference of two such distributions must be a solution to the wave
equation with either past or future compact support and therefore it
must vanish.

To deduce the regularity of $E^{\pm}_k(x)$ for sufficiently large
$k$ one may proceed as follows. Note that each $E^{\pm}_k(x)$ is
again Lorentz invariant and hence so is the difference
$E_k:=E^-_k-E^+_k$. Furthermore, $E_k(-x^0,x^i)=E_k(x^0,x^i)$ and
a detailed analysis of Lorentz invariant distributions (as in
\cite{Garding+1959} Section 8) shows that we may write
\[
E_k(x)=\mathrm{sign}\left(x^0\right)\rho(-\sigma(x))
\]
for a unique distribution $\rho$ on $\mathbb{R}$ whose support lies
in $\mathbb{R}_{\ge 0}$. Next we note that both $E^{\pm}_k$ are
homogeneous of degree $2k+2-d$ (cf.\ \cite{Hoermander} Thm.\ 7.1.16)
and hence $\rho$ is homogeneous of degree $k+1-\frac{d}{2}$. This
means that $\rho(s)=e_ks^{k+1-\frac{d}{2}}$ on $s>0$ and $\rho(s)=0$
on $s<0$. Here $e_k$ is a constant, which must be real because
$E_k$ is real. When $2k\ge d-3$ then $k+1-\frac{d}{2}$ is not an integer
$\le -1$, so $\rho$ is uniquely determined by homogeneity and its
restriction to $\mathbb{R}\setminus\left\{0\right\}$ (\cite{Hoermander}
Thm.\ 3.2.3). By inspecting the supports we therefore find
\[
E^{\pm}_k(x)=e_k\theta\left(\pm x^0\right)\theta(-\sigma(x))|\sigma(x)|^{k+1-\frac{d}{2}}
\]
when $2k\ge d-3$. The fact that $E^{\pm}_k\in C^{k+1-d}$ when
$2k\ge d-1$ can now be shown by induction from the bottom line of
Equation (\ref{Eqn_Fk1}).

A different approach to the distributions $E^{\pm}_k(x)$ is given
in full detail in \cite{Baer+2007}, which also proves that the
Hadamard parametrix based on these distributions gives rise to the
unique advanced ($-$) and retarded ($+$) fundamental solutions,
according to the choice of sign. Comparing our formulae with those
of this reference\footnote{The notations of reference \cite{Baer+2007}
relate to ours as follows: $\gamma(x)=2\sigma(x)$ and
$R_{\pm}(\alpha)=\frac{1}{k!}E^{\pm}_k$ when $\alpha=2k+2$ and
$n=d$.} allows us to determine the constants $e_k$ as
\begin{equation}\label{Eqn_ek}
e_k=\left(2^{k+\frac{d}{2}}\pi^{\frac{d}{2}-1}\Gamma\left(k+2-\frac{d}{2}\right)\right)^{-1}.
\end{equation}
(These may also be found by a direct Fourier transformation.)

Feynman and anti-Feynman fundamental solutions are only unique up to
a smooth function. To obtain them from the Hadamard parametrix
construction one first defines the distributions
$v_k^{\pm}(s):=\lim_{\epsilon\rightarrow 0^+}k!(s\mp i\epsilon)^{-k-1}$
on $\mathbb{R}$, which are homogeneous of degree $-k-1$ and whose wave
front sets are given by
\[
WF(v_k^{\pm})=\left\{(0,\varsigma)\in\mathbb{R}^{\times 2}| \mp\varsigma>0\right\}.
\]
These distributions have well-defined pull-backs $w^{\pm}_k(\xi)$ to
$\mathbb{R}^d\setminus\left\{0\right\}$ under the map
$\xi\mapsto p(\xi)$ (cf.\ \cite{Hoermander} Thm.\ 8.2.4). The
pull-backs are Lorentz invariant, homogeneous of degree $-2k-2$ and they
have
\[
WF(w^{\pm}_k)\subset\left\{(\xi,x)\in (\mathbb{R}^d\setminus\left\{0\right\})
\times\mathbb{R}^d|\ p(\xi)=0,\ x^{\mu}=a\eta^{\mu\nu}\xi_{\nu},\ \mp a>0\right\}.
\]
If we let $\hat{E}^{F,\pm}_k$ be any extensions of $w^{\pm}_k$ to
$\mathbb{R}^d$, then they are automatically tempered (by homogeneity
outside the origin) and it is straightforward to verify that the
distributions $E^{F,\pm}_k$ satisfy Equations (\ref{Eqn_Fk1}) and
(\ref{Eqn_Fk2}). Moreover, we have
\begin{equation}\label{Eqn_WFFeynman}
WF(E^{F,\pm}_k)\subset T_0^*\mathbb{R}^d\cup
\left\{(x,\xi)\in \mathbb{R}^d\times\mathbb{R}^d|\ \sigma(x)=0,
\xi_{\mu}=a\eta_{\mu\nu}x^{\nu},\ \pm a>0\right\}
\end{equation}
(cf.\ \cite{Hoermander} Thm.\ 8.1.8). Note that different choices of
extension of $w^{\pm}_k$ differ by a distribution supported at
$\xi=0$, which has a smooth inverse Fourier transform. In fact, the
distributions $E^{F,\pm}_k$ are uniquely determined up to smooth
functions by the first line of Equation (\ref{Eqn_Fk1}) and the
condition that their wave front set is contained in the right-hand
side of Equation (\ref{Eqn_WFFeynman}). This can be shown by induction,
using the fact that the difference of two solutions $E^{F,\pm}_k$ of
Equation (\ref{Eqn_Fk1}) solves the wave equation with a smooth
source term. It then follows from the Propagation of Singularities
Theorem \cite{Duistermaat+1972} that the wave front set estimates
can only hold if the difference is a smooth function. The regularity
of $E^{F,\pm}_k$ for sufficiently large $k$ is a bit harder to
see directly, but we will return to this momentarily.

The Hadamard series that is used to characterise the singularities of
Hadamard states for a real scalar QFT arises as follows. We consider
the differences $F_k:=-i(E^{F,+}_k-E^-_k)$, $k\ge 0$, which satisfy
\[
P_0F_k\sim\left\{\begin{array}{ll}
kF_{k-1}&k\ge 1\\
0&k=0\end{array}\right.
\]
and
\[
WF(F_k)\subset\left\{(x,\xi)\in\mathbb{R}^d\times\mathbb{R}^d|\
\sigma(x)=0,\ \xi_{\mu}=a\eta_{\mu\nu}x^{\nu},\ \xi_0<0\right\}.
\]
The wave front set estimate can be proved by induction, using the
Propagation of Singularities Theorem to propagate singularities in
the past light cone to singularities in the future light cone, where
$E^-_k$ vanishes and only the singularities of $E^{F,+}_k$ can
occur. It can also be shown by induction that the distributions $F_k$
are uniquely determined up to smooth functions by their wave front set
estimate and the condition that $F_k(x)-F_k(-x)=iE_k(x)$.

A convenient expression for the $F_k$ can be obtained from a Wick
rotation of the Euclidean distributions $F^e_k$. For this purpose
we consider the holomorphic function
$s^c(z,x^i):=-z^2+\sum_{i=1}^{d-1}(x^i)^2$ of $z:=x^0+i\tau$ and we
define
\begin{equation}\label{Eqn_HadFk2}
F^l_k(x)=\lim_{\tau\rightarrow 0^+}s^c(z,x')^{k+1-\frac{d}{2}}
(c_k+d_k\log s^c(z,x')),
\end{equation}
with $c_k,d_k$ as in Equation (\ref{Eqn_ck}) (cf.\ \cite{Moretti1999}).
In this formula all logarithms and all fractional powers are defined
as holomorphic functions with the principal branch cut along the
non-positive real axis. Note that the range of $s^c$ does not
intersect the branch cut of the logarithm as long as $\tau\not=0$
and that taking instead the limit $x^0\rightarrow 0$ would yield
$F^e_k(\tau,x^i)$ (as long as $\tau\not=0$).

Performing derivatives before we take the limit $\tau\rightarrow 0^+$
and proceeding as in the proof of Proposition \ref{Prop_FkEucl} one
may verify by direct computation that
\begin{eqnarray}\label{Eqn_Fk1'}
P_0F^l_k&=&\left\{\begin{array}{ll}
kF^l_{k-1}+(d-2-4k)d_k\sigma^{k-\frac{d}{2}}&k\ge 1,\ k\not=\frac{d}{2}-1\\
kF^l_{k-1}&k=\frac{d}{2}-1\ge1\\
0&k=0\end{array}\right..
\end{eqnarray}
It is apparent from these equations that any singularities must lie
on the light cone and the wave front sets can only contain lightlike
vectors, which must be future pointing by standard arguments
(\cite{Hoermander} Thm.\ 8.1.6). This proves the desired wave front
set estimate.

The fact that $F^l_k\in C^{k+1-d}$ when $2k\ge d-1$ can again be
shown by induction, as in the proof of Proposition \ref{Prop_FkEucl},
or it can be verified by direct computation. Furthermore, when $k$ is
large enough one can easily see that $F^l_k(x)-F^l_k(-x)$ is Lorentz
invariant, odd under spacetime reflection in $x=0$ and homogeneous of
degree $2k+2-d$. Since there is only one such distribution, up to a
multiplicative factor, it follows that $F^l_k(x)-F^l_k(-x)$ is a
multiple of $E_k$ when $k$ is large enough. A direct computation
shows that
\begin{eqnarray}
F^l_k(x)-F^l_k(-x)&=&e'_k\theta(\pm x^0)\theta(-\sigma(x))|\sigma(x)|^{k+1-\frac{d}{2}}
,\nonumber\\
e'_k&=&\left\{\begin{array}{ll}
2\pi i^{2k+1-d}d_k&d\ \mathrm{even},\\
2i^{2k-d}c_k&d\ \mathrm{odd},
\end{array}\right.\nonumber
\end{eqnarray}
and a comparison of the coefficients $c_k,d_k$ in Equation
(\ref{Eqn_ck}) with the coefficients $e_k$ in Equation (\ref{Eqn_ek})
shows that\footnote{Here we rewrite $c_k$ using
$\Gamma\left(\frac12-n\right)=(-1)^n\pi\Gamma\left(\frac12+n\right)^{-1}$.}
$e'_k=ie_k$ and hence $F^l_k(x)-F^l_k(-x)=iE_k(x)$ when $k$ is large
enough. The same equality then holds for all $k$, by Equations
(\ref{Eqn_Fk1}, \ref{Eqn_Fk1'}). Taken altogether this proves that the
expressions for $F^l_k$ above satisfy $F^l_k\sim F_k=-i(E^{F,+}_k-E^-_k)$.
Note that the regularity of $F^l_k$ and of $E^{\pm}_k$ for sufficiently
large $k$ also implies the desired regularity for $E^{F,+}_k$ and hence
for $E^{F,-}_k$, by complex conjugation.

We now consider the Hadamard series based on the $F^l_k$:
\begin{equation}
H^{(N)}:=-i\sum_{k=0}^Nu_k\tilde{F}_k,\quad
\tilde{F}_k:=\rho^*(F_k\otimes 1),\nonumber
\end{equation}
where again $\rho(x,y):=(\exp_y^{-1}(x),y)$. By Moretti's Theorem $H^{(N)}$
approximates a bi-solution to the Klein-Gordon equation. A two-point
distribution $\omega_2$ is called Hadamard if and only if $\omega_2-H^{(N)}$
is $C^{(N+2-d)}$ for all $N\ge \frac{d-3}{2}$. The factor $-i$ is
needed to ensure that $H^{(N)}$ can be made into a distribution of
positive type by adding a suitable $C^{(N+3-d)}$ function.

\subsection{Infinitesimal analytic continuation of the Hadamard series}\label{SSec_ACHad}

In Subsection \ref{SSec_infAC} we have made a detailed comparison of the
geometry near the Cauchy surface $\Sigma$ of a spacetime $M$ with a static
bifurcate Killing horizon and a (weak) wedge reflection $(\Sigma,\iota)$,
with the geometry near $\overline{\mu}(\Sigma)$ in $M':=\overline{M^+_{R_H}}$
when $R_H\equiv\kappa^{-1}$. The purpose of this subsection is to establish a
comparison between the Hadamard series in these two manifolds. We consider
the same geometric situation as in Section \ref{SSec_infAC}, with a Cauchy
surface $\Sigma\subset M$ satisfying the properties of Definition
\ref{Def_StaticBHSpac} and a coordinate neighbourhood $U\subset\Sigma$. By
shrinking $U$ if necessary, we may assume that there are convex normal
neighbourhoods $V\subset M$ and $V'\subset M'$ such that $U\subset V$ and
$U':=\overline{\mu}(U)\subset V'$. We let $x^{\mu}$ and $(x')^{\mu}$ be
Gaussian normal coordinates near $U$ and $U'$, which are related as in
Section \ref{SSec_infAC} and we let $y^{\mu},(y')^{\mu}$ be copies of
the same coordinates.

Whereas the Hadamard coefficients $u_k$ depend on the choice of a
second-order partial differential operator $P$, the distributional factors
$\tilde{F}_k$ in the Hadamard series for a fundamental solution or
two-point distribution only depend on the local geometry. We will first
consider the Hadamard coefficients:
\begin{proposition}\label{Prop_ACHadCoeff}
Consider the operators $K:=-\Box_g+V$ on $M$ and $K':=-\Box_{g'}+V'$
on $M'$, where $V$ and $V'$ are smooth functions which are stationary,
\[
\xi^{\mu}\partial_{\mu}V=0,\quad \overline{\xi}_R^{\mu}\partial'_{\mu}V'=0,
\]
and such that $\iota^*V=V=\overline{\mu}^*V'$ on $\Sigma$. Let
$u_k$ be the Hadamard coefficients for $K$ on $M$ and $u'_k$ those for
$K'$ on $M'$. Then the following equality holds on $U^{\times 2}$:
\[
\partial^l_{x^0}\partial^m_{y^0}u_k=i^{l+m}(\overline{\mu}^{\times 2})^*
\left(\partial^l_{(x')^0}\partial^m_{(y')^0}u'_k\right).
\]
\end{proposition}
\begin{proof*}
First we will show that
\[
\partial_{x^0}^lV=i^n\overline{\mu}^*\left(\partial_{(x')^0}^lV'\right)
\]
on $U$. For $l=0$ this is true by assumption. We proceed by induction
over $l$, exploiting the fact that $V$ and $V'$ are stationary. Indeed,
\begin{eqnarray}
0&=&\partial_{x^0}^l(\xi^{\mu}\partial_{\mu})V-
i^l\overline{\mu}^*\left(\partial_{(x')^0}^l
(\overline{\xi_R}^{\mu}\partial'_{\mu})V'\right)\nonumber\\
&=&\xi^0\partial_{x^0}^{l+1}V-i^l\overline{\mu}^*\left(
\overline{\xi_R}^0\partial_{(x')^0}^{l+1}V'\right)\nonumber
\end{eqnarray}
on $U$, where the last equality follows from the induction hypothesis
for $l'\in\left\{0,\ldots,l\right\}$ and Corollary \ref{Cor_CCurvature}.
Since $\xi^0=\overline{\mu}^*(\overline{\xi_R})^0=v\not=0$ on the dense
subset $\Sigma\setminus\mathcal{B}$ of $\Sigma$, the claim for $l+1$
follows and the proof by induction is complete.

From Proposition \ref{Prop_complexGnormal2} we find that
\[
-\partial_{x^0}^l\det g_{\mu\nu}=i^l\overline{\mu}^*\left(
\partial_{(x')^0}^l\det (\overline{g}_R)_{\mu\nu}\right)
\]
on $U$, and similarly for
$\det\left(\partial_{x^{\mu}}\partial_{y^{\nu}}\sigma(x,y)\right)$,
due to Corollary \ref{Cor_CSynge}. From Equation (\ref{Eqn_VVM}) we have
\[
u_0(x,y)=\frac{1}{\sqrt[4]{g(x)g(y)}}
\sqrt{(-1)^{d+1}\det\left(\partial_{x^{\mu}}\partial_{y^{\nu}}\sigma(x,y)\right)}
\]
and similarly for $u'_0(x',y')$, but with $(-1)^d$ instead of $(-1)^{d+1}$.
Combining these results yields the statement of the Proposition for $u_0$
and $u'_0$, i.e.\ for $k=0$. We will now proceed by induction over $k$, so
we may assume that the claim has been shown for all $k'\in\{0,\ldots,k\}$
for some $k\ge 0$ and we aim to prove it for $k+1$.

From the induction hypothesis, Corollary \ref{Cor_CCurvature} and our
results on $V$ and $V'$ we may conclude that on $U^{\times 2}$:
\[
\partial^l_{x^0}\partial^m_{y^0}Ku_k=i^{l+m}(\overline{\mu}^{\times 2})^*
\left(\partial^l_{(x')^0}\partial^m_{(y')^0}K'u'_k\right).
\]
To proceed we need to recall some notations from the proof of
Proposition \ref{Prop_CRiemann}. Given $x^{\mu}=(0,x^i)$ and
$y^{\mu}=(0,y^i)$ in $U$ we define the geodesics
$r\mapsto\gamma_0^{\mu}(r,y^i)$ and $s\mapsto\gamma_0^{\mu}(s,x^i)$
in $V$. For some $\epsilon>0$ we may define the map
$\gamma^{\mu}:(-\epsilon,\epsilon)^{\times 2}\times[0,1]\to V$ such that
$t\mapsto\gamma^{\mu}(r,s,t)$ is the unique geodesic in $V$ between
$\gamma_0(r)$ and $\gamma_1(s)$. We define a map
$(\gamma')^{\mu}:(-\epsilon,\epsilon)^{\times 2}\times[0,1]\to V'$ in
complete analogy, using the points $x':=\overline{\mu}(x)$ and
$y':=\overline{\mu}(y)$. We have shown in the proof of Proposition
\ref{Prop_CRiemann} that
\[
\partial_s^l\partial_r^m\gamma^{\mu}(0,0,t)=i^{l+m+c}
\partial_s^l\partial_r^m(\gamma')^{\mu}(0,0,t),
\]
where $c=-1$ when $\mu=0$ and $c=0$ otherwise. In particular, we
have for $t=0$:
\[
\partial_r^m\gamma_0^{\mu}(0)=i^{m+c}\partial_r^m(\gamma')^{\mu}(0).
\]

Combining this with our previous results we find
\begin{eqnarray}
\partial^l_{x^0}\partial^m_{y^0}(Ku_k)(x(t),y)&=&
\partial^l_s\partial^m_r(Ku_k)(\gamma(r,s,t),\gamma_0(r))|_{r=s=0}\nonumber\\
&=&\partial^l_{(x')^0}\partial^m_{(y')^0}(K'u'_k)(x'(t),y'),\nonumber
\end{eqnarray}
where we have written $x(t)$ for the unique geodesic in $U$ from $y$ to $x$,
$x'(t)$ for the corresponding geodesic in $U'$. The last equality uses the
chain rule and a matching up of factors $i$ on both sides, where we note that
a $\gamma^0$ or $\gamma_0^0$ comes with an extra derivative $\partial_{x^0}$
or $\partial_{y^0}$, so the additional factors of $i$ cancel out. Note that
the derivatives with respect to $x^0$ and $y^0$ also force use to vary the
curve $x(t)$, for which purpose we needed to use $\gamma^{\mu}$.

The same argument actually works for all $u_{k'}$ with $k'\le k$ and in
particular for $k=0$. All these results can then be inserted into the formula
for $u_{k+1}$ and $u'_{k+1}$ given in Lemma \ref{Lem_HadCoeff}. Differentiating
under the integral sign then proves the claim for $k+1$, so the proof is
complete.
\end{proof*}

Now we turn to the distributional parts of the Hadamard series. We choose
orthonormal frames $(e_{\alpha})^{\mu}$ and $(e'_{\alpha})^{\mu}$ on $V$ and
$V'$ as in Lemma \ref{Lem_CFrame}, and we use these to define Riemannian
normal coordinates on $V^{\times 2}$ and $(V')^{\times 2}$, respectively,
denoting the coordinate changes by $\rho$ and $\rho'$. We then consider the
distributions $F^l_k$ of Equation (\ref{Eqn_HadFk2}) and $F^e_k$ of Equation
(\ref{Eqn_HadFk1}) and the corresponding distributions
$\tilde{F}^l_k:=\rho^*(F^l_k\otimes 1)$ and
$\tilde{F}^e_k:=(\rho')^*(F^e_k\otimes 1)$.

Away from the diagonal of $U^{\times 2}$ one may easily show that in Gaussian
normal coordinates
\[
\partial_{x^0}^l\partial_{y^0}^m\tilde{F}^l_k(x,y)
=i^{l+m}(\overline{\mu}^{\times 2})^*(\partial_{(x')^0}^l
\partial_{(y')^0}^m\tilde{F}^e_k)(x,y)
\]
because of Corollary \ref{Cor_CSynge} and the fact that $\tilde{F}^l_k$ and
$\tilde{F}^e_k$ are locally given by the same expression in terms of
$\sigma$ and $\bar{\sigma}_R$, respectively. However, it is necessary to have
a more detailed understanding of this infinitesimal analytic continuation also
at the diagonal. For this purpose, we will first consider the Lorentzian
case, which is better behaved regarding restrictions to the Cauchy
surface $\Sigma$. We will use the distribution $\delta$ on $U^{\times 2}$,
defined by $\delta(f):=\int_Uf|_{\Delta}$, where $f|_{\Delta}(x):=f(x,x)$.
\begin{theorem}\label{Thm_ContF}
For $\tilde{F}^l_k$ the following equalities hold:
\begin{eqnarray}
\tilde{F}^l_k|_{U^{\times 2}}&=&\sigma_U^{k+1-\frac{d}{2}}(c_k+d_k\log(\sigma_U))
\nonumber\\
\partial_{x^0}\tilde{F}^l_k|_{U^{\times 2}}&=&
-\partial_{y^0}\tilde{F}^l_k|_{U^{\times 2}}=\left\{\begin{array}{ll}
0&k\ge 1\\
\frac{-i}{2}\delta&k=0
\end{array}\right.\nonumber\\
\partial_{x^0}\partial_{y^0}\tilde{F}^l_k|_{U^{\times 2}}&=&\left\{
\begin{array}{ll}
\frac{-1}{2}(\partial_{x^0}\partial_{y^0}\sigma)\tilde{F}^l_{k-1}|_{U^{\times 2}}
+(\partial_{x^0}\partial_{y^0}\sigma)|_{U^{\times 2}}d_k\sigma_U^{k-\frac{d}{2}}
&k\ge 1,\ k\not=\frac{d}{2}-1\\
\frac{-1}{2}(\partial_{x^0}\partial_{y^0}\sigma)\tilde{F}^l_{k-1}|_{U^{\times 2}}
&k\ge 1,\ k=\frac{d}{2}-1
\end{array}\right.\nonumber
\end{eqnarray}
where the distributional restrictions are well-defined in the sense of
microlocal analysis and $\sigma_U$ is the squared geodesic distance of
$(\Sigma,h_{ij})$ on $U$. Furthermore,
\[
(\partial_{x^0}+\partial_{y^0})^2\tilde{F}^l_0|_{U^{\times 2}}=
C(d)\sigma_U^{\epsilon+\frac12-\frac{d}{2}}\frac{-(\partial_{x^0}+\partial_{y^0})^2\sigma|_{U^{\times 2}}}
{(\sigma_U)^{\epsilon+\frac12}}
\]
for any $\epsilon\in\left(0,\frac12\right)$, where $C(d)=\left(1-\frac{d}{2}\right)c_0+d_0$.
\end{theorem}
The reason why we treat $\partial_{x^0}\partial_{y^0}\tilde{F}^l_0$ differently is to facilitate
the comparison with the Euclidean case later on.
\begin{proof*}
The distributions $\tilde{F}^l_k$ have wave front sets which are
contained in the lightlike vectors. In particular, they do not
intersect the conormal bundle $N^*U^{\times 2}$, because $U$ is
spacelike. By a standard result in microlocal analysis
(\cite{Hoermander} Theorem 8.2.4) the restrictions of all
$\tilde{F}^l_k$ and all of their derivatives are well-defined, so
it remains to compute them. For this, we use the limit in Equation
(\ref{Eqn_HadFk2}), which holds in the sense of the H\"ormander
pseudo-topology so that it commutes with restrictions. (This can be
seen most easily by a slight adaptation of the proof of Theorem 8.1.6
in \cite{Hoermander}. See also \cite{Dabrowski+2013} for a discussion
of the H\"ormander topology.)

Note that
\[
\sigma_{\tau}(x,y):=(\rho^*(s^z\otimes 1))(x,y)=\sigma(x,y)+\frac12\tau^2
+i\tau v^0(x,y),
\]
so we may write
\[
\tilde{F}^l_k(x,y)=\lim_{\tau\rightarrow 0^+}\sigma_{\tau}(x,y)^{k+1-\frac{d}{2}}
(c_k+d_k\log\sigma_{\tau}(x,y)).
\]
This limit again commutes with restrictions, because we have only tensored
in a constant function $1$ and applied a smooth change of coordinates $\rho$
to obtain $\tilde{F}^l_k=\rho^*(F^l_k\otimes 1)$. Furthermore, for
$x,y\in U$ we have $\sigma_{\tau}(x,y)=\sigma_U(x,y)+\frac12\tau^2$
(cf.\ the proof of Corollary \ref{Cor_CSynge}). As $(\sigma_U)^r$ is locally
integrable for all $r>\frac{1-d}{2}$ the formula for
$\tilde{F}^l_k|_{U^{\times 2}}$ immediately follows from the dominated
convergence theorem. The convergence of the limit even works for all
continuous test-functions $f\in C_0^0(U)$, i.e.\ in the sense of measure
theory. The same is true when $k$ is greater than or equal to the number
of normal derivatives, where we use Corollary \ref{Cor_CSynge} to obtain
the correct formulae.

For the remaining cases, we may take the normal derivatives before
taking the limit $\tau\rightarrow 0^+$. Note that $v^0|_{U^{\times 2}}\equiv 0$,
by Corollary \ref{Cor_CauchyGeodesics} and
$\partial_{x^0}\sigma|_{U^{\times 2}}=\partial_{y^0}\sigma|_{U^{\times 2}}=0$
by Corollary \ref{Cor_CSynge}. Furthermore, if $\gamma(t)$ is the unique
geodesic through $y\in U$ with tangent vector $n^{\mu}(y)$, then
$v^0(\gamma(s),\gamma(t))=s-t$, from which it follows that
\begin{eqnarray}
\partial_{x^0}v^0(x,y)|_{x=y}&=&\partial_sv^0(\gamma(s),y)|_{s=0}=1,\nonumber\\
\partial_{y^0}v^0(x,y)|_{x=y}&=&\partial_tv^0(y,\gamma(t))|_{t=0}=-1,\nonumber\\
\partial_{x^0}\partial_{y^0}v^0(x,y)|_{x=y}&=&
\partial_s\partial_tv^0(\gamma(s),\gamma(t))|_{s=t=0}=0.\nonumber
\end{eqnarray}
Taking the derivatives and restrictions, we now find
\begin{eqnarray}
\partial_{x^0}\tilde{F}^l_0|_{U^{\times 2}}&=&
\lim_{\tau\rightarrow 0^+}C(d)\sigma_{\tau}^{-\frac{d}{2}}i\tau\partial_{x^0}v^0|_{U^{\times 2}}\nonumber\\
\partial_{y^0}\tilde{F}^l_0|_{U^{\times 2}}&=&
\lim_{\tau\rightarrow 0^+}C(d)\sigma_{\tau}^{-\frac{d}{2}}i\tau\partial_{y^0}v^0|_{U^{\times 2}}\nonumber\\
\partial_{x^0}\partial_{y^0}\tilde{F}^l_1|_{U^{\times 2}}&=&
\lim_{\tau\rightarrow 0^+}\sigma_{\tau}^{-\frac{d}{2}}\left\{(a+b\log\sigma_{\tau})i\tau\partial_{x^0}v^0
i\tau\partial_{y^0}v^0-(\frac{c_0}{2}+d_1\delta_{d,2}+\frac{d_0}{2}\log\sigma_{\tau})\right\}|_{U^{\times 2}}\nonumber
\end{eqnarray}
for some constants $a,b$ and with $\delta_{d,2}$ denoting the Kronecker delta. The
limits can be computed using the technical Lemma \ref{Lem_App} in the appendix,
leading to the stated results.

It remains to prove the final equation for the case $k=0$. By a straightforward computation,
using the fact that $(2-d)d_0=0$ for all $d$, we find that
\[
(\partial_{x^0}+\partial_{y^0})^2\tilde{F}^l_0|_{U^{\times 2}}=C(d)\lim_{\tau\rightarrow 0^+}
(\sigma_U+\frac12\tau^2)^{\epsilon+\frac12-\frac{d}{2}}\frac{\Phi}{(\sigma_U+\frac12\tau^2)^{\epsilon+\frac32}}
\]
\begin{eqnarray}
\Phi&:=&-\frac{d}{2}\tau^2((\partial_{x^0}+\partial_{y^0})v^0|_{U^{\times 2}})^2
-\left(\sigma_U+\frac12\tau^2\right)(\partial_{x^0}+\partial_{y^0})^2\sigma_{\tau}|_{U^{\times 2}}.\nonumber
\end{eqnarray}
The function $\Phi$ has a Taylor series in $x^i,y^i,\tau$ which vanishes up to (and including)
third order at any point with $x^i=y^i$ and $\tau=0$, because $\sigma$ and its first-order
derivatives vanish on the diagonal $\Delta$ and we have
$((\partial_{x^0}+\partial_{y^0})P\sigma)|_{\Delta}=\partial_0((P\sigma)|_{\Delta})$ for any
partial differential operator $P$. It follows that the quotient
\[
\frac{\Phi}{(\sigma_U+\frac12\tau^2)^{\epsilon+\frac32}}
\]
is continuous at points where $x^i=y^i$ and $\tau=0$. As $\tau\rightarrow 0^+$ it converges
uniformly on compact sets to $\frac{-(\partial_{x^0}+\partial_{y^0})^2\sigma|_{U^{\times 2}}}
{\sigma_U^{\epsilon+\frac12}}$. The power
$(\sigma_U+\frac12\tau^2)^{\epsilon+\frac12-\frac{d}{2}}$ converges to
$\sigma_U^{\epsilon+\frac12-\frac{d}{2}}$ in the sense of measure theory, by the dominant
convergence theorem. It follows that the product also converges to the expression
claimed in the theorem.
\end{proof*}

We now turn to the Euclidean case and obtain a similar result:
\begin{theorem}\label{Thm_ContFe}
Let $V'\subset M'$ be a causally convex normal neighbourhood and let
$U'\subset V'\cap\overline{\mu}(\Sigma)$ be a relatively compact
open set. Then there is a $\delta>0$ such that the Gaussian normal
coordinate $(x')^0$ is a well-defined coordinate near $\overline{U'}$ for all
$|(x')^0|\le\delta$. For $F^e_k$ as in Proposition \ref{Prop_FkEucl} and
$k\in\mathbb{N}_0$ we have:
\begin{eqnarray}
\tilde{F}^e_k|^+_{(U')^{\times 2}}&=&\sigma_{U'}^{k+1-\frac{d}{2}}
(c_k+d_k\log(\sigma_{U'}))\nonumber\\
\partial_{(x')^0}\tilde{F}^e_k|^+_{(U')^{\times 2}}&=&
-\partial_{(y')^0}\tilde{F}^e_k|^+_{(U')^{\times 2}}=
\left\{\begin{array}{ll}
0& k\ge 1\\
\frac12\delta_0&k=0
\end{array}\right.\nonumber\\
\partial_{(x')^0}\partial_{(y')^0}\tilde{F}^e_k|^+_{(U')^{\times 2}}&=&
\left\{\begin{array}{ll}
\begin{array}{l}
\frac{-1}{2}(\partial_{(x')^0}\partial_{(y')^0}\bar{\sigma}_R)\tilde{F}^e_{k-1}|^+_{(U')^{\times 2}}\\
\qquad +(\partial_{(x')^0}\partial_{(y')^0}\bar{\sigma}_R)|^+_{(U')^{\times 2}}d_k\sigma_{U'}^{k-\frac{d}{2}}
\end{array}
&k\ge 1,\ k\not=\frac{d}{2}-1\\
\frac{-1}{2}(\partial_{(x')^0}\partial_{(y')^0}\bar{\sigma}_R)\tilde{F}^e_{k-1}|^+_{(U')^{\times 2}}
&k\ge 1,\ k=\frac{d}{2}-1
\end{array}\right.\nonumber
\end{eqnarray}
where $\sigma_{U'}$ is Synge's world function restricted to $U'$ and
$|^+_{(U')^{\times 2}}$ denotes the distributional limit as
$(y')^0=-(x')^0\rightarrow 0^+$. The expressions on the right-hand side
of the first line are locally integrable functions. Furthermore,
\[
(\partial_{(x')^0}+\partial_{(y')^0})^2\tilde{F}^e_0|_{(U')^{\times 2}}=
C(d)\sigma_{U'}^{\epsilon+\frac12-\frac{d}{2}}
\frac{(\partial_{(x')^0}+\partial_{(y')^0})^2\bar{\sigma}_R|_{(U')^{\times 2}}}{(\sigma_{U'})^{\epsilon+\frac12}}
\]
for any $\epsilon\in\left(0,\frac12\right)$, where $C(d)=\left(1-\frac{d}{2}\right)c_0+d_0$.
\end{theorem}
\begin{proof*}
Because $\overline{U'}$ is compact it is obvious that an $\epsilon>0$
exists such that the Gaussian normal coordinate is well-defined. We will
first show that
\begin{equation}\label{Eqn_UnifLim}
\lim_{(x')^0,(y')^0\rightarrow 0}\frac{\sigma_{U'}((x')^i,(y')^i)+\frac12|(x')^0-(y')^0|^2}{\bar{\sigma}_R(x',y')}
\rightarrow 1
\end{equation}
uniformly on compact sets. Both numerator and denominator converge uniformly on
compact sets to $\sigma_{U'}((x')^i,(y')^i)$. On compact subsets of $(U')^{\times 2}$
which do not intersect the diagonal the claim then follows. Near the diagonal we
require an additional argument. We may choose Riemannian local coordinates
$(x')^i$ on $U'$ centred at a given point $p\in U'$ and extend these to Gaussian
normal coordinates on $\overline{U'}\times[-\delta,\delta]$. In these
coordinates, the metric takes the form $g=(d(x')^0)^{\times 2}+h'_{ij}d(x')^id(x')^j$,
where $h'_{ij}(x')$ is a real, symmetric matrix that defines a bounded, positive
operator on $\mathbb{R}^{d-1}$. There are constants $M>1>m>0$ such that
$m\delta_{ij}\le h'_{ij}(x')\le M\delta_{ij}$. For any
$x',y'\in\overline{U'}\times[-\delta,\delta]$ we then find that
$h'_{ij}(x')\ge c\ h'_{ij}(y')$ with $c:=\frac{m}{M}\in (0,1]$. By shrinking the
region $W$ to a smaller neighbourhood of $p$ and by shrinking $\delta$ if
necessary we can get the constants $m$, $M$ and $c$ to be arbitrarily close to $1$.

If $\map{\gamma}{[0,1]}{V'}$ is the unique geodesic in $V'$ between
$x'=((x')^0,(x')^i)$ and $y'=((y')^0,(y')^i)$, we may write it as
$\gamma=(\gamma^0,\gamma^i)$. If $\tilde{x'}=((\tilde{x'})^0,(x')^i)$ and
$\tilde{y'}=((\tilde{y'})^0,(y')^i)$ we can use the curve
$\tilde{\gamma}(s):=((1-s)(\tilde{x'})^0+s(\tilde{y'})^0,\gamma^i(s))$
(which may not be a geodesic) to derive the following estimate:
\begin{eqnarray}\label{Eqn_Estsigma00}
\bar{\sigma}_R(x',y')&=&\frac12\int_0^1dt\ g_{\mu\nu}(\gamma)\dot{\gamma}^{\mu}\dot{\gamma}^{\nu}
=\frac12\int_0^1dt\ g_{\mu\nu}(\gamma)\dot{\tilde{\gamma}}^{\mu}
\dot{\tilde{\gamma}}^{\nu}+(\dot{\gamma}^0)^2-(\dot{\tilde{\gamma}}^0)^2\nonumber\\
&\ge&\frac12\int_0^1dt\ cg_{\mu\nu}(\tilde{\gamma})\dot{\tilde{\gamma}}^{\mu}\dot{\tilde{\gamma}}^{\nu}
+(\dot{\gamma}^0)^2-(\dot{\tilde{\gamma}}^0)^2\nonumber\\
&\ge&c\bar{\sigma}_R(\tilde{x'},\tilde{y'})
+\frac12(|(x')^0-(y')^0|^2-|(\tilde{x'})^0-(\tilde{y'})^0|^2),\nonumber
\end{eqnarray}
where we used the fact that the geodesic in $\Sigma\cap V'$ between
$\tilde{x'}$ and $\tilde{y'}$ is the unique curve of minimal length
and similarly for the geodesic between $(x')^0$ and $(y')^0$ in
$(-\delta,\delta)$ with the Riemannian metric $(d(x')^0)^2$
(cf.\ \cite{ONeill} Proposition 5.16). In particular, it follows that
\begin{equation}
c\sigma_{U'}((x')^i,(y')^i)+\frac{c}{2}|(x')^0-(y')^0|^2\le\bar{\sigma}_R(x',y')\le
c^{-1}\sigma_{U'}((x')^i,(y')^i)+\frac{1}{2c}|(x')^0-(y')^0|^2.\nonumber
\end{equation}
For any compact $K\subset U'$ we can cover a neighbourhood of the compact
intersection of $K^{\times 2}$ with the diagonal by a finite number of
sufficiently small compact sets to see that the last estimate still holds
and that the constant can be chosen arbitrarily close to $1$ if $(x')^0$
and $(y')^0$ are small enough. Therefore, the limit in Equation
(\ref{Eqn_UnifLim}) converges uniformly on compact sets.

If $r>\frac{1-d}{2}$ and $(x')^0,(y')^0\rightarrow 0$, then $\bar{\sigma}_R(x,y)^r$
converges as a measure to the locally integrable function $\sigma_{U'}(x,y)^r$.
This follows from the dominated convergence theorem, the uniform convergence
above and the estimate $\sigma_{U'}((x')^i,(y')^i)+\frac12|(x')^0-(y')^0|^2
\ge\sigma_{U'}((x')^i,(y')^i)$. Hence, for $k$ greater than or equal to the
number of normal derivatives, the distributions
\[
\tilde{F}^e_k,\quad \partial_{(x')^0}\tilde{F}^e_k,\quad
\partial_{(y')^0}\tilde{F}^e_k,\quad
-\partial_{(x')^0}\partial_{(y')^0}\tilde{F}^e_k,
\]
are continuous functions of $(x')^0\not=(y')^0$ with values in the
distributions on $(U')^{\times 2}$. Moreover, the limits
$(x')^0\rightarrow 0^-$ and $(y')^0\rightarrow 0^+$ exist and are given
by the indicated locally integrable functions, where we use Corollary
\ref{Cor_CSynge} to treat the normal derivatives of $\bar{\sigma}_R$.

For $k=1$, $k\not=\frac{d}{2}-1$, we have
\begin{eqnarray}
\partial_{(x')^0}\partial_{(y')^0}\tilde{F}^e_1&=&(\partial_{(x')^0}\partial_{(y')^0}\bar{\sigma}_R)
\left(-\frac12\tilde{F}^e_0+d_1\bar{\sigma}_R^{1-\frac{d}{2}}\right)\nonumber\\
&&+(\partial_{(x')^0}\bar{\sigma}_R)(\partial_{(y')^0}\bar{\sigma}_R)\bar{\sigma}_R^{-\frac{d}{2}}
\left(\left(1-\frac{d}{2}\right)\left(d_1-\frac12c_0\right)-\frac12d_1\right)\nonumber
\end{eqnarray}
as a locally integrable function on $(U'\times(-\epsilon,\epsilon))^{\times 2}$, whereas for
$k=1=\frac{d}{2}-1$ we find
\begin{eqnarray}
\partial_{(x')^0}\partial_{(y')^0}\tilde{F}^e_1&=&(\partial_{(x')^0}\partial_{(y')^0}\bar{\sigma}_R)
\left(-\frac12\tilde{F}^e_0\right)
+(\partial_{(x')^0}\bar{\sigma}_R)(\partial_{(y')^0}\bar{\sigma}_R)\bar{\sigma}_R^{-\frac{d}{2}}
\frac{c_0}{2}.\nonumber
\end{eqnarray}
In both cases we can use the Euclidean version of Equation (\ref{Eqn_SyngeProp})
to deduce that $|\partial_{(x')^0}\bar{\sigma}_R|\le\sqrt{2\bar{\sigma}_R}$
and $|\partial_{(y')^0}\bar{\sigma}_R|\le\sqrt{2\bar{\sigma}_R}$. We may use this to
estimate the factors in the second term as
\[
|(\partial_{(x')^0}\bar{\sigma}_R)(\partial_{(y')^0}\bar{\sigma}_R)|\le
|\partial_{(x')^0}\bar{\sigma}_R|^{\frac12}(2\bar{\sigma}_R)^{\frac{3}{4}}.
\]
Arguing as before, we see that $\partial_{(x')^0}\partial_{(y')^0}\tilde{F}^e_1$
has a limit as $(x')^0\rightarrow 0^-$ and $(y')^0\rightarrow 0^+$, where the
second term vanishes and the first term yields the expression stated in the
theorem.

Now we turn to the case $k=0$ with one normal derivative. With the
constant $C(d)$ defined in Theorem \ref{Thm_ContF} we then have
\begin{eqnarray}
\partial_{(x')^0}\tilde{F}^e_0(x',y')&=&C(d)(\partial_{(x')^0}\bar{\sigma}_R(x',y'))
\bar{\sigma}_R(x',y')^{-\frac{d}{2}}\nonumber\\
&=&C(d)\frac{(\partial_{(x')^0}\bar{\sigma}_R(x',y'))}{\tau}\left(\frac{\sigma_{U'}((x')^i,(y')^i)+\frac12|(x')^0-(y')^0|^2}
{\bar{\sigma}_R(x',y')}\right)^{\frac{d}{2}}\nonumber\\
&&\tau\left(\sigma_{U'}((x')^i,(y')^i)+\frac12\tau^2\right)^{-\frac{d}{2}}\nonumber
\end{eqnarray}
where we take $\frac12\tau=(y')^0=-(x')^0>0$. The quotient to the power $\frac{d}{2}$
converges to $1$ uniformly on compact sets as $\tau\rightarrow 0^+$. Expanding
$\partial_{(x')^0}\bar{\sigma}_R(x',y')$ in a Taylor series around $(x')^0=(y')^0=0$
we see that the first quotient converges uniformly on compact sets to
\[
\frac12\partial_{(x')^0}\partial_{(y')^0}\bar{\sigma}_R(x',y')|_{(U')^{\times 2}}
-\frac12\partial_{(x')^0}^2\bar{\sigma}_R(x',y')|_{(U')^{\times 2}},
\]
which is constantly $-1$ on the diagonal. Since the last factors
converge to $\frac{-1}{2C(d)}\delta$ by Lemma \ref{Lem_App} in Appendix
\ref{App} we find that
\[
\partial_{(x')^0}\tilde{F}^e_0|^+_{(U')^{\times 2}}=\frac12\delta.
\]
The argument for $\partial_{(y')^0}\tilde{F}^e_0|^+_{(U')^{\times 2}}$ is
entirely analogous, but incurs an additional sign from the first quotient.

For the final equality, we use the fact that $(2-d)d_0=0$ for all $d$ to compute
for all $(x')^0\not=(y')^0$:
\[
(\partial_{(x')^0}+\partial_{(y')^0})^2\tilde{F}^l_0=C(d)
\bar{\sigma}_R^{\epsilon+\frac12-\frac{d}{2}}\frac{\Phi'}{\bar{\sigma}_R^{\epsilon+\frac32}}
\]
\[
\Phi':=\left\{-\frac{d}{2}((\partial_{(x')^0}+\partial_{(y')^0})\bar{\sigma}_R)^2
+\bar{\sigma}_R(\partial_{(x')^0}+\partial_{(y')^0})^2\bar{\sigma}_R\right\}
\]
with $\epsilon\in(0,\frac12)$. The function $\Phi'$ has a Taylor series in
$x^i,y^i$ and $\tau=\frac12(y')^0=-\frac12(x')^0$ which vanishes up to (and including)
third order at any point with $x^i=y^i$ and $\tau=0$. It follows that the quotient
\[
\frac{\Phi'}{\bar{\sigma}_R^{\epsilon+\frac32}}
\]
is continuous at such points. As $\tau\rightarrow 0^+$ it converges to
$\frac{(\partial_{(x')^0}+\partial_{(y')^0})^2\bar{\sigma}_R|_{(U')^{\times 2}}}
{\bar{\sigma}_R^{\epsilon+\frac12}}$ uniformly on compact sets. The power
$\bar{\sigma}_R^{\epsilon+\frac12-\frac{d}{2}}$ converges to
$\sigma_{U'}^{\epsilon+\frac12-\frac{d}{2}}$ in the sense of measure theory, by the
dominant convergence theorem. It therefore follows that the product also converges to the
expression claimed in the theorem.
\end{proof*}

\begin{remark}
To define the restrictions of distributions appearing in Theorem \ref{Thm_ContFe} it
does not seem to suffice to appeal to general results in microlocal analysis with
finite Sobolev regularity. For example, $\hat{F}^e_k(\xi)=k!|\xi|^{-2-2k}$ outside
$\xi=0$, where $|\xi|$ denotes the Euclidean norm. It follows that $F^e_k$ is in the
Sobolev space $H^{(s)}(\mathbb{R}^d)$ for all $s<2+2k-\frac{d}{2}$. In order to define
its restriction to a (time zero) hyperplane using \cite{HoermanderNLIN} Lemma 11.6.1 we
need to require that $s>\frac12$. This is possible only when $k>\frac{d-3}{4}$, so in
particular it fails for $k=0$ unless $d=2$.
\end{remark}

We can now compare the initial data of the distributions appearing in the Lorentzian
and the Euclidean version of the Hadamard series:
\begin{proposition}\label{Prop_InfACHad}
For all $k\ge 0$ we have
\begin{eqnarray}
\tilde{F}^l_k|_{U^{\times 2}}&=&(\overline{\mu}^{\times 2})^*(\tilde{F}^e_k|^+_{(U')^{\times 2}})\nonumber\\
\partial_{x^0}\tilde{F}^l_k|_{U^{\times 2}}&=&i(\overline{\mu}^{\times 2})^*
(\partial_{(x')^0}\tilde{F}^e_k|^+_{(U')^{\times 2}})\nonumber\\
\partial_{y^0}\tilde{F}^l_k|_{U^{\times 2}}&=&i(\overline{\mu}^{\times 2})^*
(\partial_{(y')^0}\tilde{F}^e_k|^+_{(U')^{\times 2}})\nonumber\\
\partial_{x^0}\partial_{y^0}\tilde{F}^l_k|_{U^{\times 2}}&=&-(\overline{\mu}^{\times 2})^*
(\partial_{(x')^0}\partial_{(y')^0}\tilde{F}^e_k|^+_{(U')^{\times 2}}).\nonumber
\end{eqnarray}
\end{proposition}
\begin{proof*}
With the exception of the case $k=0$ with two normal derivatives, this follows
immediately from the results of Theorems \ref{Thm_ContF} and \ref{Thm_ContFe},
where the sign in the case of two normal derivatives is due to Corollary
\ref{Cor_CSynge}. It remains to show the result for $k=0$ and two normal
derivatives.

We consider the Hadamard series on $M'$ for the operator $P'=-\Box_{\bar{g}_R}$.
If we choose $u'_0$ to be the Hadamard coefficient for this operator, then the last
term in Equation (\ref{Eqn_PHadSeries}) vanishes and we have
\begin{equation}\label{Eqn_Corresp}
P'(u'_0\tilde{F}^e_0)\sim (P'u'_0)\tilde{F}^e_0+u'_0\rho^*(\delta_0\otimes 1).
\end{equation}
Note that $u'_0\rho^*(\delta_0\otimes 1)=\rho^*(\delta_0\otimes 1)=\delta$, by a change
of coordinates and the fact that $u'_0=1$ on the diagonal. In Equation
(\ref{Eqn_Corresp}) we write $P'$ in terms of Gaussian normal coordinates (acting on
the first argument):
\[
P'=-\partial_{(x')^0}^2-\frac12(\det h'_{ij})^{-1}(x')
(\partial_{(x')^0}\det h'_{ij}(x'))\partial_{(x')^0}-\Box_{h'}.
\]
Since $u'_0\not=0$ is smooth we can then rewrite the equivalence as
\[
\partial_{(x')^0}^2\tilde{F}^e_0\sim
\frac{2}{u'_0}(g')^{\mu\nu}(\partial'_{\mu}u'_0)\partial'_{\nu}\tilde{F}^e_0
-\frac{1}{2\det h'_{ij}}(\partial'_0\det h'_{ij})\partial'_0\tilde{F}^e_0
-\Box_{h'}\tilde{F}^e_0
+\delta.
\]
It is then clear from Theorem \ref{Thm_ContFe} that all terms on the right-hand
side of the equivalence have a limit in Gaussian normal coordinates as
$-(x')^0=(y')^0\rightarrow 0^+$. Since both sides of the correspondence differ
by a smooth function, the same must be true for the left-hand side and
\[
\partial_{(x')^0}^2\tilde{F}^e_0|^+_{(U')^{\times 2}}\sim
\left(\frac{2}{u'_0}(h')^{ij}(\partial'_iu'_0)\partial'_j-\Box_{h'}\right)
\tilde{F}^e_0|^+_{(U')^{\times 2}}
\]
by Propositions \ref{Prop_complexGnormal2} and \ref{Prop_ACHadCoeff},
where all operators still act on $x'$ and we divided out $u'_0\not=0$.

From Theorem \ref{Thm_ContFe} we see that $\tilde{F}^e_0(y',x')=\overline{\tilde{F}^e_0(x',y')}$,
whereas $u'_0$ is symmetric by Moretti's Theorem \ref{Thm_Moretti}. We can therefore
apply the same argument as above to the case where $P'$ acts on the argument $y'$
to find that the limit of $-\partial_{(y')^0}^2\tilde{F}^e_0$ exists as
$-(x')^0=(y')^0\rightarrow 0^+$. The limit is obtained (up to equivalence $\sim$)
from the expression for $\partial_{(x')^0}^2\tilde{F}^e_0|^+_{(U')^{\times 2}}$ by
letting all operators act on $y'$ instead of $x'$. Appealing to the last statement
of Theorem \ref{Thm_ContFe} and taking a linear combination one finds that also
$\partial_{(x')^0}\partial_{(y')^0}\tilde{F}^e_0$ has such a limit and that
\begin{eqnarray}
\partial_{(x')^0}\partial_{(y')^0}\tilde{F}^e_0|^+_{(U')^{\times 2}}&\sim&
\frac{C(d)}{2}\sigma_{U'}^{\epsilon+\frac12-\frac{d}{2}}
\frac{(\partial_{(x')^0}+\partial_{(y')^0})^2\bar{\sigma}_R|_{(U')^{\times 2}}}
{\sigma_{U'}^{\epsilon+\frac12}}\nonumber\\
&&+\left(\frac12\Box_{h',(x')}+\frac12\Box_{h',(y')}
-\frac{1}{u'_0}(h')^{ij}(x')(\partial'_{(x')^i}u'_0)\partial'_{(x')^j}\right.\nonumber\\
&&\left.-\frac{1}{u'_0}(h')^{ij}(y')(\partial'_{(y')^i}u'_0)\partial'_{(y')^j}
\right)\tilde{F}^e_0|^+_{(U')^{\times 2}}.\nonumber
\end{eqnarray}

Similar arguments apply in the Lorentzian setting with the operator $P=-\Box_g$.
In that case, we use Theorem \ref{Thm_ContF} instead of \ref{Thm_ContFe}, the
restrictions to $U^{\times 2}$ are less problematic and the term
$\rho^*(\delta_0\otimes 1)$ is absent from the beginning, because we use the Hadamard
series for a solution instead of a fundamental solution. Using Proposition
\ref{Prop_ACHadCoeff}, Corollary \ref{Cor_CSynge} and the earlier results of this
proposition one finds
\[
\partial_{x^0}\partial_{y^0}\tilde{F}^l_0|_{U^{\times 2}}\sim-(\overline{\mu}^{\times 2})^*
(\partial_{(x')^0}\partial_{(y')^0}\tilde{F}^e_0|^+_{(U')^{\times 2}}).
\]
To show that we even have equality it suffices to compute both sides away from
the diagonal, where they are smooth functions. This computation is straightforward
and the equality follows from Corollary \ref{Cor_CSynge}. This completes the proof.
\end{proof*}

\section{The Hartle-Hawking-Israel state in static black holes}\label{Sec_HHI}

We now turn to the rigorous construction of the HHI state in spacetimes with
a static bifurcate Killing horizon and a wedge reflection. Assuming that $\kappa$
is constant singles out a particular radius $R_H=\kappa^{-1}$ and hence a
particular inverse temperature
\[
\beta_H=2\pi R_H=\frac{2\pi}{\kappa},
\]
the inverse Hawking temperature. The Riemannian metric $(\bar{g}_{R_H})_{ab}$
on $M':=\overline{M^+_{R_H}}$ is smooth and we consider the elliptic operator
\[
\bar{K}:=-\Box_{\bar{g}_{R_H}}+V
\]
on $M'$, where $V$ denotes the unique stationary potential on $M'$ that extends the
one on $M^+_{R_H}$ (cf.\ Lemma \ref{Lem_SmoothV'}). Just like $K_{R_H}$ in Section
\ref{SSec_DoubleKMS}, $\bar{K}$ is a symmetric and positive operator on the dense
domain $C_0^{\infty}(M')$ in $L^2(M')$ and it has a self-adjoint Friedrichs extension
$\hat{\bar{K}}$ which is strictly positive, because $\hat{\bar{K}}\ge V$. We may
therefore consider the Euclidean fundamental solution $\bar{G}:=\hat{\bar{K}}^{-1}$,
which defines a distribution density on $(M')^{\times 2}$.

Because $\mathcal{B}=M'\setminus M^+_{R_H}$ is a submanifold of codimension two we
may identify $L^2(M^+_{R_H})=L^2(M')$. It is clear that $\bar{K}$ extends $K_{R_H}$,
which is defined on $C_0^{\infty}(M^+_{R_H})$, and hence the form domain of
$\hat{\bar{K}}$ extends that of $\hat{K}_{R_H}$. However, $K_{R_H}$ is in general
not essentially self-adjoint, so it is not obvious if $\hat{\bar{K}}=\hat{K}_{R_H}$.
We will now prove that this is in fact the case, starting with a lemma:
\begin{lemma}\label{Lem_Cutoff}
For every $\epsilon>0$ there is a $\chi_{\epsilon}\in C^{\infty}(0,\infty)$ such
that $\chi\equiv 1$ near $0$, $\chi\equiv0$ near $[\epsilon,\infty)$,
$\chi'_{\epsilon}\le 0$ everywhere and
\[
\int_0^{\infty}r\chi'_{\epsilon}(r)^2dr\le\epsilon.
\]
\end{lemma}
\begin{proof*}
For any $\delta\in(0,\frac12)$ we may choose a $\chi_{\delta}\in C_0^{\infty}(0,\infty)$
with support in $(0,1)$ and such that $0\le\chi_{\delta}\le 1$ everywhere and
$\chi_{\delta}\equiv 1$ on $[\delta,1-\delta]$. We set
$\tilde{\chi}_{\delta}(r):=\sqrt{2}\left(\frac{r}{\epsilon}\right)^{\epsilon-1}
\chi_{\delta}\left(\frac{r}{\epsilon}\right)$, so
$\tilde{\chi}_{\delta}\in C_0^{\infty}(0,\infty)$ has support in $(0,\epsilon)$ and
$\tilde{\chi}_{\delta}\ge 0$ everywhere. We now use a simple substitution to estimate:
\begin{eqnarray}
\int_0^{\infty}r\tilde{\chi}_{\delta}(r)^2dr&=&
\int_0^{\infty}2\epsilon^2 r^{2\epsilon-1}\chi_{\delta}(r)^2dr
\le\int_0^12\epsilon^2 r^{2\epsilon-1}dr=\epsilon,\nonumber\\
c(\chi_{\delta})&:=&\int_0^{\infty}\tilde{\chi}_{\delta}(r)dr
=\int_0^{\infty}\sqrt{2}\epsilon r^{\epsilon-1}\chi_{\delta}(r)dr\nonumber\\
&\ge&\int_{\delta}^{1-\delta}\sqrt{2}\epsilon r^{\epsilon-1}dr
=\sqrt{2}\left((1-\delta)^{\epsilon}-\delta^{\epsilon}\right).\nonumber
\end{eqnarray}
As $\lim_{\delta\to 0^+}\sqrt{2}\left((1-\delta)^{\epsilon}-\delta^{\epsilon}\right)=\sqrt{2}$
we can choose $\delta>0$ and $\chi_{\delta}$ such that $c(\chi_{\delta})\ge 1$.
Then $\chi_{\epsilon}(r):=\frac{1}{c(\chi_{\delta})}\int_r^{\infty}\tilde{\chi}_{\delta}(s)ds$
has all the desired properties.
\end{proof*}

\begin{proposition}\label{Prop_Friedrichs}
$\hat{\bar{K}}=\hat{K}_{R_H}$ and consequently $\bar{G}=G_{R_H}$.
\end{proposition}
\begin{proof*}
We need to show that $\hat{K}_{R_H}$ extends $\hat{\bar{K}}$ as a quadratic
form, since the converse is clear. In particular, we need to show
that for every $f\in C_0^{\infty}(M')$, which is a form core of $\hat{\bar{K}}$,
there is a sequence $f_n\in C_0^{\infty}(M'\setminus\mathcal{B})$ such that
$f=\lim_{n\to\infty}f_n$ and
$\lim_{m,n\to\infty}\langle f_m-f_n,K_{R_H}(f_m-f_n)\rangle=0$. This entails
that
\[
\langle f,\hat{K}_{R_H}f\rangle=\lim_{m,n\to\infty}\langle f_m,K_{R_H}f_n\rangle
=\lim_{m,n\to\infty}\langle f_m,\bar{K}f_n\rangle=\langle f,\hat{\bar{K}}f\rangle,
\]
so $\hat{K}_{R_H}=\hat{\bar{K}}$ on the form core of the latter and the
desired extension property follows. Equivalently, we need to show that
$f=\lim_{n\to\infty}f_n$ and
$\hat{K}_{R_H}^{\frac12}f=\lim_{n\to\infty}\hat{K}_{R_H}^{\frac12}f_n$.
The linearity of this problem allows us to use a partition of unity
argument, so it suffices to consider $f$ supported in a region $U$, where $U$
ranges over a set of coordinate neighbourhoods which cover $M'$. We may assume
that $U$ contains some point on $\mathcal{B}$, otherwise the claim is trivial.

Near any point $p\in\mathcal{B}$ we can find a coordinate neighbourhood
$U$ on which we can choose local coordinates $(X,Y,x^i)$ as in the proof
of Lemma \ref{Lem_HawkingR}, and we denote the corresponding polar
coordinates by $(\tau,r,x^i)$. For any $n\in\mathbb{N}$ we fix a function
$\chi_{n^{-1}}$ as in Lemma \ref{Lem_Cutoff} with $\epsilon=n^{-1}$ and
we note that the function $\chi_{n^{-1}}(r)$ on $U$ is smooth even at
$r=0$, because $\chi_{n^{-1}}\equiv 1$ near $r=0$. For any
$f\in C_0^{\infty}(U)$ we may now define
\[
f_n(X,Y,x^i):=(1-\chi_{n^{-1}}(r))f(X,Y,x^i).
\]
We note that $f_n\in C_0^{\infty}(U\setminus\mathcal{B})$, because the
first factor vanishes near $\mathcal{B}$ and the second has compact
support in $U$. As $n$ increases, the support of $\chi_{n^{-1}}(r)$
shrinks in the radial direction, but its values remain uniformly bounded
by $1$. From this it follows that
\[
\lim_{n\to\infty}f-f_n=0,\qquad
\lim_{n\to\infty}\sqrt{V}(f-f_n)=0,\qquad
\lim_{n\to\infty}\partial_{x^i}(f-f_n)=0.
\]
Using the polar coordinates we have
\[
|v|^{-1}\partial_{\tau}f=\frac{1}{R_H}\left(-\frac{Y}{|v|}\partial_Xf+\frac{X}{|v|}\partial_Yf\right)
\]
away from $\mathcal{B}$. Since $\lim_{r\to0^+}\frac{r}{|v|}=\kappa^{-1}$
and $r^{-1}X$ and $r^{-1}Y$ are bounded we see that
$|v|^{-1}\partial_{\tau}f$ remains bounded on $U$ and hence
\[
\lim_{n\to\infty}\partial_{\tau}(f-f_n)=0
\]
as before. Also
\[
\partial_rf=\frac{X}{r}\partial_Xf+\frac{Y}{r}\partial_Yf
\]
remains bounded on $U$, but now the derivative of $\chi_{n^{-1}}$
enters in the limit
\[
\lim_{n\to\infty}\partial_r(f-f_n)=
\lim_{n\to\infty}\chi'_{n^{-1}}f+\chi_{n^{-1}}\partial_rf.
\]
The second term vanishes in the limit, by the boundedness of
$\partial_rf$ and the shrinking support of $\chi_{n^{-1}}$. The
first term also vanishes in the limit, due to the fact that the
integration measure satisfies $\sqrt{\det (\bar{g}_{R_H})_{\mu\nu}}\le Cr$
for some $C>0$ on the compact support of $f$ and
$\int r\chi'_{n^{-1}}(r)^2\le n^{-1}$ by Lemma \ref{Lem_Cutoff}.

Putting all this into the definition of $\bar{K}$ and $\hat{\bar{K}}$
we see that $\lim_{n\to\infty}(f-f_n)=0$ and also
$\lim_{n\to\infty}\hat{\bar{K}}^{\frac12}(f-f_n)=0$. Using the
positivity of $\bar{K}$ we can then estimate
\begin{eqnarray}
\|\hat{K}_{R_H}^{\frac12}(f_n-f_m)\|^2&=&
\langle (f_n-f)-(f_m-f),\bar{K}((f_n-f)-(f_m-f))\rangle\nonumber\\
&\le&2\|\hat{\bar{K}}^{\frac12}(f_n-f)\|^2+2\|\hat{\bar{K}}^{\frac12}(f_m-f)\|^2.\nonumber
\end{eqnarray}
The right-hand side vanishes in the limit $m,n\to\infty$, showing that
$f$ is indeed in the domain of $\hat{K}_{R_H}^{\frac12}$. It follows
that $\hat{K}_{H_R}=\hat{\bar{K}}$ and $\bar{G}=G_{R_H}$.
\end{proof*}

Recall that the imaginary time reflection
$R_{\tau}:(\tau,x)\mapsto(-\tau,x)$ is a diffeomorphism of $M^+_{R_H}$,
which is isometric. Near $\mathcal{B}$ we can express $R_{\tau}$ in terms
of the coordinates $(X,Y,x^i)$ introduced in the proof of Lemma
\ref{Lem_HawkingR}, where it is given by a reflection of $Y$. It is then
easy to see that $R_{\tau}$ extends in a unique way to a diffeomorphism
of $M'$, which remains isometric. Furthermore, it leaves the hypersurface
$\overline{\mu}(\Sigma)$ pointwise fixed, whereas it sends the normal
derivative $n^{\mu}$ of this hypersurface to $-n^{\mu}$. By definition of
Gaussian normal coordinates near $\overline{\mu}(\Sigma)$ it follows that
$R_{\tau}$ is given locally by a reflection in the Gaussian normal
coordinate.

Using $\bar{G}$ on $M'$ we will now prove that it gives rise to a pure Hadamard
state $\omega^{HHI}$ on the Lorentzian side, which restricts to the double
$\beta$-KMS state $\omega^{(\beta),d}$ of Theorem \ref{Thm_DoubleKMS} in
the exterior regions. This is our main result:
\begin{theorem}\label{Thm_Main}
Consider a spacetime $M$ with a static bifurcate Killing horizon, a wedge reflection
and a globally constant surface gravity $\kappa$ and let $\Sigma$ be a Cauchy surface
as in Definition \ref{Def_StaticBHSpac}. There exists a unique state $\omega^{HHI}$
on the Weyl algebra of $M$ with a Hadamard two-point distribution $\omega^{HHI}_2$
extending $\omega^{(\beta),d}_2$. This state is pure, quasi-free, Hadamard, invariant
under the Killing flow, it extends $\omega^{(\beta),d}$ and $\omega^{HHI}_2$ is
determined by the initial data
\begin{eqnarray}
\omega^{HHI}_{2,00}(f,f')&=&\lim_{\tau\rightarrow 0^+}
\bar{G}(-\tau,\overline{\mu}_*f;\tau,\overline{\mu}_*f')\nonumber\\
\omega^{HHI}_{2,10}(f,f')&=&i\lim_{\tau\rightarrow 0^+}
\partial_{(x')^0}\bar{G}(-\tau,\overline{\mu}_*f;\tau,\overline{\mu}_*f')\nonumber\\
\omega^{HHI}_{2,01}(f,f')&=&i\lim_{\tau\rightarrow 0^+}
\partial_{(y')^0}\bar{G}(-\tau,\overline{\mu}_*f;\tau,\overline{\mu}_*f')\nonumber\\
\omega^{HHI}_{2,11}(f,f')&=&-\lim_{\tau\rightarrow 0^+}
\partial_{(x')^0}\partial_{(y')^0}\bar{G}(-\tau,\overline{\mu}_*f;\tau,\overline{\mu}_*f')\nonumber
\end{eqnarray}
on the Cauchy surface $\Sigma$, for all $f,f'\in C_0^{\infty}(\Sigma)$.
\end{theorem}
\begin{proof*}
Let $f,f'\in C_0^{\infty}(\Sigma)$, both supported in a compact set $K\subset\Sigma$.
Let $\epsilon>0$ be such that the Gaussian normal coordinate $x^0$ is well-defined
for $|x^0|<\epsilon$ on a neighbourhood of $K$ and such that the Gaussian normal
coordinate $(x')^0$ is well-defined for $|(x')^0|<\epsilon$ on a neighbourhood of
$\overline{\mu}(K)\subset\overline{\mu}(\Sigma)$. We may then consider the Euclidean
Green's function, smeared with $f,f'$, in Gaussian normal coordinates,
$\bar{G}((x')^0,f;(y')^0,f')$. This defines a distribution in the variables
$(x')^0,(y')^0$. To see that the limits, which give the initial data of
$\omega^{HHI}_2$, are well-defined, we note that $\bar{G}$ is smooth away from the
diagonal of $(M')^{\times 2}$, whereas its singularities on the diagonal are
characterised by the Hadamard construction. Hence, $\bar{G}((x')^0,f;(y')^0,f')$ is
smooth on $(x')^0\not=(y')^0$ and the limits are well-defined by Theorem
\ref{Thm_ContFe}.

The data $\omega^{HHI}_{2,ij}$ define a unique distributional bi-solution to
the Klein-Gordon equation by Equation (\ref{Eqn_InitValState}). These data
are smooth away from the diagonal of $\Sigma^{\times 2}$, whereas the
singularities on the diagonal coincide with those of a Hadamard state, due
to Propositions \ref{Prop_ACHadCoeff} and \ref{Prop_InfACHad}. This shows
that $\omega^{HHI}_2$ has the correct singularity structure. Note that
Theorem \ref{Thm_ContFe} also implies that the anti-symmetric part of
$\omega^{HHI}_2$ is the canonical commutator $\frac{i}{2}E$.

To prove that $\omega^{HHI}_2$ is of positive type we use the reflection
positivity of $\bar{G}$ (Proposition \ref{Prop_ReflPos}), now with the
reflection in the Gaussian normal coordinates (see the comments above
this theorem). For any
test-function $\chi\in C_0^{\infty}((0,\epsilon),\mathbb{R})$ we may set
$F(x'):=\chi((x')^0)f_1((x')^j)-i\chi'((x')^0)f_0((x')^j)$ and use the
reflection positivity of $\bar{G}$ to deduce that
\[
\bar{G}(\overline{F}(-(x')^0,(x')^j),F(x'))\ge 0.
\]
Letting $\chi$ approach a $\delta$-distribution at some
$\tau\in(0,\epsilon)$ leads to
\[
\bar{G}(\delta((x')^0+\tau)\overline{f_1}-i\partial_{(x')^0}\delta((x')^0+\tau)\overline{f_0},
\delta((y')^0-\tau)f_1-i\partial_{(y')^0}\delta((y')^0-\tau)f_0)
\ge 0,
\]
where the change of sign in the factor $i$ due to complex conjugation is
cancelled by the change of sign due to the derivative in the reflected
Gaussian normal coordinate. Taking the limit of the final estimate as
$\tau\rightarrow 0^+$ and using the definition of the initial data
of $\omega^{HHI}_2$ in Equation (\ref{Eqn_InitValState}) yields the
desired positivity.

Since $\omega^{HHI}_2$ is a Hadamard two-point distribution, it defines a
unique quasi-free Hadamard state $\omega^{HHI}$ on $M$. Its restriction to
the union of the exterior wedges is $\omega^{(\beta),d}$, because
$\bar{G}=G_{R_H}$ yield the same initial data on
$(\Sigma\setminus\mathcal{B})^{\times 2}$. To see why the infinitesimal
Wick rotation and the actual Wick rotation yield the same result, it
suffices to note that the normal and the Killing time derivatives on
$\Sigma$ are related by the smooth factor $v$, which is non-zero away
from the bifurcation surface.

Note that $\omega^{HHI}_2$ is the only Hadamard extension to $M$ of
$\omega^{(\beta),d}_2$. Because $\omega^{(\beta),d}_2$ is invariant
under the Killing flow, the same must be true for $\omega^{HHI}_2$ and
hence also for $\omega^{HHI}$.

The fact that $\omega^{HHI}$ is pure follows from the fact that
$\omega^{(\beta),d}$ is pure (Theorem \ref{Thm_DoubleKMS}) and
Proposition \ref{Prop_pure} below. The uniqueness of $\omega^{HHI}$
follows from its purity and the uniqueness of the Hadamard extension of
$\omega^{(\beta),d}_2$, by a result of Kay \cite{Kay1993}.
\end{proof*}

\begin{proposition}\label{Prop_pure}
Let $\omega$ be a quasi-free Hadamard state on a globally hyperbolic spacetime
$M$ and let $(p,\mathcal{H})$ be the one-particle structure of its two-point
distribution $\omega_2$. Let $\Sigma\subset M$ be a Cauchy surface and let
$\mathcal{B}\subset\Sigma$ be a submanifold of codimension at least 1. Then $p$
already has a dense range on $C_0^{\infty}(D(\Sigma\setminus\mathcal{B}))$,
where $D$ denotes the domain of dependence. Consequently, if the restriction
of $\omega$ to $D(\Sigma\setminus\mathcal{B})$ is pure, then $\omega$ itself
is pure.
\end{proposition}
\begin{proof*}
We can consider the initial value formulation and replace $p$ by the
continuous linear map $q:C_0^{\infty}(\Sigma)^{\oplus 2}\rightarrow\mathcal{H}$.
We denote by $\mathcal{H}'$ the closed range of $q$ on
$C_0^{\infty}(\Sigma\setminus\mathcal{B})^{\oplus 2}$ and by $\mathcal{H}^0$
the closed range of $q$ on $C_0^{\infty}(\Sigma)\oplus\{0\}$. We let $P$ denote
the orthogonal projection of $\mathcal{H}$ onto $\mathcal{H}'$ and we introduce
a real-linear isometric involution $C$ on $\mathcal{H}$ which acts as the
identity on $(\mathcal{H}^0)^{\perp}$ and which is defined on $\mathcal{H}^0$
by continuous linear extension of $Cq(f_0,0):=q(\overline{f_0},0)$. Note that
this is indeed isometric, because $\omega_{2,11}$ is symmetric. For any
$\psi\in\mathcal{H}$ and $\chi\in\mathcal{H}'$ we have
$\|\psi-\chi\|=\|C\psi-C\chi\|$. Since $\chi=P\psi$ minimises this norm we
must have $CP\psi=PC\psi$, i.e.\ $CP=PC$.

We claim that the range of $q$ on $\{0\}\oplus C_0^{\infty}(\Sigma)$ is
entirely contained in the subspace $\mathcal{H}'$. To see why this is so we
choose $f_1\in C_0^{\infty}(\Sigma)$ and we let
$\chi_n\in C_0^{\infty}(\Sigma,\mathbb{R})$ be a sequence of test-functions
which remain uniformly bounded and such that each $\chi_n\equiv 1$ on a
neighbourhood of $\mathrm{supp}(f_1)\cap\mathcal{B}$, but such that the
support of $\chi_n$ shrinks towards $\mathcal{B}$ as $n\rightarrow\infty$. Then
the sequence $(1-\chi_n)f\in C_0^{\infty}(\Sigma\setminus\mathcal{B})$ is also
uniformly bounded and it converges pointwise to $f$ almost everywhere. We now
use the fact that $\|q(0,\chi_nf_1)\|^2=\omega_{2,00}(\chi_nf_1,\chi_nf_1)$.
Because $\omega_2$ is Hadamard, $\omega_{2,00}$ is given by a locally
integrable function on $\Sigma^{\times 2}$ (cf.\ Theorem \ref{Thm_ContF}). It
therefore follows from the Dominated Convergence Theorem that
$q(0,\chi_nf_1)\rightarrow0$ as $n\rightarrow\infty$ and hence
$q(0,(1-\chi_n)f_1)\rightarrow q(0,f_1)$. Because the latter sequence remains
in $\mathcal{H}'$, the limit must also lie in this subspace, which proves the
claim.

Now we define an $\mathcal{H}'$-valued distribution on $M$ by $p'(f):=Pp(f)$.
We will show that $(p',\mathcal{H}')$ is a one-particle structure on $M$. The
distribution $p'$ satisfies the Klein-Gordon equation, just like $p$, and it
has a dense range. To establish the commutator property we use the initial
value formulation $q':C_0^{\infty}(\Sigma)^{\oplus 2}\rightarrow\mathcal{H}'$
of $p'$. For any $f,f'\in C_0^{\infty}(M)$ let
$(f_0,f_1,f'_0,f'_1)\in C_0^{\infty}(\Sigma)$ denote the corresponding initial
data. Then,
\[
q'(f_0,f_1)=Pq(f_0,0)+q(0,f_1)
\]
since $Pq(0,f_1)=q(0,f_1)$ by the previous paragraph. By the same token and
the commutator property of $p$:
\begin{eqnarray}
&&\langle q'(\overline{f_0},\overline{f_1}),q'(f'_0,f'_1)\rangle-
\langle q'(\overline{f'_0},\overline{f'_1}),q'(f_0,f_1)\rangle\nonumber\\
&=&iE(f,f')+\langle q(\overline{f'_0},0),(I-P)q(f_0,0)\rangle
-\langle q(\overline{f_0},0),(I-P)q(f'_0,0)\rangle\nonumber\\
&=&iE(f,f')+\langle q(\overline{f'_0},0),(I-P)q(f_0,0)\rangle
-\langle C(I-P)q(f'_0,0),Cq(\overline{f_0},0)\rangle\nonumber\\
&=&iE(f,f')+\langle q(\overline{f'_0},0),(I-P)q(f_0,0)\rangle
-\langle q(\overline{f'_0},0),(I-P)q(f_0,0)\rangle\nonumber\\
&=&iE(f,f').\nonumber
\end{eqnarray}
Hence, $p'$ has the desired commutator property and $(p',\mathcal{H}')$ is a
one-particle structure on $M$. Note that the two-point distribution
$\omega'_2$ corresponding to $p'$ coincides with $\omega_2$ on
$D(\Sigma\setminus\mathcal{B})$ and that it is Hadamard, because the
estimate $\omega'_2(\overline{f},f)\le\omega_2(\overline{f},f)$ allows us
to estimate the wave front set of the Hilbert space-valued distribution
$p'$ as in \cite{Strohmaier+2002}. It follows that the initial data of
$\omega_2-\omega'_2$ are smooth and that they vanish on the dense subset
$(\Sigma\setminus\mathcal{B})^{\times 2}$ of $\Sigma^{\times 2}$. They must
therefore vanish everywhere and $\omega_2=\omega'_2$. This implies that
$P=I$ and in particular $\mathcal{H}=\mathcal{H}'$.

Due to Lemma A.2 of \cite{Kay+1991}, a quasi-free state $\omega$ is pure
if and only if the one-particle structure $(p,\mathcal{H})$ of its two-point
distribution is such that $p$ already has a dense range on the real-valued
test-functions. Combining this criterion with the results we have just shown
and the assumption that the restriction of $\omega$ to
$D(\Sigma\setminus\mathcal{B})$ is pure, we see that $p$ already has a dense
range on $C_0^{\infty}(D(\Sigma\setminus\mathcal{B}),\mathbb{R})$, so that
$\omega$ is pure.
\end{proof*}

In the case where $M$ is Minkowski spacetime and $M^+$ the Rindler wedge,
the Hartle-Hawking-Israel state is well-known to be the Minkowski vacuum
(cf.\ \cite{Unruh1976}).

We have already seen in Theorem \ref{Thm_Main} that $\omega^{HHI}$ is
Hadamard and invariant under the Killing flow. It is also invariant under
the wedge reflection, in the following sense. Just as in Section
\ref{SSec_DoubleKMS} one may use the wedge reflection $I$ on $M$ to define a
complex anti-linear involution $\tau_I$ of the Weyl algebra of the entire
spacetime by setting $\tau_I(zW(f))=\overline{z}W(I^*f)$. One may then show
that $\omega^{HHI}$ is invariant under $\tau_I$ in the conjugate linear sense
of Equation (\ref{Eqn_WedgeInv}). This is because the restriction of
$\omega^{HHI}$ to $M^+\cup M^-$ is a double $\beta_H$-KMS state, which is
necessarily $\tau_I$-invariant. That this invariance extends across the
bifurcation surface $\mathcal{B}$ follows from the Hadamard property, because
the Hadamard two-point distributions $\omega^{HHI}_2(x,y)$ and
$\omega^{HHI}_2(I(y),I(x))$ differ by a smooth function and their initial
data are equal on the dense set $(\Sigma\setminus\mathcal{B})^{\times 2}$, so
they must be equal everywhere.

Any other Hadamard two-point distribution $\omega_2$ is of the form
$\omega_2=w_2+\omega^{HHI}_2$, where $w_2$ is a smooth, real-valued bi-solution
to the Klein-Gordon equation. If $\omega_2$ is invariant under the Killing flow,
then so is $w_2$. In particular, we can consider two points $x,y$ on one of the
Killing horizons $\mathfrak{h}_A$ of $M$, as was done in \cite{Kay+1991}. These
points are most conveniently expressed in terms of local coordinates $x^i,y^i$
on $\mathcal{B}$ and an affine coordinate $x^U$ along the lightlike geodesics,
starting at $\mathcal{B}$, that generate $\mathfrak{h}_A$. Using the invariance
of $w_2$ and these coordinates we find
\begin{equation}\label{Eqn_KWUnique}
w_2|_{\mathfrak{h}_A^{\times 2}}(x^U,x^i;y^U,y^i)=w_2|_{\mathcal{B}^{\times 2}}(x^i,y^i),
\end{equation}
because we can exploit the Killing flow to simultaneously transport $x=(x^U,x^i)$
and $y=(y^U,y^i)$ to  $\tilde{x}=(0,x^i)$ and $\tilde{y}=(0,y^i)$, respectively.

The Equality (\ref{Eqn_KWUnique}) is directly related to the uniqueness result
found by Kay and Wald \cite{Kay+1991}. Because they restricted attention to
observables that are generated by $U$-derivatives of compactly supported data
on the horizon $\mathfrak{h}_A$, the term involving $w_2$ does not contribute
to their expectation value in the two-point distribution $\omega_2$. On this
subalgebra, the state $\omega^{HHI}$ can therefore be characterised uniquely by
its Killing field invariance, dropping the assumption that it restricts to a
double $\beta_H$-KMS state. This uniqueness claim is interesting for physical
investigations involving phenomena near the horizon, but it is not clear under
what circumstances it extends to a similar uniqueness claim on the entire Weyl
algebra $\mathcal{A}$. That question would involve a more detailed analysis of
$w_2$ on the entire manifold $M^{\times 2}$ and of the circumstances under which
$w_2$ must vanish.

\section{Discussion}\label{Sec_Concl}

In this final section, we comment on some aspects of our result and on the
possibilities of generalising it.

Let us first note that we have made rather few assumptions about the future and
past regions of the spacetime. Indeed, the Wick rotation and wedge reflection only
require information about the left and right wedge regions and an arbitrarily small
neighbourhood of the bifurcation surface. The future and past regions are not part
of any of the complexified or Riemannian manifolds that we considered. For this
reason, the Wick rotation does not provide any direct information about the
behaviour of the state in the future or past regions. Instead, we have obtained this
information indirectly, using the Cauchy problem and causal propagation. The only
assumptions that we have made about the future and past regions is the existence of
the Killing field $\xi^a$. This assumption was only necessary to formulate the
Killing field invariance of $\omega^{HHI}$ on the entire spacetime $M$. For the
main part of the construction it seems sufficient if $\xi^a$ is only defined on the
exterior wedge regions and a neighbourhood of the bifurcation surface.

The determination of $\omega^{HHI}$ from initial data works very well for a free
field, but it is doubtful that it extends to interacting fields, whose restriction
to a Cauchy surface may not be well-defined. It is therefore unlikely that our
proof can be extended to such interacting fields and a proof of the conjecture of
\cite{Jacobson1994} would presumably require different (or additional) methods.
As a first step one might investigate whether our results can be generalised to
perturbatively interacting fields, e.g.\ using the ideas of \cite{Gibbons+1978_2}.

We have shown the existence of a Hadamard extension $\omega^{HHI}$ of a double
$\beta_H$-KMS state in the static case using a Wick rotation. In the more general,
stationary case this method of proof is no longer available, but the result could
still be true. Under what circumstances, if any, a Hadamard extension of a double
$\beta_H$-KMS state exists is at present unclear. In the even more general case
of Kerr spacetime the non-existence of a state which is invariant under the flow
of the Killing field that generates the horizon is known to follow from a
certain superradiance property, which is expected to hold \cite{Kay+1991}.

\section*{Acknowledgements}

I am indebted to Bob Wald for suggesting the problem of rigorously constructing
Hartle-Hawking-Israel states and for various helpful discussions and comments along
the way. I am also grateful to Rainer Verch, Stefan Hollands, Marc Casals and Giovanni
Collini for discussions, comments and questions. Most of the research reported on in
this paper was conducted while I was a postdoctoral researcher at the University of
Chicago and it has been presented at the joint $20^{\mathrm{th}}$ International
Conference on General Relativity and Gravitation and the $10^{\mathrm{th}}$ Amaldi
Conference on Gravitational Waves in Warsaw, 2013.

\appendix

\section{A technical lemma}\label{App}

\begin{lemma}\label{Lem_App}
Let $U$ be a convex normal neighbourhood in a Riemannian manifold $(\Sigma,h_{ij})$
of dimension $d-1$ and let $\sigma_U$ be half the squared geodesic distance on $U$.
For $d\ge 2$ we let
\[
C(d):=-(4\pi)^{-\frac{d}{2}}2^{\frac{d}{2}-1}\Gamma\left(\frac{d}{2}\right)
\]
and we use the distribution $\delta(f):=\int_U f|_{\Delta}$ on $U^{\times 2}$, where
$f|_{\Delta}(x):=f(x,x)$ is the restriction to the diagonal. Then
\[
\lim_{\tau\rightarrow 0^+}\tau\left(\sigma_U+\frac12\tau^2\right)^{-\frac{d}{2}}
=\frac{-1}{2C(d)}\delta
\]
for all continuous $f\in C_0^0(U^{\times 2})$.
\end{lemma}
Note that $C(d)=\left(1-\frac{d}{2}\right)c_0+d_0$, with $c_0$ and $d_0$ as in
Proposition \ref{Prop_FkEucl}.
\begin{proof*}
For $d\ge 2$ we will need the following identities:
\begin{eqnarray}
\mathrm{vol}(\mathbb{S}^{d-2})&=&2\pi^{\frac{d-1}{2}}\Gamma\left(\frac{d-1}{2}\right)^{-1}\nonumber\\
X(d)&:=&\int_0^{\infty}r^{d-2}(r^2+1)^{-\frac{d}{2}}dr\ =\
\frac{\sqrt{\pi}}{2}\Gamma\left(\frac{d-1}{2}\right)
\Gamma\left(\frac{d}{2}\right)^{-1}.\nonumber
\end{eqnarray}
The first identity is a standard result, which is proved by expressing
$\pi^{\frac{d-1}{2}}$ as a Gaussian integral and changing to polar coordinates.
To determine $X(d)$ one may show by partial integration that
$X(d+2)=\frac{d-1}{d}X(d)$. From a direct computation one finds $X(2)=\frac{\pi}{2}$
and $X(3)=1$. The result then follows from a proof by induction.

Let $h(y):=\det h_{ij}(y)$ in Gaussian normal coordinates and
$\tilde{h}(v,y):=\det h_{ij}(v)$ in Riemannian normal coordinates centred
on $y$, so that $\tilde{h}(0,y)=1$. Then we may compute
\begin{eqnarray}
\lim_{\tau\rightarrow 0^+}\tau\left(\sigma_U+\frac12\tau^2\right)^{-\frac{d}{2}}(f)
&=&\lim_{\tau\rightarrow 0^+}2^{\frac{d}{2}}
\int dy\ dv\ \sqrt{h}(y)\sqrt{\tilde{h}}(v,y)
f(\exp_y(v),y)\tau(|v|^2+\tau^2)^{-\frac{d}{2}}\nonumber\\
&=&\lim_{\tau\rightarrow 0^+}2^{\frac{d}{2}}
\int dy\ dv\ \sqrt{h}(y)\sqrt{\tilde{h}}(\tau v,y)
f(\exp_y(\tau v),y)(|v|^2+1)^{-\frac{d}{2}}\nonumber\\
&=&2^{\frac{d}{2}}\int dv\ (|v|^2+1)^{-\frac{d}{2}}
\int dy\ \sqrt{h}(y)f(y,y)\nonumber\\
&=&2^{\frac{d}{2}}\mathrm{vol}(\mathbb{S}^{d-2})X(d)\delta(f).\nonumber
\end{eqnarray}
Inserting the formulae for $X(d)$ and $\mathrm{vol}(\mathbb{S}^{d-2})$ we find the
desired result.
\end{proof*}

\end{document}